\documentclass{article}

\usepackage{natbib}
\usepackage[pagebackref,breaklinks]{hyperref} %
\usepackage{url}

\usepackage{amsmath,amsfonts,bm}

\newcommand{\mypar}{\textbf}

\def\eqref#1{equation~\ref{#1}}

\def\1{\bm{1}}

\DeclareMathAlphabet{\mathsfit}{\encodingdefault}{\sfdefault}{m}{sl}
\SetMathAlphabet{\mathsfit}{bold}{\encodingdefault}{\sfdefault}{bx}{n}

\usepackage[capitalize,nameinlink]{cleveref}

\usepackage[utf8]{inputenc} %
\usepackage[T1]{fontenc}    %

\usepackage[dvipsnames,table]{xcolor} %

\usepackage{graphicx}       %
\usepackage{booktabs}       %
\usepackage{multicol}       %
\usepackage{multirow}       %
\usepackage{makecell}       %
\usepackage{colortbl}       %
\usepackage{caption}        %
\usepackage{subcaption}     %

\usepackage{wrapfig}        %
\usepackage{adjustbox}      %
\usepackage{titletoc}       %

\usepackage{mathtools}
\usepackage{amsmath,amsthm} %
\usepackage{amsfonts}       %
\usepackage{amssymb}        %
\usepackage{nicefrac}       %
\usepackage{microtype}      %
\usepackage{relsize}

\usepackage{comment}        %
\usepackage{array}

\usepackage{doi}            %
\usepackage[normalem]{ulem} %
\usepackage{placeins}       %
\usepackage[detect-all]{siunitx} %
\usepackage[all]{nowidow}   %

\usepackage{bbding}         %
\definecolor{cold}{HTML}{008acd}
\def\frozen{\raisebox{-.15ex}{\color{cold}\SnowflakeChevron}}
\DeclareUnicodeCharacter{2744}{\frozen}%

\usepackage{pifont}         %

\definecolor{vlightgray}{gray}{0.87}

\definecolor{tblue}{HTML}{1F77B4}
\definecolor{torange}{HTML}{FF7F0E}
\definecolor{tgreen}{HTML}{2CA02C}
\definecolor{tred}{HTML}{FF0000}

\definecolor{linkcolor}{HTML}{991408}  %
\definecolor{citecolor}{HTML}{2E7E2A}  %
\definecolor{filecolor}{HTML}{131877}  %
\definecolor{menucolor}{HTML}{727500}  %
\definecolor{runcolor} {HTML}{137776}  %
\definecolor{urlcolor} {HTML}{0a2bbf}  %
\definecolor{linkblue}{HTML}{0057B8}%

\makeatletter
\DeclareRobustCommand*{\corrmark@}{%
  \begingroup
  \setbox\z@\hbox{$\m@th\ast$}%
  \setbox\tw@\hbox{1}%
  \resizebox{!}{\ht\tw@}{\raisebox{\dp\z@}{\usebox\z@}}%
  \endgroup}
\newcommand*{\corrmark}{{}\corrmark@}
\makeatother

\newcommand{\appref}[1]{\hyperref[#1]{Appendix~\ref*{#1}}}%

\renewcommand*{\appref}[1]{\hyperref[#1]{Appx.~\ref*{#1}}}

\def\Snospace~{\S{}}%

\newcommand{\best}[1]{\ifmmode\mathbf{#1}\else\textbf{#1}\fi}
\newcommand{\secbest}[1]{\underline{#1}}
\newcommand{\runnerup}[1]{\secbest{#1}}
\newcommand{\na}[1]{{\color{gray}N/A}}
\newcommand{\tabpm}[1]{{\color{gray}\,\tiny{$\pm$\,#1}}}

\newcommand{\hider}[1]{}

\hypersetup{colorlinks=true, allcolors=linkblue}

\usepackage{enumitem}

\usepackage{etoolbox}       %
\newtoggle{arxiv}%
\toggletrue{arxiv}%
\iftoggle{arxiv}{
  \usepackage[parfill]{parskip}

  \usepackage[%
    a4paper,
    inner=25mm,
    outer=25mm,%
    top=25mm,
    bottom=25mm,
    marginparsep=5mm,
    marginparwidth=40mm,
  ]{geometry}

  \newlength{\defbaselineskip}
  \setlength{\defbaselineskip}{\baselineskip}
  \setlength{\parskip}{6pt}%
  \setlength{\parindent}{0pt}%

  \RequirePackage[T1]{fontenc}
  \RequirePackage[tt=false, type1=true]{libertine}
  \RequirePackage[varqu]{zi4}
  \RequirePackage[libertine]{newtxmath}

  \usepackage{fancyhdr}
  \pagestyle{fancy}
  \fancyhead[L]{BIOTRAM}
  \fancyhead[R]{Boehly, et al. (2026)}

  \usepackage{titling}
  \predate{}%
  \postdate{\vspace{-0.1em}}%
  \setlength{\droptitle}{-6em}%

  \usepackage{authblk}%
  
  \renewcommand*{\Affilfont}{\small}
  \setlength{\affilsep}{0.5em} %
  \pretitle{\begin{flushleft}\LARGE}
  \posttitle{\par\end{flushleft}\vskip 0.5em}
  \preauthor{\begin{flushleft}}%
  \postauthor{\par\end{flushleft}}
}{
}

\newcommand{\AbZeroAhead}{four}

\newcommand{\AbTopAhead}{three}
\newcommand{\AbTopTied}{one}

\newcommand{\AbMidBucketsBehind}{three}
\newcommand{\AbundTopCutPct}{7}
\newcommand{\AbundTopCutVs}{the Transformer}
\newcommand{\AbundTopCutSig}{within seed noise}

\newcommand{\AbSeenAhead}{four}

\newcommand{\AbUnseenAhead}{one}
\newcommand{\AbUnseenTied}{one}
\newcommand{\AbUnseenBehind}{two}
\newcommand{\ExtBottles}{fifteen}

\newcommand{\ExtSpecimensL}{4,057}
\newcommand{\ExtBioJSDL}{0.219}
\newcommand{\ExtBioBetterL}{thirteen}

\newcommand{\ExtTubeJSDL}{0.224}

\newcommand{\ExtReadShareJSDL}{0.277}
\newcommand{\ExtUniformJSDL}{0.255}

\newcommand{\ExtGLMMJSDL}{0.203}

\newcommand{\ExtBioJSDH}{0.218}
\newcommand{\ExtBioBetterH}{twelve}

\newcommand{\ExtTubeJSDH}{0.217}

\newcommand{\ExtReadShareJSDH}{0.237}

\newcommand{\ExtGLMMJSDH}{0.197}

\newcommand{\ExtNewBins}{534}
\newcommand{\ExtVocab}{74,365}

\newcommand{\ExtBioSeeds}{twelve}
\newcommand{\BioJSD}{0.1194}

\newcommand{\UniformJSD}{0.3174}
\newcommand{\ReadsJSD}{0.3586}
\newcommand{\McLarenJSD}{0.3435}

\newcommand{\GLMMSpatialJSD}{0.2141}
\newcommand{\GLMMTemporalJSD}{0.2366}
\newcommand{\GLMMLaneJSD}{0.2267}

\newcommand{\GLMMVersion}{1.1.13}

\newcommand{\ExtGLMMMAEL}{4.15}
\newcommand{\ExtBioMAEL}{2.78}

\newcommand{\ExtGLMMPairPL}{0.069}

\newcommand{\ExtGLMMMAEH}{3.98}
\newcommand{\ExtBioMAEH}{2.54}

\newcommand{\ExtGLMMPairPH}{0.045}

\newcommand{\DeltaSpatial}{+0.0419}
\newcommand{\DeltaTemporal}{+0.0662}
\newcommand{\DeltaLane}{+0.0517}

\newcommand{\UnseenPct}{19}
\newcommand{\NBottles}{1,188}
\newcommand{\NBins}{132,184}
\newcommand{\NRows}{414,640}
\newcommand{\NSpecimens}{800,875}
\newcommand{\NFamilies}{1,283}
\newcommand{\MedBins}{288}
\newcommand{\MaxBins}{1,534}
\newcommand{\MedInd}{375}
\newcommand{\MaxInd}{16,900}
\newcommand{\NClean}{585}
\newcommand{\EtaSq}{80.4}
\newcommand{\PctSingleton}{60}
\newcommand{\PctZero}{50.9}
\newcommand{\PctAllThree}{46.8}
\newcommand{\PctOneOfThree}{29.4}
\newcommand{\Decades}{4.2}

\newcommand{\RpsEmptiest}{5,896}
\newcommand{\RpsFullest}{720}
\newcommand{\RpsRatio}{8}
\newcommand{\NTest}{89}
\newcommand{\NVal}{87}
\newcommand{\NTrain}{409}
\newcommand{\NExcess}{603}
\newcommand{\NExcessUsed}{455}

\newcommand{\RsqReads}{23}
\newcommand{\SpearReads}{0.49}
\newcommand{\AUCReads}{0.78}
\newcommand{\PctTopOne}{13.8}
\newcommand{\NSites}{578}
\newcommand{\YearFirst}{2014}
\newcommand{\YearLast}{2025}
\newcommand{\PctInsecta}{73.7\%}
\newcommand{\PctCollembola}{21.1\%}
\newcommand{\PctArachnida}{5.1\%}

\newcommand{\SeenMAE}{0.853}
\newcommand{\SeenMAEBase}{0.941}

\newcommand{\MatchedZeroGain}{13.5}
\newcommand{\MatchedNonZeroGain}{7.5}
\newcommand{\MatchedJSDGap}{-0.0044}

\newcommand{\MatchedMAEbtlGainPct}{5.8}

\newcommand{\TabICLParamPct}{35}
\newcommand{\TabICLCountRel}{level with}
\newcommand{\TabICLMAEGapPct}{0.6}

\newcommand{\ElastGenStackU}{1.00}

\newcommand{\ElastGenMergeU}{0.60}

\newcommand{\ElastGenFinalU}{0.40}
\newcommand{\ElastGenSeeds}{3}

\newcommand{\ElastReadsModuleU}{0.29}

\newcommand{\ElastReadsStackU}{0.71}

\newcommand{\ElastReadsMergeU}{0.40}

\newcommand{\ElastReadsFinalU}{0.31}

\newcommand{\JacOwnLo}{0.37}
\newcommand{\JacOwnHi}{0.43}
\newcommand{\JacCross}{0.002}
\newcommand{\JacRatioLo}{190}
\newcommand{\JacSeeds}{12}

\newcommand{\AblBlocks}{20}

\newcommand{\AblFullSeeds}{four}
\newcommand{\AblAnchorTestJSD}{0.1198}
\newcommand{\AblAnchorMAEbtl}{0.700}
\newcommand{\AblAnchorCountAUC}{0.9136}

\newcommand{\AblPrimerTestJSD}{+0.0127}

\newcommand{\AblExcessTestJSD}{+0.0117}

\newcommand{\AblBarcodeTestJSD}{+0.0014}
\newcommand{\AblBarcodeTestJSDP}{0.101}

\newcommand{\AblTaxonomyTestJSD}{+0.0034}
\newcommand{\AblTaxonomyTestJSDP}{0.016}

\newcommand{\AblRateTransformerTestJSD}{+0.0054}

\newcommand{\AblReadsInModuleTestJSD}{+0.0039}

\newcommand{\AblLearnedEncoderTestJSD}{+0.0085}

\newcommand{\SegScored}{631,161}

\newcommand{\SegWingedPct}{22}
\newcommand{\SegWingCarvedPct}{48}
\newcommand{\SegWingCarvedHiPct}{71}
\newcommand{\SegCarvedAllPct}{13}
\newcommand{\SegExAId}{LPGEQ49117-24}
\newcommand{\SegExAOrder}{Lepidoptera}
\newcommand{\SegExAEnv}{222.1}
\newcommand{\SegExABody}{12.6}
\newcommand{\SegExACarvedPct}{94}
\newcommand{\SegExBId}{CRBCA996-24}
\newcommand{\SegExBOrder}{Plecoptera}
\newcommand{\SegExBEnv}{14.5}
\newcommand{\SegExBBody}{2.4}
\newcommand{\SegExBCarvedPct}{83}
\newcommand{\ScatterSeed}{200}
\newcommand{\ScatterRows}{26,934}
\newcommand{\ScatterBottles}{89}

\newcommand{\ScatterSlopeBio}{0.95}
\newcommand{\ScatterRtwoBio}{0.89}
\newcommand{\ScatterSlopeRead}{0.86}
\newcommand{\ScatterRtwoRead}{0.70}

\newcommand{\CalibMidPIT}{0.520}
\newcommand{\CalibCov}{0.967}
\newcommand{\CalibNaivePIT}{0.804}
\newcommand{\CalibSeeds}{12}

\newcommand{\PctNoGenus}{76.9}
\newcommand{\PctNoSpecies}{87.5}

\newcommand{\OODSpatialGap}{-0.0090}
\newcommand{\OODSpatialDir}{ahead of}

\newcommand{\OODLaneGap}{-0.0047}
\newcommand{\OODLaneDir}{ahead of}

\newcommand{\OODTemporalGap}{-0.0099}
\newcommand{\OODTemporalDir}{ahead of}

\newcommand{\MechSumR}{0.627}
\newcommand{\MechSumSd}{0.007}
\newcommand{\MechFrR}{0.593}
\newcommand{\MechFrSd}{0.013}
\newcommand{\MechMassR}{0.444}
\newcommand{\MechMassSd}{0.025}

\newcommand{\MechSeeds}{12}
\newcommand{\MechTwinSumR}{0.461}
\newcommand{\MechTwinSumSd}{0.013}
\newcommand{\MechTwinFrR}{0.397}

\newcommand{\MechTwinMassR}{0.411}

\newcommand{\MechTwinSeeds}{12}

\newcommand{\TrackJSDGap}{+0.0036}

\newcommand{\SepGenMAEbtlAhead}{two}
\newcommand{\SepGenMAEbtlTied}{two}

\newcommand{\SepGenJSDAhead}{three}
\newcommand{\SepGenJSDTied}{one}

\newcommand{\SepReadsMAEbtlAhead}{four}

\newcommand{\SepReadsJSDAhead}{four}

\newcommand{\SepReadsSpearAhead}{two}
\newcommand{\SepReadsSpearTied}{two}

\newcommand{\SepReadsAUCTied}{three}

\newcommand{\NMatchedBaselines}{four}

\newcommand{\TunedNBaselines}{four}
\newcommand{\TunedBioJSD}{0.1200}
\newcommand{\TunedBioMAE}{0.699}
\newcommand{\TunedBestBaselineMAE}{0.723}
\newcommand{\TunedBestBaselineJSD}{0.1234}
\newcommand{\TunedMAEGainPct}{3.2}
\newcommand{\TunedJSDGap}{-0.0034}
\newcommand{\TunedZeroGain}{5.4}
\newcommand{\TunedNonZeroGain}{3.9}

\newcommand{\TunedSepSpearAhead}{one}
\newcommand{\TunedSepSpearTied}{three}

\newcommand{\TunedSepAUCBehind}{four}

\newcommand{\TunedTabICLCountRel}{level with}
\newcommand{\TunedTabICLMAEGapPct}{0.7}
\newcommand{\TunedTabICLMAE}{0.704}
\newcommand{\TunedTabICLCompRel}{ahead of}
\newcommand{\SegNImages}{639,751}
\newcommand{\SegNBins}{78,259}
\newcommand{\SizeImagedPct}{50.2}
\newcommand{\MassNSpecimens}{26,114}
\newcommand{\MassNBins}{13,287}
\newcommand{\AlloNGenera}{332}
\newcommand{\AlloBeta}{1.37}
\newcommand{\AlloBetaLo}{1.27}
\newcommand{\AlloBetaLoOrder}{Coleoptera}
\newcommand{\AlloBetaHi}{1.47}
\newcommand{\AlloBetaHiOrder}{Diptera}
\newcommand{\AlloHeldR}{0.815}
\newcommand{\AlloHeldRsq}{0.644}
\newcommand{\AlloHeldMAE}{0.90}

\newcommand{\AlloOrderMAE}{1.61}

\newcommand{\AlloGlobalMAE}{1.77}

\newcommand{\AreaMassR}{0.88}
\newcommand{\AreaMassNBins}{587}
\newcommand{\SeqAreaR}{0.86}
\newcommand{\SeqAreaRMSE}{0.84}
\newcommand{\SeqAreaNNR}{0.82}
\newcommand{\SeqAreaNNRMSE}{0.98}
\newcommand{\SeqAreaGenusR}{0.74}
\newcommand{\SeqAreaGenusRMSE}{1.12}
\newcommand{\SeqAreaGenusNNR}{0.70}
\newcommand{\SeqAreaGenusNNRMSE}{1.28}
\newcommand{\SeqSigmaRMax}{0.10}
\newcommand{\PrimerNGenomes}{10,780}
\newcommand{\PrimerNGenomeBins}{9,396}

\newcommand{\PrimerFMean}{0.64}
\newcommand{\PrimerRMean}{0.58}
\newcommand{\PrimerFPoorPct}{24}
\newcommand{\PrimerRPoorPct}{30}
\newcommand{\PrimerFRCorr}{0.15}
\newcommand{\PrimerFExactPct}{91.5}
\newcommand{\PrimerRExactPct}{97.3}

\newcommand{\PrimerMitoBins}{1,554}
\newcommand{\PrimerMitoPct}{1.2}
\newcommand{\PrimerBarcodeFPct}{91.5}
\newcommand{\PrimerBarcodeFIndPct}{97.9}

\newcommand{\PrimerRealFPct}{92.6}
\newcommand{\PrimerRealFIndPct}{99.4}
\newcommand{\PrimerImputedFPct}{7.4}
\newcommand{\PrimerImputedFIndPct}{0.6}
\newcommand{\PrimerImputedRPct}{98.8}
\newcommand{\PrimerAraneaeF}{0.34}
\newcommand{\PrimerAraneaePoorPct}{72}
\newcommand{\PrimerHymenopteraF}{0.43}
\newcommand{\PrimerHymenopteraPoorPct}{52}
\newcommand{\PrimerDipteraF}{0.57}

\newcommand{\PrimerLepidopteraF}{0.68}

\newcommand{\PrimerRpsR}{0.21}
\newcommand{\PrimerRpsNBins}{44,182}
\newcommand{\PrimerRpsRBoth}{0.33}
\newcommand{\PrimerRpsNBoth}{223}
\newcommand{\PrimerImpNTrain}{7,663}
\newcommand{\PrimerImpNTest}{1,508}
\newcommand{\PrimerImpFRsqPct}{23}
\newcommand{\PrimerImpFR}{0.49}
\newcommand{\PrimerImpFRho}{0.41}
\newcommand{\PrimerImpFMAE}{0.20}
\newcommand{\PrimerTaxFRsqPct}{26}
\newcommand{\PrimerImpRRsqPct}{16}
\newcommand{\PrimerImpRR}{0.41}
\newcommand{\PrimerImpRRho}{0.38}
\newcommand{\PrimerImpRMAE}{0.21}
\newcommand{\PrimerTaxRRsqPct}{9}
\newcommand{\PrimerImpRFamRsq}{0.17}
\newcommand{\PrimerTaxRFamRsq}{0.07}
\newcommand{\PrimerImpRSd}{0.09}
\newcommand{\PrimerMitoRSd}{0.26}
\newcommand{\BarcodeLenMedian}{653}
\newcommand{\AmpliconStart}{237}
\newcommand{\TabICLParams}{28.5M}
\newcommand{\TabICLScratchJSD}{0.180}
\newcommand{\TabICLPretrainedJSD}{0.128}
\newcommand{\TabICLAUC}{0.924}
\newcommand{\TabICLMAElog}{0.226}
\newcommand{\TabICLMAEZero}{0.249}

\newcommand{\XbinAttnJSD}{0.0033}
\newcommand{\XbinAttnSeeds}{6/6}
\newcommand{\XbinAttnP}{0.0067}
\newcommand{\XbinSelfMaskedJSD}{-0.0004}

\newcommand{\XbinSelfMaskedSeeds}{3/6}
\newcommand{\XbinSelfMaskedP}{1}

\newcommand{\XbinNoReadsSeeds}{3/6}
\newcommand{\XbinNoReadsP}{1}

\newcommand{\XbinNoDepthJSD}{-0.0020}
\newcommand{\XbinNoDepthSeeds}{6/6}
\newcommand{\XbinNoDepthP}{0.045}
\newcommand{\XbinDepthOnlyJSD}{-0.0001}
\newcommand{\XbinDepthOnlySeeds}{3/6}
\newcommand{\XbinDepthOnlyP}{1}
\newcommand{\LossWeightComp}{8}
\newcommand{\LossWeightPres}{0.15}
\newcommand{\LossWeightRps}{0.05}

\newcommand{\LossTuneModels}{five}

\newcommand{\LossTuneSearchRuns}{1{,}570}
\newcommand{\LossTuneSearchSeeds}{five}
\newcommand{\LossTuneConfirmSeeds}{eight}
\newcommand{\LossTuneConfigs}{61}
\newcommand{\LossTuneSpendLo}{61}
\newcommand{\LossTuneSpendHi}{65}
\newcommand{\LossTuneSpendBio}{62}
\newcommand{\LossTuneBoundaryMax}{four}
\newcommand{\LossTuneGridComp}{\{2, 8, 32\}}
\newcommand{\LossTuneGridPres}{\{0.0375, 0.15, 0.6\}}
\newcommand{\LossTuneGridRpo}{\{0.0125, 0.05, 0.2\}}
\newcommand{\LossTuneInteractN}{four}
\newcommand{\LossTuneBioJSD}{0.1200}
\newcommand{\LossTuneBaseJSD}{0.1234}
\newcommand{\LossTuneBaseJSDModel}{the MLP}
\newcommand{\LossTuneJSDP}{8.4{\times}10^{-7}}
\newcommand{\LossTuneBioMAE}{0.699}
\newcommand{\LossTuneBaseMAE}{0.723}
\newcommand{\LossTuneBaseMAEModel}{the Transformer}
\newcommand{\LossTuneMAEP}{0.025}
\newcommand{\LossTuneBioAUC}{0.9213}
\newcommand{\LossTuneBaseAUC}{0.9162}
\newcommand{\LossTuneBaseAUCModel}{the DeepSets model}
\newcommand{\LossTuneAUCP}{1.6{\times}10^{-6}}
\newcommand{\PctTopBucket}{0.1}
\newcommand{\ExcessBinPct}{3.2}
\newcommand{\ExcessHighOccPct}{40}
\newcommand{\DataVolMin}{54}
\newcommand{\DataVolMax}{864}
\newcommand{\DataVolSteps}{six}
\newcommand{\DataVolSeeds}{three}
\newcommand{\DataVolBioMin}{0.1841}
\newcommand{\DataVolBioMax}{0.1198}
\newcommand{\DataVolTabICLMax}{0.1227}
\newcommand{\DataVolTabICLP}{0.001}

\newcommand{\DataVolBehindTo}{216}
\newcommand{\DataVolDeepAheadFrom}{648}
\newcommand{\ExtBioAheadOf}{DeepSets, the 1-D CNN and the MLP}
\newcommand{\ExtBioLevelWith}{the Transformer and TabICLv2}
\newcommand{\ExtBioBehind}{the hurdle GLMM}
\newcommand{\LadderMin}{200k}
\newcommand{\LadderMax}{128M}

\newcommand{\LadderBioMin}{3.2M}

\newcommand{\LadderPromotedSize}{10.5M}

\newcommand{\LadderBioWorst}{0.121}
\newcommand{\LadderBaseBest}{0.123}
\newcommand{\LadderBaseBestModel}{the 1-D CNN}
\newcommand{\LadderSpanMax}{0.0062}
\newcommand{\LadderGapMin}{0.0018}
\newcommand{\LadderGapMax}{0.0052}

\newcommand{\CostBioFlops}{1.4{\times}10^{10}}

\newcommand{\CostDeepFlopsLo}{6.1{\times}10^{9}}
\newcommand{\CostDeepFlopsHi}{9{\times}10^{9}}

\newcommand{\CostTrainRatio}{4.4}

\newcommand{\CostTabICLFlops}{5.4{\times}10^{15}}
\newcommand{\CostTabICLRatio}{4{\times}10^{5}}

\newcommand{\BatchBottles}{eight}

\title{BIOTRAM: Biochemistry-informed Transformer for\\Relative Abundance via Metabarcoding}

\usepackage{multicol}
\newlength{\affilindent}
\makeatletter
\newcommand\AB@affilmark[1]{\textsuperscript{#1}}
\renewcommand\AB@affilnote[1]{\protect\Affilfont\protect\AB@affilmark{#1}}
\renewcommand\@author{
    \AB@authlist\\[\affilsep]
    \makebox[\linewidth][c]{%
    \begin{minipage}{0.95\textwidth}
    \begin{multicols}{2}
        \raggedright
        \global\affilindent\z@
        \begingroup
            \let\\\relax
            \renewcommand\AB@affilmark[1]{%
                \settowidth\@tempdima{\textsuperscript{##1}}%
                \ifdim\@tempdima>\affilindent \global\affilindent\@tempdima\fi}%
            \setbox\z@\hbox{\AB@affillist}%
        \endgroup
        \setlength{\leftskip}{\affilindent}%
        \setlength{\parindent}{-\affilindent}%
        \setlength{\parskip}{0pt}%
        \let\\\par
        \renewcommand\AB@affilmark[1]{\makebox[\affilindent][r]{\textsuperscript{##1}}}%
        \AB@affillist
    \end{multicols}
    \end{minipage}
    }%
    }
\makeatother

\author[1,2]{Emma~Boehly}
\author[2]{Scott~C.~Lowe}
\author[1,3]{Ken~A.~Thompson}
\author[2]{Anthony~Fuller}
\author[4]{Will~Armstrong}
\author[1,2]{Iuliia~Zarubiieva}
\author[5]{Neelu~Madan}
\author[1,2]{Tiancheng~Gao}
\author[1,3]{Sameer~Padhye}
\author[1,3]{Jayme~Sones}
\author[1,3]{Kate~H.~J.~Perez}
\author[6]{Bess~Hardwick}
\author[7]{Brendan~Furneaux}
\author[1,3]{Evgeny~V.~Zakharov}
\author[1,3]{Dirk~Steinke}
\author[1,3]{Jeremy~R.~deWaard}
\author[1,3]{Stephanie~L.~deWaard}
\author[1,3]{Jaclyn~McKeown}
\author[1,3]{Norman~Monkhouse}
\author[1,3]{Cole~Pace}
\author[1,3]{Sujeevan~Ratnasingham}
\author[1,3]{Raina~Schmuck}
\author[1,3]{Amy~Thompson}
\author[8]{Jorge~L.~León-Cortés}
\author[9]{Luciano~Cagnolo}
\author[9]{Julieta~Soledad~Arena}
\author[9]{Jeremías~Domínguez~Masciale}
\author[10,11]{Errol~Douwes}
\author[12]{Elli~Groner}
\author[13]{Masae~Iwamoto~Ishihara}
\author[13]{Shunsuke~Matsuoka}
\author[14]{Mailyn~A.~Gonzalez}
\author[15]{Alexandre~Antonelli}
\author[16]{Renata~S.~Sousa-Lima}
\author[17]{Armin~Niessner}
\author[18]{Patrícia~Guedes}
\author[19]{Elizabeth~G.~Biro}
\author[20]{Zuania~Colón-Piñeiro}
\author[21]{Dezene~P.~W.~Huber}
\author[22,23]{Santosh~Mazumdar}
\author[24,25]{Magda~Bou~Dagher~Kharrat}
\author[26,27]{Eddie~P.~Mondejar}
\author[28,29]{Pablo~Luis~Tubaro}
\author[30]{Eugene~Marais}
\author[31]{Marie-Eugénie~Maggia}
\author[28]{Darío~Alejandro~Lijtmaer}
\author[1]{Sean~W.~J.~Prosser}
\author[8]{Andrea~Murillo-Vázquez}
\author[32]{Paola~Pulido-Santacruz}
\author[33]{Paula~Leitman}
\author[34]{Thiago~Arnt~Dorigo}
\author[1]{Nao~Ito}
\author[1]{Paul~Knoops}
\author[1]{Maria~L.~Kuzmina}
\author[1]{Hikaru~Morishita}
\author[13]{Keishi~Kimoto}
\author[35]{Dimby~Raharinjanahary}
\author[15]{Rhian~J.~Smith}
\author[19]{Solny~A.~Adalsteinsson}
\author[14,36]{Kevin~G.~Borja-Acosta}
\author[37]{Paul~Amouroux}
\author[38]{Angela~M.~Mendoza-Henao}
\author[14]{Angélica~Diaz-Pulido}
\author[16]{Anamaria~Dal~Molin}
\author[39,40,41]{Luke~L.~Powell}
\author[42]{Diego~Inclan}
\author[16]{Alice~Calvente}
\author[43]{David~Dunson}
\author[44,45]{Brian~Fisher}
\author[46]{Tomas~Roslin}
\author[7]{Otso~Ovaskainen}
\author[1,3]{Paul~D.~N.~Hebert}
\author[1,2,\corrmark]{Graham~W.~Taylor}

\affil[1]{University of Guelph}
\affil[2]{Vector Institute}
\affil[3]{Centre for Biodiversity Genomics}
\affil[4]{Gaiametric}
\affil[5]{Aalborg University}
\affil[6]{University of Helsinki}
\affil[7]{University of Jyväskylä}
\affil[8]{El Colegio de la Frontera Sur}
\affil[9]{Universidad Nacional de Córdoba}
\affil[10]{eThekwini Municipality} %
\affil[11]{University of KwaZulu-Natal}
\affil[12]{Dead Sea and Arava Science Center}
\affil[13]{Kyoto University}
\affil[14]{Instituto Humboldt} %
\affil[15]{Royal Botanic Gardens, Kew} %
\affil[16]{Universidade Federal do Rio Grande do Norte}
\affil[17]{Hochschule für Forstwirtschaft Rottenburg} %
\affil[18]{Faculdade de Ciências da Universidade de Lisboa}%
\affil[19]{Washington University in St. Louis}%
\affil[20]{University of Florida}
\affil[21]{University of Northern British Columbia}
\affil[22]{Bangladesh Forest Research Institute}
\affil[23]{Alliance for Environment and Community Development}
\affil[24]{Université Saint-Joseph} %
\affil[25]{European Forest Institute}
\affil[26]{Mindanao State University -- Iligan Institute of Technology} %
\affil[27]{Premier Research Institute of Science and Mathematics}
\affil[28]{Museo Argentino de Ciencias Naturales} %
\affil[29]{Universidad de Buenos Aires}
\affil[30]{Gobabeb Namib Research Institute}
\affil[31]{Andes Amazon Fund}
\affil[32]{Universidad del Rosario}
\affil[33]{Independent Researcher}
\affil[34]{Universidade do Estado do Rio de Janeiro}
\affil[35]{Madagascar Biodiversity Center} %
\affil[36]{Universidad Pedagógica y Tecnológica de Colombia}
\affil[37]{Universidad Mayor}
\affil[38]{Asociación Conservación Osa}
\affil[39]{CIBIO-InBIO, Universidade do Porto} %
\affil[40]{BIOPOLIS Program in Genomics, CIBIO} %
\affil[41]{Biodiversity Initiative, Houghton, Michigan}
\affil[42]{Universidad Central del Ecuador}
\affil[43]{Duke University}
\affil[44]{University of California, Berkeley}
\affil[45]{California Academy of Sciences}
\affil[46]{Swedish University of Agricultural Sciences}
\affil[\corrmark]{Correspondence: {\smaller\texttt{gwtaylor@uoguelph.ca}}}

\date{}

\begin{document}
\maketitle

\begin{abstract}%
Arthropod monitoring at scale begins with metabarcoding of bulk samples: trap collections of 
hundreds to thousands of specimens processed as one unit, extraction of DNA from the bulked 
specimens, then sequencing of the DNA from all specimens together.
Metabarcoding reveals which taxa such a sample contains, but due to factors such as differing DNA 
amplification rates and body sizes, its read counts do not neatly correspond to how abundant the 
different taxa are relative to one another.
This lack of relative abundance limits the data's utility for the multitude of ecological analyses 
that depend on it.
We propose BIOTRAM: a neural network that estimates per-taxon specimen counts by dividing the 
observed read counts by a learned read yield per specimen, which it factors into the DNA each 
specimen contributes and how efficiently that DNA is amplified.
On a paired dataset of \NBottles{} bottles (800\,k specimens) processed both by metabarcoding 
and by comprehensive individual barcoding for ground-truthing, BIOTRAM reduces the Jensen--Shannon 
divergence between the estimated and the true composition from \ReadsJSD{} for raw read proportions 
(the usual proxy for abundance) to \TunedBioJSD{}. 
Trained on the same objective and inputs as \TunedNBaselines{} generic deep architectures, all with 
tuned loss weights, it reaches a count error \TunedMAEGainPct\% below that of the strongest of them 
and the lowest composition divergence. It also generalizes to out-of-domain data better, beating all 
baselines on bottles from unseen regions, later collection dates and unseen sequencing batches, and 
transferring to data from an unseen laboratory. %
Structuring the mapping from reads to counts around the steps of the experimental pipeline thus 
yields more accurate abundance estimates than leaving that mapping unconstrained.
\end{abstract}

\section{Introduction}\label{sec:intro}

\begin{figure}[ht!]\centering
\includegraphics[width=\textwidth]{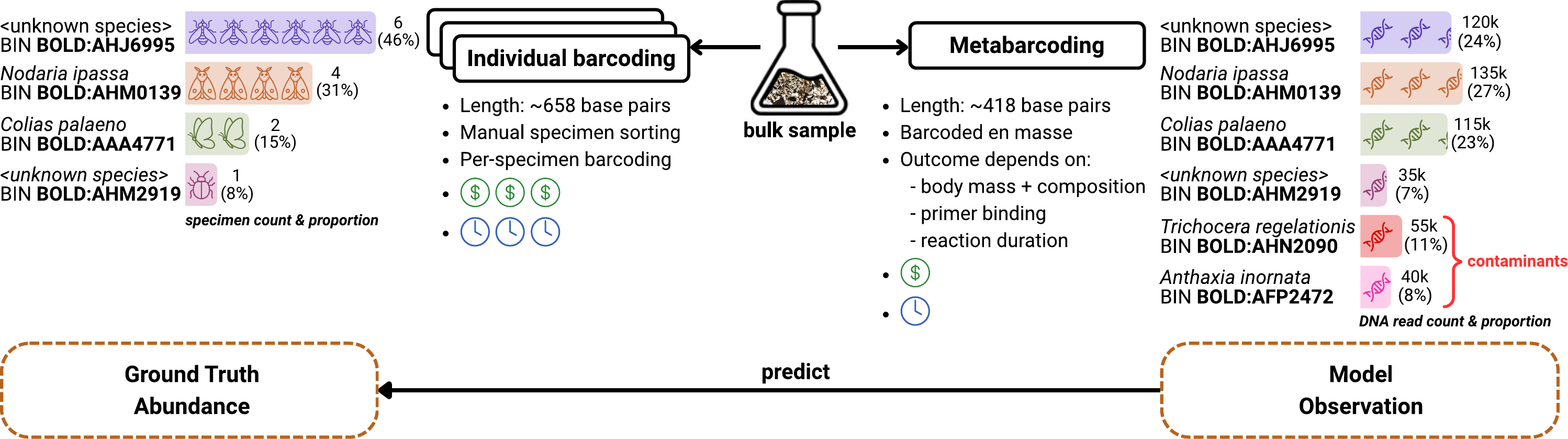}
\caption{\textbf{Individual barcoding vs. metabarcoding for estimating species abundance.} Individual 
barcoding (manual sorting and per-specimen sequencing) gives accurate species counts but is slow and 
costly; metabarcoding (bulk amplification of the sample's DNA) is fast and cheaper, but its read 
counts are not linearly related to specimen abundances.}
\label{fig:overview}
\end{figure}

Insects are by far the most diverse multicellular taxon on Earth \citep{Colwell2026-si} and are 
fundamentally important for human existence given their foundational role in ecosystem processes 
\citep{Yang2014-xw}. Yet, they are undergoing extensive global declines \citep{Wagner2021-wx, 
Hallmann2017-ll}; efforts to quantify this are limited by the paucity of reliable data 
\citep{Didham2020-nr}. Biodiversity indices used to track populations rely on estimates of both 
species richness (number of species) and abundance (number of specimens per species) 
\citep{Hallmann2017-ll,van-Klink2020-ie}, while methods that provide only species lists but not 
abundances are of limited use. Malaise traps---tent-like structures that passively capture flying 
insects and direct them into a bottle of preservative---are widely used for capturing insects for 
study \citep{Karlsson2020-oz}. Although identification of their contents by morphology was common in 
the past, many species cannot be differentiated by morphology alone, and identifying species by 
manual or visual inspection is highly time consuming, expensive, and requires significant taxonomic expertise.
Consequently DNA-based identification systems are increasingly used to identify captured species.
Identifying specimens individually using DNA barcoding provides high-quality estimates of both richness and relative abundance, but is still relatively tedious and costly.
Metabarcoding, a method wherein each bulk sample is processed as a unit by extracting and sequencing DNA from the entire collection, scales far 
better than DNA barcoding but does not produce individual specimen labels---only the number of amplicon reads---and so cannot directly estimate abundance \citep{Yu2012-rs,Ji2013-jz} 
(see \cref{fig:overview}).

Rather than abundance, the output of metabarcoding is a vector of sequence read counts, one count per taxon. 
Besides the number of individuals, four key factors influence how many reads each taxon receives: 
(1) how much DNA each individual contributes to the bulk DNA mixture, (2) how efficiently its 
barcode sequence is amplified during PCR, (3) how long the PCR reaction runs before reagents are 
exhausted, and (4) what fraction of the amplified product is sequenced.
Factors 3 \& 4 vary by protocol, and factors 1 \& 2 vary across taxa by orders of magnitude, so read 
counts track specimen counts only loosely. In bottles whose specimens were also barcoded one by one, 
raw read counts, taken as relative abundances, are on average further from the true composition than 
a guess that every taxon is equally abundant (\cref{sec:results}).

Machine learning is changing how metabarcode reads are interpreted: foundation models 
trained on barcodes support identification and clustering even for unseen taxa 
\citep{Millan-Arias2026-pm, Gao2024-fe, Safari2025_BarcodeMAE}. Abundance has resisted the same treatment: identification models learn from reference libraries of millions of individually
barcoded specimens \citep{CanadianInvertebrates,bioscan5m}, but abundance requires bulk samples
whose composition is known specimen by specimen---the tedious, costly work that metabarcoding bypasses. The reads-to-specimens mapping also differs by taxon,
and per-taxon correction factors \citep{McLaren2019-ga} cannot be calibrated for taxa
that are rare or absent from calibration samples.\looseness=-1 

BIOTRAM, a \textbf{bio}chemistry-informed \textbf{t}ransformer for \textbf{r}elative \textbf{a}bundance via \textbf{m}etabarcoding, is instead structured around the steps of the experimental pipeline: it estimates specimen counts by dividing each taxon's reads by a learned read yield per specimen, factored into per-specimen DNA contribution and amplification efficiency, so it can produce estimates for taxa absent from training. Its taxonomic unit is the Barcode Index Number (BIN), a cluster of near-identical barcodes proxying for species that usually agrees with species assigned by morphology~\citep{Ratnasingham2013-lr}. Our contributions are as follows:
\begin{enumerate}[label=(\arabic*)]
    \item \textbf{A paired DNA barcoding-metabarcoding dataset} (\cref{sec:data}). \NBottles{} bottles processed with both 
    metabarcoding and (for ground-truthing) individual barcoding, yielding \NRows{} matched (reads, count) pairs across 
    \NBins{} unique BINs and \NSites{} sites---more than an order of magnitude larger than existing data.%
    \item \textbf{A bespoke architecture built on biological and chemical knowledge} (\cref{subsec:model}). Rather than a na\"ive model, we propose a staged model based on the biochemical properties of the metabarcoding process.
    \item \textbf{Latent predictions of primer binding efficiency and body size} (\cref{subsec:anchors}).
    Our model produces physically interpretable latent estimates of both properties, which directly shape the relationship between specimen counts and read counts per BIN and bottle. 
    \item \textbf{Controlled evaluation under distribution shift} (\cref{sec:results}).
    We compare BIOTRAM against deep baselines trained with the same objective and inputs, generalizing 
    to novel collection locations (spatial blocking), novel time-points (temporal cut-off), and across 
    metabarcode sequencing methods (data from another laboratory).
\end{enumerate}

\section{Related work}
Metabarcoding studies have long sought relative abundances from bulk samples, mostly with regression
or other statistical models \citep{Ji2013-jz, Sickel2023-RecAdv, Lamb2019-Meta-bar_Meta-an}.
For Malaise traps, \citet{FurneauxRoslin_FlyInSoup} analyzed a paired barcode-metabarcode dataset of
45 bulk samples from Sweden and Finland with mixed models, and found strong correspondence between
read counts and abundance; they also modeled which BINs are present as individuals, even when
undetected by metabarcoding, and how BIN-specific abundances affect detection.
\citet{IwaszkiewiczEggebrecht_MetabarcodingAccuracy} applied a hierarchical Bayesian model that
incorporates spike-in data to 15 Swedish samples with comprehensive barcode and metabarcode data.
Both studies found close correspondence between read counts and relative abundances from barcoding,
but were fit to a few dozen samples from northern Europe, vs.~\NBottles{} bottles from \NSites{}
global sites here. We refit the hurdle mixed model of \citet{FurneauxRoslin_FlyInSoup} to our data
as a baseline (\cref{sec:results,app:baselines}).

Metabarcoding studies of other taxa also estimate relative abundance, including that of plants \citep{Matesanz2019-PlantBG}, freshwater macroinvertebrates \citep{Elbrecht2015-og}, and fishes \citep{Littlefair2026-FishAbund}. Metagenomic workflows such as SPIKEPIPE \citep{JiHuotari_SPIKEPIPE} avoid PCR bias but cost far more than metabarcoding. To our knowledge, none of these methods uses a neural network trained on paired data at this scale.\looseness=-1

\section{Dataset}\label{sec:data}

Training requires metabarcoding data paired with ground-truth specimen counts---these come from separating and barcoding every specimen in a bottle, which is an expensive process.
See \cref{fig:overview} for an overview of the respective individual barcoding and metabarcoding data pipelines.
We assembled such data for Malaise-trap bottles from the LIFEPLAN project \citep{deWaard_LIFEPLANmbc, Hardwick2024_LIFEPLAN} and the Global Malaise Program \citep{Geiger2016-qp}, as described below. Permissions to use each bottle were arranged with the data owners (\cref{app:data_provenance}).

\mypar{Barcoding.}
DNA was extracted from every bottle by mild lysis, a chemical extraction that draws DNA from the 
specimens while leaving them intact, and every specimen was then barcoded individually.  
Metabarcoding amplified a ${\sim}418$\,bp stretch of the cytochrome \textit{c} oxidase subunit I gene (COI; BF3/BR2 primers;
\citealp{Elbrecht2017-fn,Elbrecht2019-jy}), in three PCR replicates, which were sequenced on an Illumina NextSeq 2000 and processed with the Metabarcoding Analysis Pipeline (MAP;
\citealp{Prosser2026-yp}). MAP clustered the reads into operational taxonomic units (OTUs), each matched to a BIN in a reference library (\cref{app:data_assembly}). OTUs with no BIN match within 2.3\% sequence identity were assigned to their nearest BIN, because new BINs can only be created within BOLD. 
Individual barcoding yields a per-BIN specimen count for every bottle,
which BIOTRAM learns to predict from the metabarcoding data alone. Individual barcoding fails for under 10\% of specimens~\citep{Orsholm2026-mm}, so some BINs that metabarcoding detects but barcoding misses are present, not false detections.

At deployment, a monitoring program has only the metabarcoding results, without individual sorting 
or prior calibration. We therefore drop the BINs detected by individual barcoding but absent from
metabarcoding, and keep the true zeros of BINs seen only in metabarcoding, which the model is 
trained and evaluated to predict.

The paired dataset spans \NBottles{} bottles from \NSites{} sites collected between \YearFirst{} and 
\YearLast{}. It contains \NRows{} rows, one per BIN per bottle, across \NBins{} BINs and
\NSpecimens{} counted specimens. A typical bottle holds \MedBins{} BINs and \MedInd{} specimens
(medians; maxima \MaxBins{} and \MaxInd{}), and the dataset covers \NFamilies{}
families. %

As our goal is to predict the distribution of species within a bottle, the bottle is the model's unit of processing, and we partition the data by bottle.
Specimens are almost entirely arthropods (Insecta \PctInsecta{}, Collembola \PctCollembola{}, Arachnida \PctArachnida{}).
Most BINs lack genus and species labels, so we use BINs instead of named species.

\mypar{Reads vs.~counts.}\label{par:readsVcounts}
We call $\smash[t]{n_i^{(b)}}$ the specimen count of BIN $i$ in bottle $b$ and $\smash[t]{p_i^{(b)}}=\smash[t]{n_i^{(b)}}/\sum_j \smash[t]{n_j^{(b)}}$ its relative abundance; the vector $\smash[t]{p^{(b)}}$ is the bottle's composition.
Read shares are the usual proxy for composition, but they are biased~\citep{McLaren2019-ga, FurneauxRoslin_FlyInSoup, IwaszkiewiczEggebrecht_MetabarcodingAccuracy}, and in our data they track it poorly:
on a $\log(1+x)$ scale, read counts explain \RsqReads{}\% of the variance in specimen counts,
rank BINs within a bottle at a mean Spearman correlation of \SpearReads{}, and separate present BINs from false detections at an AUC of \AUCReads{}. The most-sequenced BIN of a bottle is also its most abundant in only \PctTopOne{}\% of bottles.

\mypar{Excess specimens.} In \NExcess{} of the \NBottles{} bottles, technicians counted at least one 
highly abundant morphospecies by eye instead of barcoding each specimen (Figure~\ref{fig:data_pipeline}), so these bottles lack 
complete per-BIN ground truth. We evaluate only on the \NClean{} bottles without this \textit{excess} 
and use \NExcessUsed{} of the others, those least affected by it, for training only (\cref{app:excess}).

\section{Methods}\label{sec:methods}

\begin{figure}[tb!]
\centering
\includegraphics[width=\textwidth]{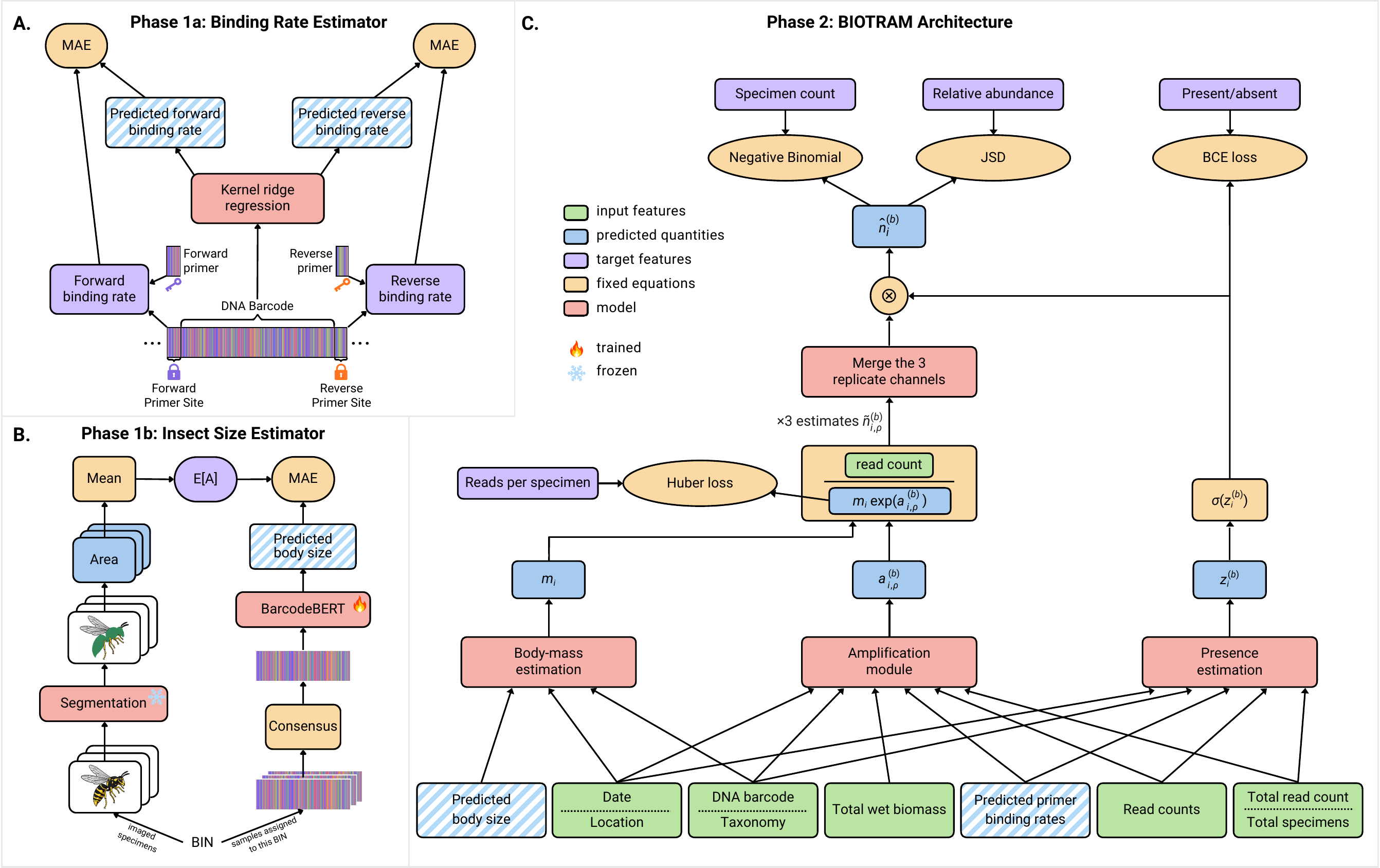}
\caption{\textbf{BIOTRAM method overview.} 
(\textbf{A})~\textit{Binding Rate Estimator} predicts each primer's binding rate from its barcode, 
trained against rates estimated for longer sequences with known DNA at primer sites.
(\textbf{B})~\textit{Insect Size Estimator} predicts a BIN's average specimen size from a 
representative barcode sequence, trained on estimates found from paired image data. 
(\textbf{C})~\textit{BIOTRAM's architecture} integrates barcode, taxonomic, and sample metadata to jointly 
predict each BIN's body-mass term and presence; an attention module corrects raw read counts for 
amplification bias across replicates before the corrected estimates are merged into a final 
abundance prediction, trained with negative binomial, Jensen--Shannon divergence (JSD), binary
cross-entropy (BCE) and Huber losses (\cref{subsec:loss}).
For submodule component design, see \cref{fig:arch}.
}
\label{fig:arch_simp}
\end{figure}

We assume that the read count of BIN $i$ in bottle $b$ for replicate $\rho$ is, up to a bottle-level constant that absorbs the replicate's total read count, the product of three quantities:
(1) the specimen count \smash{$n_i^{(b)}$}, which we predict;
(2) the amplifiable DNA per specimen, which on average scales with the BIN's expected body mass, the body-mass term $m_i$;
and (3) the log amplification \smash{$a_{i,\rho}^{(b)}$} of its barcode in replicate $\rho$ before the reagents ran out:
\begin{equation}%
\mathrm{reads}_{i,\rho}^{(b)} \;\propto\; n_i^{(b)} \cdot m_i \cdot \exp\!\big(a_{i,\rho}^{(b)}\big).
\label{eq:rate}
\end{equation}
See \cref{app:rate} for our derivation of this forward relationship.

\subsection{Body size and primer binding efficiency}\label{subsec:anchors}
Two per-BIN inputs inform the terms of  \cref{eq:rate} and can be estimated from data outside 
metabarcoding: body size, which the body-mass term $m_i$ should track, and primer binding efficiency, 
which shapes the log amplification $\smash[t]{a_{i,\rho}^{(b)}}$.
We estimate both from public specimen images and reference sequences,  freeze the estimators, and give their outputs to BIOTRAM and every baseline.
\Cref{subsec:anchors_eval} evaluates these models and \cref{app:primer,app:bodysize} detail them.

\mypar{Body size.}
Larger individuals have more surface in contact with the DNA extraction fluid, so more of their DNA enters the amplification process.
To estimate specimen sizes, we used the size of photographed insects as a proxy (\cref{fig:arch_simp}B).
From the public specimen images in the Barcode of Life Data System (BOLD), we assembled \SegNImages{} 
images associated with BIN labels and segmented each image using pretrained vision models (\cref{app:bodysize}).
The pipeline segments the whole organism and then carves away spread wings, which add projected area 
but little tissue (\cref{fig:segmentation}).
For each BIN, we averaged the projected body area over specimens, giving an expected size for 
\SegNBins{} BINs, representing \SizeImagedPct\% of our dataset's BINs. To generalize this model 
beyond BINs with imaged examples, we fine-tuned a pretrained barcode model, BarcodeBERT
\citep{Millan-Arias2026-pm}, to estimate the mean log area of a BIN and its within-BIN spread from 
a single barcode within the BIN; BIOTRAM reads these predicted values for every BIN.

\mypar{Primer binding efficiency.}
The PCR process consists of copying the barcode region between two primers, short fixed sequences 
that bind to the DNA on either side of it (\cref{fig:arch_simp}A).
The sequence at these binding sites varies across species, and a primer binds better the more 
closely the site matches it. These sites are not shown in reads, because amplification overwrites 
them with the primer sequences, so we take them from two external sources: full-length barcode 
records, which extend beyond the amplicon and give the forward site for \PrimerBarcodeFPct\% of 
BINs, and mitogenome records, which give both the forward and the reverse sites for another 
\PrimerMitoPct\%. We score each retrieved 20-nucleotide site to estimate a binding efficiency 
between 0 and 1; for the remaining BINs, we predict the efficiency from the amplicon with a ridge 
and kernel ridge regression (\cref{app:primer}).

\subsection{Sources of variation in reads per specimen}
We quantify what distorts the relationship between reads and specimen counts: how much is attributable
to the BIN itself (e.g., primer binding, body size), the reaction as a whole, and detection noise. 
Each source of variation motivates a specific component of the model.

\mypar{Most of the variation is per-BIN, but it cannot be looked up.} The number of reads one 
specimen produces spans \Decades{} orders of magnitude across BINs, and BIN identity explains
\EtaSq{}\% of that variation. The classical remedy corrects each BIN by its own properties
\citep{McLaren2019-ga}, a structure our model keeps, but \PctSingleton{}\% of BINs appear in a
single bottle and \UnseenPct{}\% of test rows carry a BIN absent from training, which a lookup table
cannot handle. We, instead, compute the per-BIN quantities from the barcode sequence, since primer
binding and body size are conserved among relatives.

\mypar{False detections.} \PctZero{}\% of the BINs detected in a bottle never occur: contamination, 
cross-talk between bottles sequenced together, traces of environmental DNA and clustering artifacts 
all leave reads where no specimen was present \citep{Schnell2015-zv,Zinger2019-dm}. A dedicated 
presence head, trained with its own loss, gates the abundance estimate, so a detected BIN can be 
reported absent.

\mypar{Replicate disagreement.} Only \PctAllThree{}\% of detections appear in all three PCR
replicates and \PctOneOfThree{}\% in just one. When little of a BIN's DNA is present, whether it
amplifies is partly chance, so replicate disagreement is itself evidence of how much DNA was there.
The model takes the three replicate counts as inputs, not their sum.

\mypar{Bottle-level saturation.} A PCR amplifies exponentially until its fixed reagent budget is
spent, then plateaus. How long that takes depends on the whole bottle. Because BINs amplify at 
different rates, the ratio between two BINs' reads depends on every BIN in the bottle 
\citep{Suzuki1996-ed}: the emptiest quarter of bottles yields a median of
\RpsEmptiest{} reads per specimen, roughly \RpsRatio{}$\times$ the \RpsFullest{} of the fullest
quarter. Sequencing then samples only a fraction of the amplified product. Both effects require
bottle-level terms.

\subsection{The model}\label{subsec:model}
 BIOTRAM predicts $m_i$ and \smash{$a_{i,\rho}^{(b)}$}, together with 
the probability that the BIN is present at all ($\sigma(\smash[t]{z_i^{(b)}})$), and inverts \cref{eq:rate} against the observed read count,
\begin{equation}%
\begin{aligned}
\log \tilde{n}_{i,\rho}^{(b)} &\;=\; \log\!\big(1 + \mathrm{reads}_{i,\rho}^{(b)}\big) \;-\; \log T_{\rho}^{(b)}
  \;-\; a_{i,\rho}^{(b)} \;-\; \log m_i \;+\; \log \nu^{(b)}, \\
\log \hat{n}_{i}^{(b)} &\;=\; \mathcal{M}_i\!\big(\log \tilde{n}_{i,1}^{(b)},\, \log \tilde{n}_{i,2}^{(b)},\,
  \log \tilde{n}_{i,3}^{(b)};\ \log \nu^{(b)}\big) \;+\; \log \sigma\!\big(z_i^{(b)}\big),
\end{aligned}
\label{eq:inversion}
\end{equation}
with $\smash[t]{T_{\rho}^{(b)}}$ the replicate's total reads, $\smash[t]{\nu^{(b)}}$ the
bottle's total specimen count, $\smash[t]{z_i^{(b)}}$ the logit that BIN $i$ is truly
present, $\sigma$ the sigmoid function, and $\mathcal{M}_i$ a small learned network that merges the 
three replicate channels, starting from their mean. The channels are each BIN's smallest, median and
largest read count, so they do not depend on the arbitrary order of the replicates.

Two trained networks predict the per-BIN quantities $m_i$ and $\smash[t]{z_i^{(b)}}$. Each computes a frozen
BarcodeMamba embedding of the barcode \citep{Gao2024-fe,Gao2025_BarcodeMambaPlus} through its own 
attention layer, adds a learned taxonomy embedding and the bottle's location and collection date 
(through a pretrained geographic encoder \citep{Klemmer2025-sn} and a seasonal encoding, respectively) and
passes the result through an MLP. The frozen embedding is what lets an unseen BIN inherit from its
relatives. The body mass network also reads the barcode-predicted body size of \cref{subsec:anchors}, 
the presence network reads the primer efficiencies, the bottle's total specimen count and the BIN's 
replicate read counts, so $\sigma(\smash[t]{z_i^{(b)}})$ is a read-informed presence probability.

The amplification term $\smash[t]{a_{i,\rho}^{(b)}}$ comes from a self-attention module over every BIN of
the bottle, run once per replicate channel. Each BIN is characterized by its barcode, its two primer
efficiencies (each with a binary input marking whether it was computed from a retrieved binding site or predicted; \cref{subsec:anchors}), its predicted 
body size, and its read count and read share within the bottle. Every BIN
also receives a summary of the bottle: the mean of all its BINs' representations,
weighted by a softmax over their amplification demand, estimated from the predicted body
size and the two primer efficiencies with a learned temperature, along with the bottle's
covariates and the replicate's total reads. The BINs expected to consume most 
of the shared reagents thus dominate the context from which every amplification term is estimated.
A \emph{read-blind} variant estimates $\smash[t]{a_{i,\rho}^{(b)}}$ without the BIN's own read (\cref{app:matched}), 
and \cref{app:ablation} measures what each input contributes. The count is read out as the mean of a 
negative binomial whose dispersion is predicted per row. Apart from the replicate merge, the 
arithmetic that turns the predicted quantities into a count is fixed by \cref{eq:inversion}
(overview in \cref{fig:arch_simp}C, full architecture in \cref{fig:arch}).

\subsection{Training objective}\label{subsec:loss}
Let $\mathcal{B}$ be the bottles in a batch and $\mathcal{R}$ the pairs $(i,b)$ of the whole batch, 
and write $\smash[t]{n_i^{(b)}}$ for the observed count, $\hat{n}_i^{(b)}$ for
\cref{eq:inversion}'s prediction, and $\sigma(\smash[t]{z_i^{(b)}})$ for the
predicted presence probability. Training minimizes a negative-binomial count likelihood (NB), which
accommodates the over-dispersion of ecological counts, whose variance exceeds their mean, plus a per-bottle divergence on composition,
a presence loss, and a term tying the predicted reads per specimen to its observed value:
\begin{equation}%
\mathcal{L} \;=\; \mathcal{L}_{\mathrm{NB}} \;+\; \lambda_{\mathrm{comp}}\,\mathcal{L}_{\mathrm{comp}}
\;+\; \lambda_{\mathrm{pres}}\,\mathcal{L}_{\mathrm{pres}} \;+\; \lambda_{\mathrm{rps}}\,\mathcal{L}_{\mathrm{rps}},
\label{eq:loss}
\end{equation}
{\allowdisplaybreaks
\begin{align}%
\mathcal{L}_{\mathrm{NB}} &\;=\; -\frac{1}{|\mathcal{R}|}\sum_{(i,b)\in\mathcal{R}}
  \log \mathrm{NB}\big(n_i^{(b)} \,;\, \hat{n}_i^{(b)},\, \alpha_i^{(b)}\big),
  \qquad \operatorname{Var}\big[n_i^{(b)}\big] \;=\; \hat{n}_i^{(b)} + \alpha_i^{(b)}\,\big(\hat{n}_i^{(b)}\big)^{2},
  \label{eq:loss-nb} \\
\mathcal{L}_{\mathrm{comp}} &\;=\; \frac{1}{|\mathcal{B}|}\sum_{b\in\mathcal{B}}
  \operatorname{JSD}\big(p^{(b)} \,\big\|\, \hat{p}^{(b)}\big),
  \qquad p_i^{(b)} = \frac{n_i^{(b)}}{\sum_{j\in b} n_j^{(b)}}, \quad
  \hat{p}_i^{(b)} = \frac{\hat{n}_i^{(b)}}{\sum_{j\in b} \hat{n}_j^{(b)}},
  \label{eq:loss-comp} \\
\mathcal{L}_{\mathrm{pres}} &\;=\; -\frac{1}{|\mathcal{R}|}\sum_{(i,b)\in\mathcal{R}}
  \Big[\, y_i^{(b)}\log\sigma(z_i^{(b)}) + \big(1-y_i^{(b)}\big)\log\big(1-\sigma(z_i^{(b)})\big) \Big],
  \label{eq:loss-pres} \\
\mathcal{L}_{\mathrm{rps}} &\;=\; \frac{1}{\sum_{(i,b)} w_i}\sum_{(i,b)\in\mathcal{R}}
  w_i\, H_{\delta}\!\left(\frac{\varepsilon_i^{(b)} - \bar{\varepsilon}}{s_i}\right),
  \qquad \varepsilon_i^{(b)} \;=\; \big(\bar{a}_i^{(b)} + \log m_i\big) - \ell_i .
  \label{eq:loss-rps}
\end{align}}

The NB variance keeps a taxon whose count genuinely varies between bottles from having that variance 
absorbed as bias in the mean.
$\mathcal{L}_{\mathrm{comp}}$ scores the within-bottle composition, with $\operatorname{JSD}(P\|Q) = \tfrac{1}{2}\operatorname{KL}(P\|M) +
\tfrac{1}{2}\operatorname{KL}(Q\|M)$ at $M = \tfrac{1}{2}(P+Q)$; it is also the reported metric and
the checkpoint-selection criterion. We use JSD rather than KL because it is symmetric --- over-predicting 
a rare BIN and under-predicting an abundant one are penalized alike --- and because it stays finite on 
the \PctZero{}\% of rows whose truth is zero, which $\operatorname{KL}(\smash[t]{p^{(b)}}\|\smash[t]{\hat{p}^{(b)}})$, weighted by
the true abundance, does not directly penalize.
$\mathcal{L}_{\mathrm{pres}}$ trains the presence probability $\sigma(z_i^{(b)})$ of
\cref{eq:inversion} directly on the presence label $y_i^{(b)} = \mathbf{1}[n_i^{(b)} \ge 1]$, so that a BIN with reads but no individuals 
is treated as a false detection rather than a small count.
Unlike the other terms, $\mathcal{L}_{\mathrm{rps}}$ supervises an intermediate quantity rather than the
output. By \cref{eq:rate}, $\bar{a}_i^{(b)} + \log m_i$ is the log number of reads per specimen of BIN $i$, where $\bar{a}_i^{(b)}$ is the mean of $a_{i,\rho}^{(b)}$ over the three replicates. Its
target $\ell_i$ is the mean of $\log(\mathrm{reads}_i^{(b)}/n_i^{(b)})$ over the training
bottles, with $\smash[t]{\mathrm{reads}_i^{(b)}}$ summed over the replicates, and $s_i$ is the 
standard deviation of that quantity across those bottles. Only BINs observed in at least three 
training bottles contribute to the loss ($w_i = 1$, otherwise $w_i = 0$), and $\bar{\varepsilon}$ is 
the $w$-weighted mean of $\smash[t]{\varepsilon_i^{(b)}}$ over the batch. $H_{\delta}$ is the Huber 
loss with threshold $\delta = 1$. 
Training batches are whole bottles (\cref{app:model}).

\section{Results}\label{sec:results}

We evaluate BIOTRAM against a broad range of baselines, on random splits, under three kinds of
distribution shift and on an external dataset with a different laboratory workflow.

\mypar{Evaluation design and metrics.}\label{sec:eval}
Data is partitioned at the bottle level. The \NClean{} excess-free bottles are split
\NTrain{}\,/\,\NVal{}\,/\,\NTest{} into train/val/test, and \NExcessUsed{} of the \NExcess{} bottles 
with excess specimens (\cref{app:excess}) join the training set only, for \DataVolMax{} training 
bottles in all. The baselines span linear models, nearest neighbours, tree-based models, 
zero-inflated and hurdle models, deep set and attention-based architectures. All receive the same 
inputs and are each tuned on their own hyperparameter search space against Jensen--Shannon 
divergence (JSD) between the predicted and the true within-bottle composition on the validation set, 
our primary metric. From the same predicted counts, we also score the mean absolute error (MAE; per 
bottle, on rows with no individuals, MAE$_{=0}$, on the rest, MAE$_{>0}$, and on log counts, 
MAE-log), presence AUC, Spearman correlation and two summary statistics used frequently in ecology 
(\cref{app:metrics}). Two models are \emph{level} when a two-sided Welch test over seeds cannot 
separate them.

Every deep model, BIOTRAM included, is trained on the objective of \cref{eq:loss}, with its three
auxiliary weights tuned for it under one protocol and budget (\cref{app:losstune}).
Its term $\mathcal{L}_{\mathrm{rps}}$ supervises the log reads per specimen, which by \cref{eq:rate}
is $\bar{a}_i^{(b)} + \log m_i \simeq \log(\mathrm{reads}_i^{(b)}/n_i^{(b)})$. A baseline has no amplification term,
but its predicted count $\hat{n}_i^{(b)}$ gives the same quantity as
$\log(\mathrm{reads}_i^{(b)}/\hat{n}_i^{(b)})$, so the term applies to every model.

\begin{table}[t]\centering\footnotesize
\caption{\textbf{In-domain test performance.}
Test split of the excess-free bottles; every model sees the same inputs.
The hurdle mixed model of \citet{FurneauxRoslin_FlyInSoup} is fitted on its own likelihood with its published predictors (\cref{app:baselines}); the deep models are trained on the objective of \cref{eq:loss}, each with its auxiliary weights tuned for it (\cref{app:losstune}).
JSD scores the predicted composition, Spearman the ranking of the counts over all test rows, presence AUC how well the predicted counts separate present from absent BINs, and MAE/bottle the error on the counts within each bottle, averaged over bottles; MAE$_{=0}$ and MAE$_{>0}$ are the count error over all BINs truly absent or present.
Values are means over seeds $\pm$ one standard deviation across seeds (the hurdle mixed model is a single fit). We highlight the \best{best} and \runnerup{second best} model.
}
\label{tab:tuned}
\adjustbox{max width=\textwidth}{
\footnotesize
\begin{tabular}{l lll lll}
\toprule
Model & JSD $\downarrow$ & Spearman $\uparrow$ & Pres.\ AUC $\uparrow$ & MAE/bottle $\downarrow$ & MAE$_{=0}$ $\downarrow$ & MAE$_{>0}$ $\downarrow$ \\
\midrule
Hurdle GLMM \citep{FurneauxRoslin_FlyInSoup} & 0.1947\phantom{\tabpm{0.000}} & 0.609\phantom{\tabpm{0.000}} & 0.8419\phantom{\tabpm{0.000}} & 1.498\phantom{\tabpm{0.000}} & 0.730\phantom{\tabpm{0.000}} & 2.286\phantom{\tabpm{0.000}} \\
\midrule
Transformer @ 9.6M & 0.1257\tabpm{0.0019} & \runnerup{0.741}\tabpm{0.003} & 0.9156\tabpm{0.0017} & 0.723\tabpm{0.020} & 0.295\tabpm{0.023} & \runnerup{1.395}\tabpm{0.048} \\
MLP @ 11.1M & \runnerup{0.1234}\tabpm{0.0006} & \runnerup{0.741}\tabpm{0.001} & 0.9154\tabpm{0.0007} & 0.737\tabpm{0.025} & 0.303\tabpm{0.013} & 1.409\tabpm{0.047} \\
DeepSets @ 10.1M & 0.1256\tabpm{0.0010} & \runnerup{0.742}\tabpm{0.002} & \runnerup{0.9162}\tabpm{0.0013} & 0.752\tabpm{0.036} & 0.325\tabpm{0.020} & 1.453\tabpm{0.110} \\
1-D CNN @ 11.5M & 0.1235\tabpm{0.0008} & 0.737\tabpm{0.001} & 0.9145\tabpm{0.0008} & 0.762\tabpm{0.016} & 0.321\tabpm{0.010} & 1.473\tabpm{0.036} \\
\midrule
\textbf{BIOTRAM} & \best{0.1200}\tabpm{0.0009} & \runnerup{0.741}\tabpm{0.001} & 0.9136\tabpm{0.0005} & \best{0.699}\tabpm{0.017} & \runnerup{0.279}\tabpm{0.016} & \best{1.341}\tabpm{0.024} \\
\midrule
TabICLv2 (in-context, 28.5M) & 0.1238\tabpm{0.0001} & \best{0.751}\tabpm{0.001} & \best{0.9239}\tabpm{0.0005} & \runnerup{0.704}\tabpm{0.003} & \best{0.249}\tabpm{0.002} & 1.413\tabpm{0.006} \\
\bottomrule
\end{tabular}}
\end{table}

\mypar{Overall accuracy.} \Cref{tab:tuned} reports accuracy on the held-out test bottles. Among
the deep models trained on our data, BIOTRAM is the most accurate on both the counts and the
composition. Its mean absolute error per bottle is \TunedBioMAE{}, which is \TunedMAEGainPct\% lower
than that of the strongest baseline (\TunedBestBaselineMAE{}), and its JSD is \TunedBioJSD{}, against
\TunedBestBaselineJSD{} for the best baseline. Both differences are significant against each of the
\TunedNBaselines{} baselines (two-sided Welch test over seeds). The first row of the table is the
hurdle mixed model of \citet{FurneauxRoslin_FlyInSoup}, the established statistical approach to this
problem. It is fitted on its own likelihood rather than on \cref{eq:loss}, so we show it for
reference, outside the matched comparison. Simple predictors do much worse (\cref{tab:main}). A
uniform composition, with every taxon equally abundant, scores \UniformJSD{} on JSD and the raw read
shares score \ReadsJSD{}. The classical per-BIN correction of %
\citet{McLaren2019-ga} reaches only \McLarenJSD{} JSD, because \PctSingleton{}\% of BINs appear in a
single bottle and most test rows therefore have no correction to apply.

The ranking metrics show a different pattern. On rank correlation, BIOTRAM is ahead of
\TunedSepSpearAhead{} of the baselines and level with \TunedSepSpearTied{}, and on presence AUC it 
trails \TunedSepAUCBehind{} of them. This follows from the mechanism. Dividing the measured reads by 
an estimated amplification and body mass (\cref{eq:inversion}) constrains the scale of each 
prediction, which is what MAE measures, but it offers no comparable help in deciding which of two 
taxa in a bottle is more abundant. That ranking is largely decided by presence, which the read 
counts already carry and which an unconstrained network can fit directly.

TabICLv2 \citep{qu2026_TabICLv2} was kept apart since it works differently. It is pretrained on 
synthetic tabular datasets and is not trained on our data at all; instead, it predicts each test 
bottle by attending to examples from the training set. It has the best presence AUC, MAE-log and 
MAE$_{=0}$ of any model. BIOTRAM is \TunedTabICLCountRel{} it on count accuracy (MAE \TunedBioMAE{} 
against \TunedTabICLMAE{}) and \TunedTabICLCompRel{} it on composition, with \TabICLParamPct\% of its
parameters. Attending to the training set also makes TabICLv2 expensive to run: it needs
\textit{$\CostTabICLRatio$ times more compute per bottle} than BIOTRAM (\cref{app:baselines}).

\begin{table*}[t]\centering\small
\caption{\textbf{Out-of-domain test performance.}
Each regime holds out one stratum of bottles from training; the collaterals rows are the external dataset of \cref{app:external}.
BIOTRAM is compared with the best baseline of each row, among the four matched deep baselines and the hurdle mixed model (\cref{app:shift}); $\Delta$ is each model's change in JSD from its own random-split score, MAE/bottle is the count error of \cref{tab:tuned}, and the \best{best} value of each metric is in bold.
}
\label{tab:shift}
\adjustbox{max width=\textwidth}{
{\small
\setlength{\tabcolsep}{4pt}
\begin{tabular}{@{}l l ccc lccc@{}}
\toprule
 & & \multicolumn{3}{c}{BIOTRAM} & \multicolumn{4}{c}{Best baseline} \\
\cmidrule(lr){3-5}\cmidrule(lr){6-9}
Regime & Simulated domain change & JSD $\downarrow$ & $\Delta$ $\downarrow$ & MAE/bottle $\downarrow$ & Model & JSD $\downarrow$ & $\Delta$ $\downarrow$ & MAE/bottle $\downarrow$ \\
\midrule
Random & in-domain reference & \best{0.1194} & -- & \best{0.700} & 1-D CNN & 0.1238 & -- & 0.759 \\
Spatial & a region never sampled in training & \best{0.1612} & +0.0419 & \best{0.977} & Transformer & 0.1702 & +0.0445 & 1.076 \\
Lane & a new sequencing batch & \best{0.1711} & +0.0517 & \best{0.670} & Transformer & 0.1758 & +0.0501 & 0.848 \\
Temporal & bottles collected later than any in training & \best{0.1856} & +0.0662 & \best{0.896} & Transformer & 0.1954 & +0.0697 & 1.029 \\
\midrule
Collaterals dataset & different location, protocol (lysis) & 0.2189 & +0.0995 & \best{2.781} & Hurdle GLMM & \best{0.2026} & +0.0078 & 4.146 \\
Collaterals dataset & different location, protocol (homogenate) & 0.2177 & +0.0984 & \best{2.544} & Hurdle GLMM & \best{0.1965} & +0.0018 & 3.981 \\
\bottomrule
\end{tabular}
}
}
\end{table*}

\mypar{Error structure and distribution shift.} By true abundance, BIOTRAM is most accurate on the 
rows with no individuals, ahead of \AbZeroAhead{} of the matched baselines, and on counts above 100, 
ahead of \AbTopAhead{} of them and level with \AbTopTied{}: only \PctTopBucket\% of rows, but they 
carry a large share of each bottle's individuals.
It is also ahead of \AbSeenAhead{} of them on rows whose BIN was seen in training, but behind at 
least one of them on \AbMidBucketsBehind{} of the four buckets between 1 and 100, and behind 
\AbUnseenBehind{} of them on BINs never seen in training (\cref{app:errors}).
A random split shares sites, seasons and sequencing batches between training and testing, which
deployment does not, so we also retrain with a region never seen, bottles collected later than any in 
training, or a new sequencing batch held out (\cref{tab:shift}).
The temporal shift is the hardest, adding \DeltaTemporal{} in JSD against \DeltaSpatial{} for the
spatial and \DeltaLane{} for the batch shift. Against each regime's best baseline (\cref{tab:shift}),
BIOTRAM is \OODSpatialDir{} it on the spatial split
($\OODSpatialGap$ JSD), \OODLaneDir{} it on the batch split ($\OODLaneGap$) and \OODTemporalDir{} it
on the temporal one ($\OODTemporalGap$; \cref{app:shift}).
On \ExtBottles{} Malaise-trap bottles from Sweden (\ExtSpecimensL{} specimens, each barcoded 
individually), metabarcoded by another laboratory with its own protocol (unpublished data; 
\cref{app:external}), BIOTRAM stays closer to the true composition than the read shares and is level 
with \ExtBioLevelWith{}; the best baseline there is the hurdle mixed model of
\citet{FurneauxRoslin_FlyInSoup}, closer on composition but with the largest count error of any
model (\cref{tab:shift}).
Beyond point estimates, BIOTRAM's NB predictive distribution is conservative: its central 90\% 
intervals cover \CalibCov{} of held-out rows (\cref{app:scatter}).

\mypar{Robustness of the comparison.}\label{subsec:scaling} \Cref{tab:tuned} rests on three choices
for every deep model: loss weights tuned for it, a size close to BIOTRAM's and the full training set. We 
vary each in turn to test whether BIOTRAM's lead depends on them. 
With one set of loss weights shared by every model (\cref{app:losstune}), only the architecture
differs: BIOTRAM's MAE then separates from \SepReadsMAEbtlAhead{} of the \NMatchedBaselines{} deep baselines
and its JSD from \SepReadsJSDAhead{}, with an MAE \MatchedMAEbtlGainPct\% below the strongest of them,
against \TunedMAEGainPct\% with tuned weights (\cref{app:matched}).
Across a capacity ladder from \LadderMin{} to \LadderMax{} parameters, retuned at every rung, no 
model moves by more than \LadderSpanMax{} JSD, and BIOTRAM at its worst size (\LadderBioWorst{}) is 
still ahead of \LadderBaseBestModel{} at its best (\LadderBaseBest{}).
Only training-set size reverses the lead: on nested subsets of the training bottles,
BIOTRAM trails every other model up to \DataVolBehindTo{} bottles and leads the four deep baselines
from \DataVolDeepAheadFrom{} on, and every model is still improving at the full set, BIOTRAM most
steeply (\cref{app:scaling}). Its structure thus pays off once there are enough bottles to learn the
per-BIN quantities it relies on.

\mypar{External anchors.}\label{subsec:anchors_eval} Mean log area and mean log dry mass correlate at
$r = \AreaMassR$ across the \AreaMassNBins{} weighed BINs, BarcodeBERT predicts the mean log area of
held-out BINs from the barcode alone at $r = \SeqAreaR$ (against $\SeqAreaNNR$ for the nearest
training barcode), and $\log(F \cdot R)$ correlates with a BIN's empirical 
$\log(\mathrm{reads}/\mathrm{specimen})$ at $r = \PrimerRpsR$. The reverse primer site comes only
from mitogenomes (\PrimerMitoPct\% of BINs) and is otherwise imputed, which can be a concern when 
deploying the model on new communities (\cref{app:primer_eval}).

\mypar{Ablations.}\label{subsec:ablations} A hierarchical ablation over \AblBlocks{} capability 
blocks, each removed from the reported configuration and compared seed-for-seed against a control, 
finds that the two largest effects are data and chemistry rather than architecture: the primer
binding-efficiency block (\AblPrimerTestJSD{} test JSD) and the extra training bottles with excess
(\AblExcessTestJSD{}). The amplification module's cross-BIN attention costs \AblRateTransformerTestJSD{},
and giving it the BIN's own read is worth \AblReadsInModuleTestJSD{} on its own (full results in
\cref{app:ablation}).

\mypar{Do the predicted quantities mean what they are named?} The model's distinguishing
property is its intermediate quantities, which should correspond to physical quantities and explain
its predictions. Since the primer efficiency and body mass receive the corresponding external 
estimates as input features, correlating them with those estimates would be circular.
Instead, we check, for BIOTRAM at the development weights of \cref{tab:matched}, how well the 
predicted log reads per specimen, averaged per BIN over the test bottles ($\bar{a}_i + \log m_i$; 
\cref{eq:rate}), correlates with its empirical value: it reaches Pearson $+\MechSumR \pm \MechSumSd$ 
over \MechSeeds{} seeds, and the read-blind variant, whose amplification term never sees the read 
and so cannot reproduce it mechanically, still reaches $+\MechTwinSumR \pm \MechTwinSumSd$ 
(\cref{app:latents}).

\section{Discussion}\label{sec:limits}

We have presented a new dataset that pairs metabarcoding read counts with specimen counts, and 
designed BIOTRAM, a machine learning model built specifically to map between those
two quantities, trained and evaluated on that dataset.

\mypar{Limitations.} The number of training bottles, not model capacity, limits accuracy: larger
variants of BIOTRAM do not improve it, no baseline gains more than \LadderSpanMax{} JSD across the
capacity ladder, and every model is still improving at the full training set
(\cref{app:scaling}). The counts identify only the product of amplification and body mass, and
the amplification module sees the BIN's own read, so the likelihood alone pins down neither term.
What separates them is what each component reads: of the two, only the amplification module sees
the primer features, the reads and the rest of the bottle. The latent quantities are therefore
cleanest to interpret in the read-blind variant (\cref{tab:matched}), at a cost in accuracy. Body
size varies within a BIN with life stage, sex and season, the image-based estimate depends on
specimen pose, and the weighed-specimen dataset is small. The largest counts carry the least certain
training labels: excess-affected taxa are \ExcessBinPct{}\% of BINs but \ExcessHighOccPct{}\% of
the rows with more than a hundred specimens, and in the bottles with excess the training labels
count only the individually barcoded specimens, leaving the excess out (\cref{app:excess}). The deep
models, BIOTRAM included, also receive the bottle's total specimen count $\smash[t]{\nu^{(b)}}$, here
the sum of its barcoded counts, which is expected to be known at deployment. BIOTRAM does not yet use
spike-ins, known quantities of specimens or DNA added to a sample, which improve quantification in
metabarcoding and metagenomics \citep{ChenHu_SpikeInControl, HarrisonCalder_AbsoluteAbundance,
IwaszkiewiczEggebrecht_MetabarcodingAccuracy, JiHuotari_SPIKEPIPE}. Finally, the pretrained
geographic encoder \citep{Klemmer2025-sn} contributes little, suggesting that general-purpose
location embeddings do not transfer readily to biodiversity prediction.

\mypar{Conclusions.} BIOTRAM treats the biochemistry of metabarcoding as part of the model.
Every quantity it predicts has a physical meaning, two are anchored to independent measurements, and,
apart from a learned merge of the three replicates, the arithmetic that converts them into counts is 
fixed. With every model's loss weights tuned, it reaches the lowest composition divergence and the 
lowest count error of the architectures we compared. Every model we tested degrades substantially 
under spatial, temporal and batch shift. Primer binding efficiency is among the largest gains in our 
ablation, which argues for single-specimen barcodes that retain the primer binding sites 
(\cref{app:primer}).

\FloatBarrier
\subsection*{Acknowledgements}

Resources used in preparing this research were provided, in part, by the Province of Ontario, the Government of Canada through CIFAR, and \href{https://vectorinstitute.ai/partnerships/current-partners/}{companies sponsoring} the Vector Institute.
GWT and SCL are supported by the INSPIRE (Integrated Network for the Surveillance of Pathogens: Increasing REsilience and capacity in Canada's pandemic response) project funded through the Canada Biomedical Research Fund (grant CBRF2-2023-00008), the Biomedical Research Infrastructure Fund, and the Ontario Research Fund.
This work was supported by the \href{http://abcresearchcenter.org/}{AI and Biodiversity Change (ABC) Global Center}, which is funded by the US National Science Foundation under \href{https://www.nsf.gov/awardsearch/showAward?AWD_ID=2330423&HistoricalAwards=false}{Award No. 2330423} Natural Sciences and Engineering Research Council of Canada (NSERC) and the Canada CIFAR AI Chairs program. GWT is also supported by the Canada Research Chairs program (grant~CRC-2021-00561).
This study was enabled by support to PDNH from the Government of Canada through Genome Canada and Ontario Genomics (OGI-208 and OGI-233), by the New Frontiers in Research Fund (NFRFT-2020-00073), and Canada First Research Excellence Fund (CFREF)–Food from Thought. Infrastructure required to carry out the work was acquired with support from the Canada Foundation for Innovation (CFI) and the Gordon and Betty Moore Foundation. A Major Science Infrastructure award (MSI 42450) from CFI plays a key role in sustaining the CBG’s informatics and sequencing platforms.

\subsection*{Author Contributions}

\begingroup %
\setlength{\parindent}{0pt}
\leavevmode\setlength{\parskip}{0pt}%
\textbf{Conceptualization:} Emma Boehly, Scott C. Lowe, Ken A. Thompson, Will Armstrong, Graham W. Taylor.

\textbf{Methodology:} Emma Boehly, Scott C. Lowe, Ken A. Thompson, Anthony Fuller, Will Armstrong, Neelu Madan, Tiancheng Gao, Graham W. Taylor.

\textbf{Software:} Emma Boehly, Ken A. Thompson, Will Armstrong, Sameer Padhye.

\textbf{Validation:} Emma Boehly, Scott C. Lowe, Graham W. Taylor.

\textbf{Formal analysis:} Emma Boehly, Ken A. Thompson, Will Armstrong.

\textbf{Investigation:} Emma Boehly, Scott C. Lowe, Ken A. Thompson, Will Armstrong, Sameer Padhye, Jayme Sones, Kate H. J. Perez, Bess Hardwick, Brendan Furneaux, Evgeny V. Zakharov, Dirk Steinke, Jeremy R. deWaard, Stephanie L. deWaard, Jaclyn McKeown, Norman Monkhouse, Cole Pace, Sujeevan Ratnasingham, Raina Schmuck, Amy Thompson, Jorge L. León-Cortés, Luciano Cagnolo, Julieta Soledad Arena, Jeremías Domínguez Masciale, Errol Douwes, Elli Groner, Masae Iwamoto Ishihara, Shunsuke Matsuoka, Mailyn A. Gonzalez, Alexandre Antonelli, Renata S. Sousa-Lima, Armin Niessner, Patrícia Guedes, Elizabeth G. Biro, Zuania Colón-Piñeiro, Dezene P. W. Huber, Santosh Mazumdar, Magda Bou Dagher Kharrat, Eddie P. Mondejar, Pablo Luis Tubaro, Eugene Marais, Marie-Eugénie Maggia, Darío Alejandro Lijtmaer, Sean W. J. Prosser, Andrea Murillo-Vázquez, Paola Pulido-Santacruz, Paula Leitman, Thiago Arnt Dorigo, Nao Ito, Paul Knoops, Maria L. Kuzmina, Hikaru Morishita, Keishi Kimoto, Dimby Raharinjanahary, Rhian J. Smith, Solny A. Adalsteinsson, Kevin G. Borja-Acosta, Paul Amouroux, Angela M. Mendoza-Henao, Angélica Diaz-Pulido, Anamaria Dal Molin, Luke L. Powell, Diego Inclan, Alice Calvente, David Dunson, Brian Fisher, Tomas Roslin, Otso Ovaskainen, Paul D. N. Hebert, Graham W. Taylor.

\textbf{Resources:} Ken A. Thompson, Sameer Padhye, Jayme Sones, Kate H. J. Perez, David Dunson, Brian Fisher, Tomas Roslin, Otso Ovaskainen, Graham W. Taylor.

\textbf{Data curation:} Emma Boehly, Ken A. Thompson, Jayme Sones, Kate H. J. Perez, Bess Hardwick, Brendan Furneaux, Dirk Steinke.

\textbf{Writing -- original draft:} Emma Boehly, Scott C. Lowe, Ken A. Thompson, Graham W. Taylor.

\textbf{Writing -- review \& editing:} Emma Boehly, Scott C. Lowe, Ken A. Thompson, Anthony Fuller, Iuliia Zarubiieva, Brendan Furneaux, Dirk Steinke, Dezene P. W. Huber, Marie-Eugénie Maggia, David Dunson, Brian Fisher, Tomas Roslin, Otso Ovaskainen, Graham W. Taylor.

\textbf{Visualization:} Emma Boehly, Scott C. Lowe, Iuliia Zarubiieva, Graham W. Taylor.

\textbf{Supervision:} Scott C. Lowe, Paul D. N. Hebert, Graham W. Taylor.

\textbf{Project administration:} Ken A. Thompson, Iuliia Zarubiieva, Bess Hardwick, Rhian J. Smith, David Dunson, Brian Fisher, Tomas Roslin, Otso Ovaskainen, Graham W. Taylor.

\textbf{Funding acquisition:} David Dunson, Brian Fisher, Tomas Roslin, Otso Ovaskainen, Paul D. N. Hebert, Graham W. Taylor.

\par
\endgroup

\FloatBarrier
{
\bibliography{biotram}
\bibliographystyle{iclr2027_conference}
}

\clearpage
\label{marker:appendixstart}
\appendix
\crefalias{section}{appendix}\crefalias{subsection}{appendix}\crefalias{subsubsection}{appendix}
\section*{Appendices}
{
    \hypersetup{hidelinks}%
    \startcontents[sections]
    \printcontents[sections]{l}{1}{\setcounter{tocdepth}{2}}
}
\FloatBarrier
\clearpage

\section{Data provenance}
\label{app:data_provenance}
\begin{figure}[h!]\centering
\includegraphics[width=\textwidth]{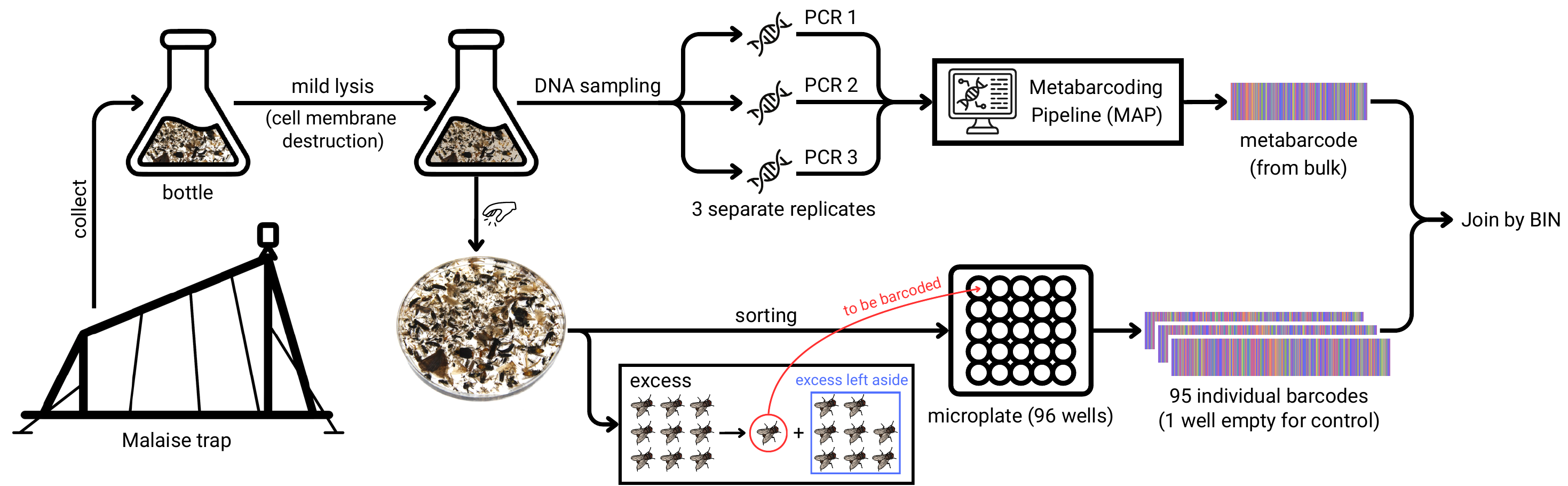}
\caption{\textbf{DNA barcoding data pipeline.} Each bottle is collected from a Malaise trap and 
undergoes mild lysis to release DNA into solution while preserving the physical specimens. From this 
lysate, a DNA sample is drawn for metabarcoding: three independent PCR replicates are amplified from 
the bulk sample and passed through the Metabarcoding Analysis Pipeline (MAP) to produce a single bulk 
metabarcode. In parallel, the physical sample is fully sorted and its specimens are plated onto 
96-well microplates for individual barcoding, yielding up to 95 individual barcodes per plate 
(one well is left empty as a negative control); where many specimens appear to belong to the same 
species, only some representatives are barcoded, while the remaining specimens are counted and set 
aside as excess. Individual barcode and metabarcode records are then joined by Barcode Index Number 
(BIN).}
\label{fig:data_pipeline}
\end{figure}

We identified a set of Malaise trap samples (bottles) that met two conditions: (i) the bottle was 
metabarcoded upon receipt via mild lysis and (ii) the bottle was entirely sorted for single-specimen 
barcoding, with any `excess' (see below) recorded. Samples were derived from the LIFEPLAN project 
\citep{deWaard_LIFEPLANmbc, Hardwick2024_LIFEPLAN} and the Global Malaise Program \citep{Geiger2016-qp}. 
Permissions to use each bottle were arranged with the data owners. Figure~\ref{fig:data_pipeline} 
summarizes the DNA barcoding data pipeline. 

\section{Core dataset generation and assembly}
\label{app:data_assembly}
Raw metabarcode data were obtained as paired-end gzipped fastq files (i.e., .R1.fq.gz \& .R2.fq.gz) 
accompanied by a plate map. Plate maps were converted into MAP parameter files and were analyzed with 
MAP v1.0 \citep{Prosser2026-yp} with default settings for Illumina data. OTUs were queried against 
the June 2026 BOLDistilled COI library \citep{Prosser2025_BOLDistilled} to provide BIN matches and 
taxonomy. 

Once tabular metabarcode results were obtained, we integrated the data with data from BOLD to produce a single table containing all relevant information from both sources.
Data import, manipulation, and file handling throughout relied on the \texttt{tidyverse} \citep{Wickham_Tidyverse} and \texttt{readxl} \citep{WickhamBryan_readxl} packages, while retrieval of specimen-level Barcode Core Data Model (BCDM) records from the Barcode of Life Data System was performed with \texttt{BOLDconnectR} \citep{Padhye_2025_BOLDconnectR1}.

Processing proceeded independently for each sequencing lane.
For every lane, filtered OTU tables and per-replicate read counts were loaded into memory.
Next, BCDM records for all process IDs associated with the lane were retrieved.
Nucleotide sequences from the BOLD SSB records were deduplicated and used to generate a FASTA file, which was then queried against BOLDistilled to obtain BIN assignments. This was done because the `true' BIN assignments on BOLD cannot be efficiently replicated for metabarcode data due to (i) the RESL algorithm being proprietary, (ii) the use of private data on BOLD, and (iii) BOLD's Barcode ID being only available via web browser and restricted to 100 sequences at a time during an exhaustive search. 
Since the reference library for metabarcode data becomes out of date every three months, metabarcoding OTUs assigned to Arthropoda were also matched to their corresponding BIN via the BOLDistilled lookup.
Following BIN assignment, the metabarcode and single-specimen barcode data were joined by BIN in a single table that summarized total specimen counts and metabarcode read statistics within each bottle.

\section{Derivation of the forward DNA amplification relationship}\label{app:rate}
\begin{figure}[htb]\centering
\includegraphics[width=0.77\textwidth]{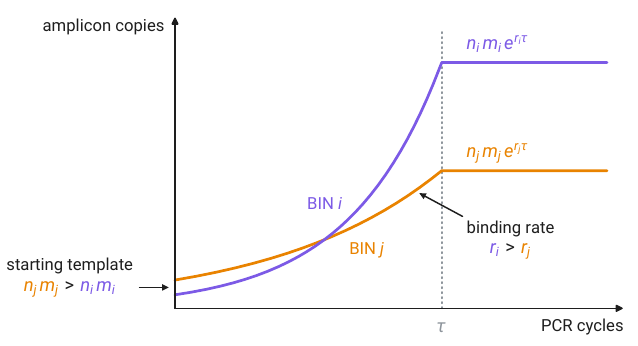}
\caption{\textbf{Amplification from different starting points at different rates.}
Two BINs amplified in the same PCR reaction (\cref{eq:pcr-start,eq:pcr-growth,eq:pcr-budget}).
BIN $j$ starts from more template ($n_j m_j > n_i m_i$: heavier specimens, DNA more accessible, or both), but the primers bind BIN $i$ better ($r_i > r_j$), so its copies grow faster and overtake those of BIN $j$.
When the shared reagents run out at cycle $\tau$, both plateau at $n\,m\,e^{r\tau}$, up to the factor $\kappa$ common to the reaction, and sequencing returns reads in proportion to these final copy numbers (\cref{eq:pcr-reads}).
Had the reaction stopped before the curves cross, BIN $j$ would have yielded more reads.
The axes are schematic: a real reaction multiplies its template by many orders of magnitude.}
\label{fig:pcr}
\end{figure}

\Cref{eq:rate} follows from the kinetics of a PCR run on a mixture of templates \citep{Suzuki1996-ed,Polz1998-jl}, which \cref{fig:pcr} traces for two BINs.
Consider one PCR replicate $\rho$ of bottle $b$, and let $x_i(t)$ be the number of copies of BIN $i$'s amplicon after $t$ cycles, treating $t$ as continuous.
We drop the indices $b$ and $\rho$ until the end of the derivation.

\mypar{Starting point.}
Each specimen releases DNA into the lysate, on average in proportion to its body mass 
\citep{Elbrecht2015-og}, and the reaction draws a fixed fraction $\kappa$ of that lysate.
BIN $i$ therefore enters the reaction with
\begin{equation}%
x_i(0) \;=\; \kappa\, n_i\, m_i
\label{eq:pcr-start}
\end{equation}
copies of template.
As $\kappa$ is shared by every BIN of the reaction, the starting points of different BINs differ 
through the composition of the bottle ($n_i$) and through the body mass of their specimens ($m_i$).

\mypar{Exponential growth.}
In each cycle, a fraction $E_i \in [0, 1]$ of BIN $i$'s templates is copied.
This efficiency depends on how well the primers bind the BIN's sequence \citep{Polz1998-jl,Elbrecht2015-og}, so it differs between BINs (\cref{app:primer}), but not from one cycle to the next while reagents are plentiful.
After $t$ cycles,
\begin{equation}%
x_i(t) \;=\; x_i(0)\,(1 + E_i)^{t} \;=\; \kappa\, n_i\, m_i\, e^{r_i t},
\qquad r_i \;=\; \ln(1 + E_i),
\label{eq:pcr-growth}
\end{equation}
so each BIN grows exponentially, at a rate $r_i$ between $0$ (the primers do not bind) and $\ln 2$ (every template is copied in every cycle).

\mypar{Plateau.}
The primers, nucleotides and polymerase are shared by every BIN in the reaction, and together they support at most some total number $X$ of copies.
Idealizing saturation as an abrupt stop shared by every BIN, amplification ends at the cycle $\tau$ at which the total product reaches this budget,
\begin{equation}%
\sum_k x_k(\tau) \;=\; \kappa \sum_k n_k\, m_k\, e^{r_k \tau} \;=\; X,
\label{eq:pcr-budget}
\end{equation}
and BIN $i$ ends with $x_i(\tau) = \kappa\, n_i\, m_i\, e^{r_i \tau}$ copies.
The duration $\tau$ is thus set by the whole bottle: a bottle with more template, or with more readily amplified BINs, exhausts the budget in fewer cycles, so each of its specimens is amplified less.
This is the bottle-level saturation of \cref{sec:methods}.

\mypar{Sequencing.}
The sequencer returns $T$ reads for the replicate, each drawn from the pooled product, so in expectation BIN $i$ receives reads in proportion to its share of that product,
\begin{equation}%
\mathrm{reads}_i \;=\; T\,\frac{x_i(\tau)}{\sum_k x_k(\tau)}
\;=\; \frac{T\kappa}{X}\; n_i\, m_i\, e^{r_i \tau}.
\label{eq:pcr-reads}
\end{equation}
The factor $T\kappa/X$ is common to every BIN of the replicate.
Restoring the indices, \cref{eq:pcr-reads} is \cref{eq:rate}, with the amplification term $a_{i,\rho}^{(b)} = r_i\,\tau_{\rho}^{(b)}$.

\mypar{Consequences for the model.}
Two BINs of the same reaction compare as
\begin{equation}%
\log \frac{\mathrm{reads}_i}{\mathrm{reads}_j} \;=\; \log \frac{n_i}{n_j} \;+\; \log \frac{m_i}{m_j} \;+\; (r_i - r_j)\,\tau,
\label{eq:pcr-ratio}
\end{equation}
so the read ratio confounds the ratio of specimen counts, which we want to recover, with the ratio 
of body masses and with a difference in amplification that grows with the duration of the reaction.
A BIN that starts from less template but binds the primers better overtakes the other if the 
reaction runs long enough (\cref{fig:pcr}), and whether it does depends, through 
\cref{eq:pcr-budget}, on every other BIN in the bottle.
The idealized amplification term factorizes into a per-BIN rate and a per-reaction duration.
BIOTRAM keeps the product form of \cref{eq:rate} but does not impose this factorization: its 
attention module predicts $a_{i,\rho}^{(b)}$ from the content of the whole bottle 
(\cref{subsec:model}), so the term can also absorb departures from the idealization 
\citep{Suzuki1996-ed}.

\section{Primer binding efficiency}\label{app:primer}

\subsection{Methods}\label{app:primer_meth}

\mypar{Retrieving the binding sites.} How well a BIN amplifies directly depends on its primer 
binding efficiency that itself depends on the sequences of its two primer binding sites. During 
metabarcoding PCR, the BF3 and BR2 primers must bind to the 20 nucleotides that flank the amplicon 
DNA sequence. Since those binding sites are replaced by the primer sequence itself, we cannot use 
the sequenced amplicons to determine their binding affinity. To accomplish this, we took two 
approaches.

The first route to obtaining primer binding sites used full mitogenomes, which contain both sites. 
We downloaded 11,080 complete mitogenomes from GenBank \citep{Sayers2025_Genbank}, restricting our 
search to Arthropoda records annotated as mitochondrial complete genomes. Records were queried and 
retrieved via the NCBI E-utilities using the \texttt{rentrez} R package \citep{Winter2017_Rentrez}, 
filtered to a minimum length of 14,503\,bp (the smallest known mitogenome), and deduplicated to one 
record per named species. Sequences were parsed and written to per-species FASTA files using 
\texttt{Biostrings} \citep{Pages2024BioStrings}. For each downloaded mitogenome, we obtained 
taxonomic information using the \texttt{taxize} R package \citep{Chamberlain2013taxize}, which 
queries the NCBI Taxonomy database. Species names were taken directly from each genome's GenBank 
organism field, so only the ranked classification required an additional network query. For each 
mitogenome, we searched for binding sites of the degenerate COI primers BF3 and BR2 
\citep{Elbrecht2019-jy}, allowing up to four mismatches per primer, using \texttt{Biostrings}. Sites 
overlapping ambiguous (N) bases were discarded, and where a primer matched more than one site in a 
genome, the best-scoring site was retained via pairwise alignment implemented in the R package 
\texttt{pwalign} \citep{Aboyoun2024pwalign}. The amplicon between the retained BF3 and BR2 sites was 
extracted for each genome and written to a FASTA file, and candidate amplicons were filtered to a 
plausible COI length (388--424\,bp) and in-frame codon length. \PrimerNGenomes{} mitogenome records 
passed these filters and contain both sites, which gives forward and reverse sites for 
\PrimerNGenomeBins{} BINs, of which \PrimerMitoBins{} were in our dataset. When several genomes 
matched to one BIN, they were averaged. 

The second route used full-length (658 bp) barcodes from single-specimen barcoding, which cover far 
more BINs, but only the forward site. The BF3/BR2 amplicon starts approximately \AmpliconStart{} 
bases into the barcode region (median length \BarcodeLenMedian{} bases in BIOSCAN records), so the 
forward site lies inside a barcode record, whereas the reverse site lies beyond its 3' end. Forward 
sites extracted from these records cover \PrimerBarcodeFPct\% of the BINs in our dataset. Forward 
sites were located in the full-length record using the same method as was used on mitogenomes.

\mypar{From site to efficiency.} Each site is scored into an efficiency between 0 and 1 by a 
function of its full 20-nucleotide sequence. For each retained amplicon, we estimated PCR 
efficiency based on the sequences of the binding sites and the primers, using the 
\texttt{CalculateEfficiencyPCR} function in the R package \texttt{DECIPHER} \citep{Wright_DECIPHER}, which calculates the Gibbs free energy, formamide melt point, and hybridization efficiency of the primer and its binding site. 

\mypar{Imputation.} BINs without a retrieved site receive predicted efficiencies. The forward 
and reverse efficiency, predicted for \PrimerImputedFPct\% and \PrimerImputedRPct\% of BINs, respectively, is an equal blend of a ridge regression on the one-hot encoded amplicon and a phylogenetic kernel 
ridge regression whose kernel is defined by the pairwise alignment identity between amplicons. Each 
BIN carries its two efficiencies, the logarithm of their product and, for each primer, a flag
marking efficiencies taken from a mitogenome or from a barcode record of documented provenance.

\mypar{A note for sequencing practice.} Primer binding sites flank the amplified fragment and are
discarded by short-read metabarcoding, yet the primer binding-efficiency block is one of the two
largest contributions in our ablation (\cref{app:ablation}). Extending the region sequenced by
single-specimen barcoding by about 50 base pairs in each direction would make this information
directly available in future datasets, and as read lengths grow, the limitation may resolve on its
own. In general, for effective estimation of abundance, the marker used for single-specimen
barcoding should remain longer than the one used for metabarcoding, ideally capturing both the
forward and the reverse binding site.

\subsection{Evaluation}\label{app:primer_eval}
\mypar{Reference values.} Over the \PrimerNGenomeBins{} BINs with a
mitogenome, the forward site has a mean efficiency of \PrimerFMean{} and the reverse site 
\PrimerRMean{}, and \PrimerFPoorPct\% and \PrimerRPoorPct\% of BINs, respectively, fall below 0.4. 
Most sites match the primer exactly (\PrimerFExactPct\% of forward and \PrimerRExactPct\% of reverse 
sites), but an exact match does not guarantee efficient binding. The forward and 
reverse efficiencies of a BIN are nearly independent (Pearson $r = \PrimerFRCorr$), and efficiency 
differs between taxa. Among the BINs of our dataset with a retrieved forward site, the mean forward 
efficiency is \PrimerAraneaeF{} in spiders (Araneae) and \PrimerHymenopteraF{} in Hymenoptera, with 
\PrimerAraneaePoorPct\% and \PrimerHymenopteraPoorPct\% of their BINs below 0.4, against 
\PrimerDipteraF{} in Diptera and \PrimerLepidopteraF{} in Lepidoptera. Over the \PrimerRpsNBins{}
BINs with at least one specimen in the excess-free bottles, $\log(F \cdot R)$ correlates with a BIN's
mean empirical $\log(\mathrm{reads}/\mathrm{specimen})$ at $r = \PrimerRpsR$, and at
$r = \PrimerRpsRBoth$ over the \PrimerRpsNBoth{} of them whose two sites both come from mitogenomes.

\mypar{Coverage.} Mitogenomes cover only \PrimerMitoBins{} of the BINs in our dataset
(\PrimerMitoPct\%), and they are the only source of reverse sites. Full-length barcode records 
supply a forward site for \PrimerBarcodeFPct\% of BINs, which carry \PrimerBarcodeFIndPct\% of the 
individuals in the excess-free bottles. In total, \PrimerRealFPct\% of BINs, carrying 
\PrimerRealFIndPct\% of those individuals, have a forward site read from a reference sequence. The 
forward efficiency is imputed for the remaining \PrimerImputedFPct\% (\PrimerImputedFIndPct\% of the 
individuals), and the reverse efficiency for \PrimerImputedRPct\% of BINs.

\mypar{Imputation accuracy.} We train the barcode-based predictor on the \PrimerImpNTrain{} mitogenome 
BINs absent from our dataset and on 60{,}000 forward sites extracted from barcode records. We then
evaluate it on the \PrimerImpNTest{} mitogenome BINs present in it. The forward predictor explains 
\PrimerImpFRsqPct\% of the variance of forward efficiency (Pearson $r = \PrimerImpFR$, Spearman 
$\rho = \PrimerImpFRho$, MAE \PrimerImpFMAE), slightly less than a shrunken taxonomic mean 
(\PrimerTaxFRsqPct\%). The reverse predictor explains \PrimerImpRRsqPct\% of the variance of reverse 
efficiency ($r = \PrimerImpRR$, $\rho = \PrimerImpRRho$, MAE \PrimerImpRMAE), against 
\PrimerTaxRRsqPct\% for the taxonomic mean, and keeps its advantage on families absent from training 
($R^2 = \PrimerImpRFamRsq$ against \PrimerTaxRFamRsq). Its predictions are strongly shrunk toward 
the mean (standard deviation \PrimerImpRSd{} for imputed reverse efficiencies, against 
\PrimerMitoRSd{} for those from mitogenomes), so an imputed value mainly carries the trend of the 
BIN's relatives.

\section{`Excess' specimens}
\label{app:excess}
During routine single-specimen barcoding (SSB) workflows, technicians regularly encounter samples with many individuals of what appears to be the same species (Figure~\ref{fig:data_pipeline}). 
In such cases, it is neither practical nor cost-effective to barcode them all. 
These specimens are counted and then archived, and technicians note the microplate(s) on which some representative specimens were barcoded and remark on their taxonomy. 
The existence of excess will always be a feature of DNA barcoding data gathered in a resource-limited environment.
This \textit{excess} is recorded as a count and a taxon label, not as BINs, so the \NExcess{} of our \NBottles{} bottles that contain excess lack complete per-BIN ground truth, even though their metabarcoding data include the DNA of every specimen.
We evaluate only on the \NClean{} bottles without excess. Some of the others are used for training
only; the allocation of the excess to specific BINs is described below.

Since the excess were metabarcoded, it is essential to accurately update the BIN specimen counts to reflect it. 
To accomplish this, we obtained a copy of the excess database from the CBG's Specimens Unit, and filtered it to the bottles included in our dataset.
To identify the BINs associated with excess specimens, we retrieved the records of all relevant 
microplates from the data previously downloaded with BOLDconnectR \citep{Padhye_2025_BOLDconnectR1}.
This BOLD-derived dataset serves as the first reference point to identify candidate BINs.
Finally, the cleaned excess dataset of bottle, taxon, and count fields were written to a standardized 
output file that served as the authoritative excess-count reference for all subsequent analyses.

After assembling a clean database of all BINs on relevant microplates and a clean excess database, we identified potential candidate BINs. After refining the manually-entered taxonomic information and outdated nomenclature, we queried the unique sequences against the BOLDistilled library using VSEARCH to assign BINs. BIN assignments were then rejoined to the data by sequence identity.

To generate a candidate list of BINs consistent with each excess record, the full taxonomic hierarchy (phylum through species) associated with every BIN recovered from a given sample was reshaped into long format and matched, case-insensitively and within sample, against the corresponding excess taxon label, allowing a label recorded at any taxonomic rank to be resolved to the appropriate BIN(s). 
Data manipulation, reshaping, and joins throughout were performed using the \texttt{tidyverse} collection of packages \citep{Wickham_Tidyverse}. 
Excess records for which no candidate BIN could be identified within the focal sample set were flagged and proceeded through a secondary path (detailed below), while all successfully matched sample--taxon--BIN combinations were retained as the candidate excess dataset used in downstream specimen identification.

We then used the model itself to inform our excess assignment. We used a model trained only on 
no-excess bottles to predict the relative abundance of taxa in bottles with excess. 
(In the final testing set, only bottles with no excess were included.)
For each BIN in the dataset, we quantified its ``model prediction factor'' (MPF) as its predicted abundance divided by its true abundance, quantifying the degree to which the model over- or under-predicted abundance for that BIN in that sample. Samples with a true abundance of zero but nonzero read support were assigned the minimum nonzero true abundance observed within the sample to avoid division by zero. 
Within each sample--taxon group, MPFs were rescaled to relative weights that summed to 1. Excess counts were assigned to each BIN in proportion to their relative weights to produce corrected ground-truth abundances.

For taxa that did not have a candidate BIN, we assigned them by matching the excess taxon against family, order, class, and phylum---in that order of precedence until at least one candidate could be found.
All BINs sharing the matched higher rank were then treated as above and the same excess allocation procedure was applied.
Results were then merged with the main dataset.

\section{Body size estimation: pipeline and evaluation}\label{app:anchors}\label{app:bodysize}
Larger individuals contribute more DNA to the bulk DNA mixture, so the typical body size of a BIN is an
important signal for our task. We estimate it in three steps, described below, and evaluate the estimates in \cref{app:bodysize_eval}.

\subsection{Segmenting body area from specimen images}
During the individual barcoding pipeline, specimens are photographed individually.
Following image-based estimates of arthropod size and biomass \citep{Schneider2021-uh}, we segmented 
\SegNImages{} of these photographs. Our segmentation pipeline combines pretrained vision models
without fine-tuning, as follows.

\begin{figure*}[t]\centering
\includegraphics[width=\textwidth]{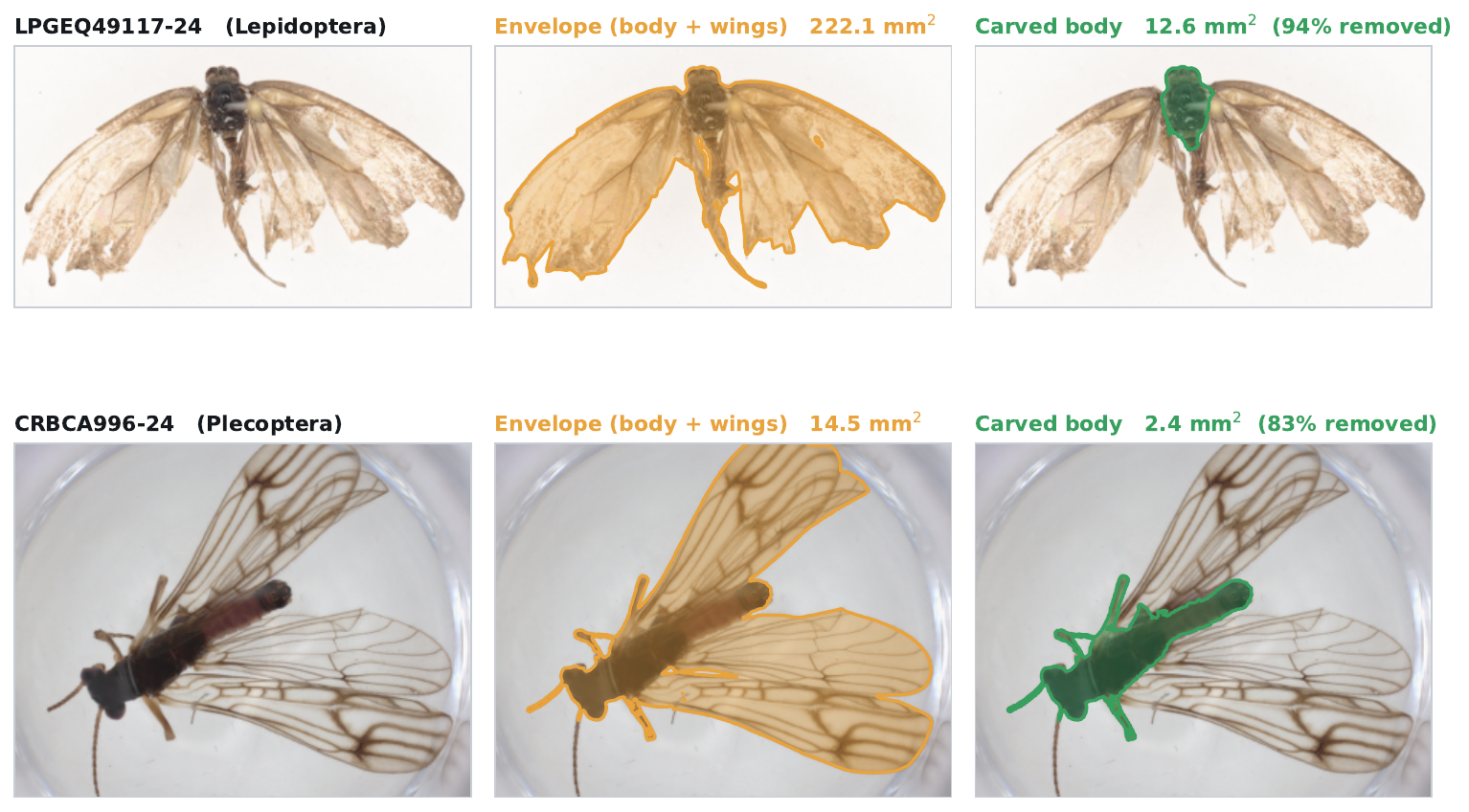}
\caption{\textbf{Segmentation and the wing carve, on two winged specimens.} Each row is one
specimen: the photograph, the \textcolor[HTML]{C6862B}{envelope} from the two full-organism models
(SAM-HQ and SAM~3), and the \textcolor[HTML]{2E8A4F}{carved body} the pipeline keeps. Masks are
drawn translucently over the photograph so the specimen stays visible underneath. \SegExAId{}
(\SegExAOrder{}) goes from \SegExAEnv{} to \SegExABody{}~mm$^2$, \SegExACarvedPct\% of the
silhouette removed; \SegExBId{} (\SegExBOrder{}) from \SegExBEnv{} to \SegExBBody{}~mm$^2$,
\SegExBCarvedPct\% removed. Areas are the calibrated values the size anchor is built from.}
\label{fig:segmentation}
\end{figure*}

\textbf{1.} We first locate the specimen with an open-vocabulary detector (Grounding DINO; \citealp{Liu2025_GroundingDINO}).
\textbf{2.} The identified bounding-box is fed to two box-prompted segmentation models (SAM-HQ and SAM~3; \citealp{Ke2023_SAMHQ,Carion2026_SAM3}) which outline its overall shape (``envelope''; middle panels of \cref{fig:segmentation}).
We keep SAM-HQ's mask, which traces finer detail, unless the two masks disagree (IoU below 0.8), in 
which case we keep the one with the higher predicted IoU; if SAM-HQ misses much of the specimen, we 
add the large regions that only SAM~3 finds. This way we prioritize the more detailed mask while 
avoiding the risk of missing parts of the specimen.
\textbf{3.} Next, we separate out the wings from the main body. To do this, we use two segmentation 
models focused on the body only:
(a) we ask a vision-language model (Qwen3-VL-8B-Instruct; \citealp{Bai2025_Qwen3VL}) for a bounding box around the compact body 
(head, thorax and abdomen, ignoring spread wings), which SAM~2 \citep{Ravi2025_SAM2} then segments.
(b) SegGPT \citep{Wang2023_SegGPT} prompted with five hand-annotated examples.
Spread wings add projected area but little tissue, so we remove them only when a wing-presence score 
is high. The score combines the vision-language model's judgement of whether the wings are spread 
with the disagreement between the envelope and the body masks. We then cut from the envelope the 
regions that the body masks exclude: all of them if the two body masks agree (IoU at least 0.8); 
otherwise the larger of the two exclusions when the vision-language model or SAM~3 confirms the 
wings, and only the regions both exclude when neither does. The pixel count of the final mask, 
converted to mm$^2$ with the image's scale (known from the imaging system or read from a scale bar), 
is the specimen's projected body area, a two-dimensional projection of its body. Averaging the 
logarithm of this area over the images of each BIN gives a per-BIN mean log area and a within-BIN 
spread for the \SegNBins{} BINs with images, which cover \SizeImagedPct\% of the BINs in our dataset.

\Cref{fig:segmentation} shows two examples of specimens with spread wings. The envelope (middle panels) is
what a general-purpose segmenter returns: the organism's outline, wings included. It is a faithful
silhouette of the specimen as a whole, but it is a poor proxy for its mass.
In particular, the wings provide a large cross-section of the image---for winged insects this is typically most of the surface area of the insect seen within the image---but contain very little tissue mass.
We refine this segmentation mask with the wing carve described above, which keeps the envelope's boundary only where the two body-only models
agree that the pixel is not wing. How much this changes is not a detail: across the \SegScored{} scored specimens,
\SegWingedPct\% carry spread wings, and on those the carve removes a median \SegWingCarvedPct\% of
the silhouette, reaching \SegWingCarvedHiPct\% at the ninetieth percentile. Over the whole set,
including specimens with folded or absent wings, the median removal is \SegCarvedAllPct\%.

\subsection{Estimating mass from segmented image surface area} Insect mass scales as a power of 2D body dimensions
\citep{Sample1993-rn}, so log mass should be linear in log area. We calibrate this relation on a 
separate collection of \MassNSpecimens{} dried and weighed specimens from \MassNBins{} BINs. Weighed 
and imaged specimens are different individuals, so the relation cannot be fitted specimen by 
specimen; instead, we match the mean log mass of each genus to its mean log area, over the 
\AlloNGenera{} genera with at least two specimens of each kind:
\begin{equation}%
\overline{\log \mathrm{mass}}_{t} \;=\; \alpha_t + \beta_t\,\overline{\log \mathrm{area}}_{t}.
\label{eq:allometry}
\end{equation}
The offset $\alpha_t$ is estimated hierarchically from class down to genus, each level shrunk toward 
its parent. The slope is shared across taxa, except in the orders with enough genera to estimate 
their own. Applying \cref{eq:allometry} to every imaged BIN gives a per-BIN dry mass.

\subsection{Generalizing to unseen BINs} Half of the BINs in our data, and any BIN new to a monitoring
program, have no image. To estimate their size as well, we fine-tune BarcodeBERT
\citep{Millan-Arias2026-pm} to predict, from the BIN's DNA barcode, both the BIN's mean log area and 
the within-BIN standard deviation of that log area, with a squared-error loss on each. Genera and 
BINs held out for evaluation are excluded from training (\cref{app:bodysize_eval}). The same 
architecture trained on the allometric masses of step (2) predicts mass instead. Because the barcode 
carries nothing about an individual, its life stage or its sex, both networks predict the typical 
size of a taxon and the spread around it, not the size of any one specimen. The reported version of 
BIOTRAM reads the barcode-predicted log area and its spread for every BIN, including BINs with 
images; the measured areas are not among its inputs. %

\subsection{Evaluation}\label{app:bodysize_eval}

\mypar{From area to mass.} The fitted slope of \cref{eq:allometry} is \AlloBeta{} overall and 
ranges from \AlloBetaLo{} (\AlloBetaLoOrder) to \AlloBetaHi{} (\AlloBetaHiOrder) across the orders 
with their own slope. All are below $1.5$, the value expected if mass scaled with the cube of a 
linear dimension and area with its square, that is, if body shape and density did not change with 
size. To test the relation on taxa it was not fitted on, we predict each genus with its family's 
parameters withheld (five-fold cross-validation). Genus mean log mass is then predicted with Pearson 
$r = \AlloHeldR$ ($R^2 = \AlloHeldRsq$) and a mean absolute error of \AlloHeldMAE{} natural-log 
units, against \AlloOrderMAE{} and \AlloGlobalMAE{} when each genus is given the mean mass of its 
order or the global mean. At the level of single BINs, the mean log area and mean log weighed mass 
correlate at $r = \AreaMassR$ across the \AreaMassNBins{} BINs that carry both measurements.

\mypar{Body size from the barcode.} On BINs held out from training whose genus is represented in
training, the barcode network predicts the mean log area with Pearson $r = \SeqAreaR$ (RMSE 
\SeqAreaRMSE{} natural-log units), against $r = \SeqAreaNNR$ (RMSE \SeqAreaNNRMSE) when each BIN is 
given the value of its nearest training barcode. On genera held out entirely it reaches 
$r = \SeqAreaGenusR$ (RMSE \SeqAreaGenusRMSE), against \SeqAreaGenusNNR{} (\SeqAreaGenusNNRMSE) for 
the nearest barcode. The predicted within-BIN spread is only weakly related to the observed one 
(Pearson $r \le \SeqSigmaRMax$ on held-out BINs), so it should be seen as a coarse taxon-level 
feature rather than a calibrated uncertainty.

\section{Detailed model architecture}\label{app:model}

\begin{figure}[htp!]\centering
\includegraphics[width=\textwidth]{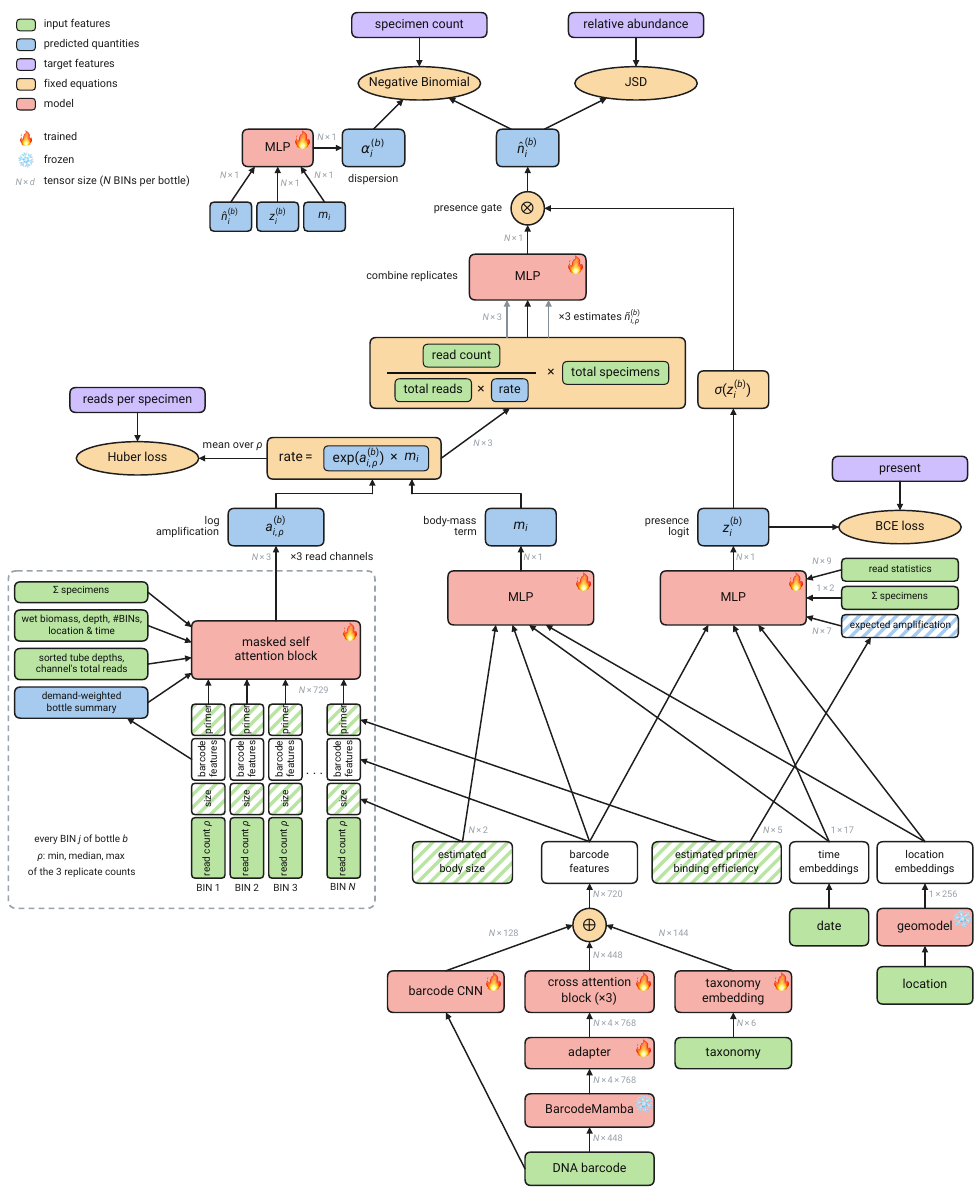}
\caption{\textbf{Full BIOTRAM architecture.}
Extends \cref{fig:arch_simp}C.
A frozen barcode embedding, a learned taxonomy
embedding and the bottle's location and date feed two per-BIN heads (the body-mass term $m_i$ and the
presence logit $\smash[t]{z_i^{(b)}}$) and a masked-attention module that runs over every BIN of the bottle, once per
replicate, and emits the amplification term $\smash[t]{a_{i,\rho}^{(b)}}$ from the BIN's identity, its primer and
size features, its own read count and the replicate's total read count. \Cref{eq:inversion} then
combines these quantities with the raw measured reads; the three per-replicate estimates are merged
and gated by $\sigma(\smash[t]{z_i^{(b)}})$. A separate MLP predicts the negative-binomial dispersion
$\smash[t]{\alpha_i^{(b)}}$ from $\smash[t]{\hat{n}_i^{(b)}}$, $\smash[t]{z_i^{(b)}}$ and $m_i$.}
\label{fig:arch}
\end{figure}

Expanding on the model overview in the main text (\cref{fig:arch_simp}), we show the full model architecture in \cref{fig:arch}.

\mypar{Batches.} A training example is a whole bottle rather than a single (bottle, BIN) row,
since both the amplification module and $\mathcal{L}_{\mathrm{comp}}$ act on a bottle's BINs jointly.
A batch holds \BatchBottles{} bottles, padded to the largest number of BINs among them and carrying
a mask; padded entries take no part in the attention and are excluded from every term of
\cref{eq:loss}.

\section{Evaluation methods}\label{app:metrics}

\begin{figure}[tb]\centering
\includegraphics[width=\textwidth]{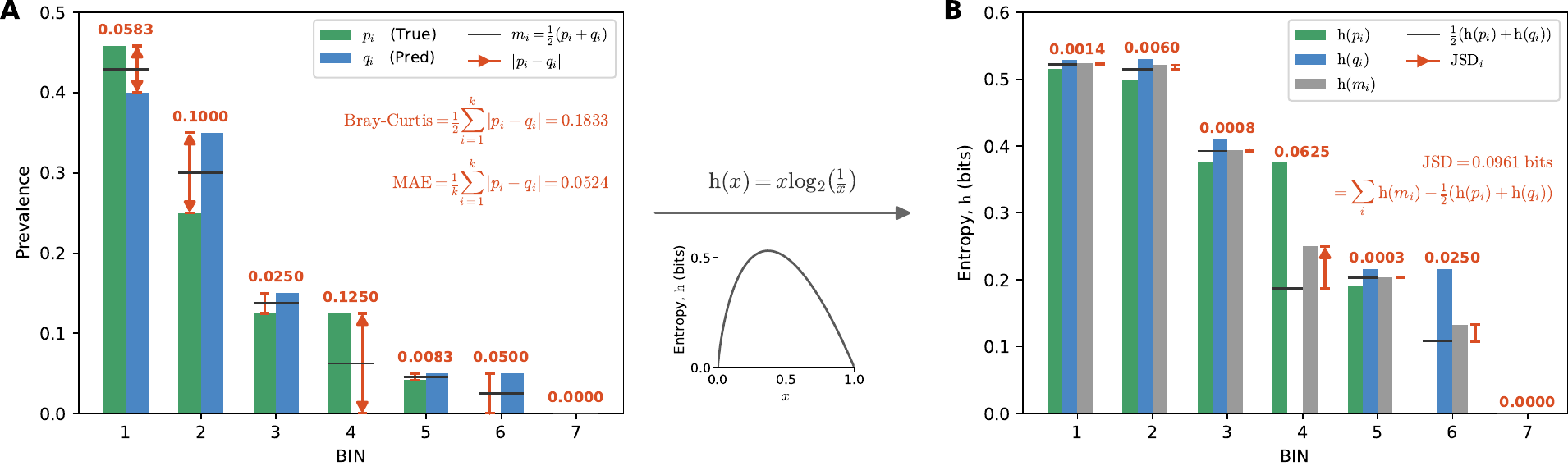}
\caption{%
\textbf{Calculation of mean absolute error (MAE), Bray--Curtis, and Jensen--Shannon divergence (JSD) 
between a true and predicted BIN composition.} For a single bottle, the figure writes $p_i$ and $q_i$
for the true and predicted composition, $p_i^{(b)}$ and $\hat{p}_i^{(b)}$ in \cref{eq:loss-comp}, and $m_i$
for their mean. (\textbf{A}): Illustrative true (green) and predicted (blue) distributions over BINs. Their mean 
distribution is also shown (black line). Bray--Curtis is half the sum of BIN-wise absolute 
differences (red) between the two distributions, whereas MAE is the mean of the absolute differences.
(\textbf{B}): To compute JSD, we take the entropy of each BIN composition. We compare the entropy of 
the mean distribution (grey bars) against the mean of the entropies for the true and predicted 
compositions (black line). JSD is their difference (red), summed over BINs. Note that larger relative 
errors (BINs 4 and 6) contribute much more to the total JSD than smaller relative errors (BINs 2 and 
1) despite having similar sized absolute difference between the predicted and true composition.
}
\label{fig:jsd-mae}
\end{figure}

BIOTRAM predicts a specimen count for each BIN in a bottle, and every metric derives from those 
same predicted counts. Some score the counts directly; others first normalize within the bottle to 
score the predicted composition. Since we care about the composition, about the counts themselves 
and about ecological usefulness, we report a set of metrics.
Our primary metrics, JSD and MAE, are illustrated in \Cref{fig:jsd-mae}.

\begin{figure}[tb]\centering
\begin{subfigure}[b]{0.35\textwidth}\centering
    \includegraphics[width=\linewidth]{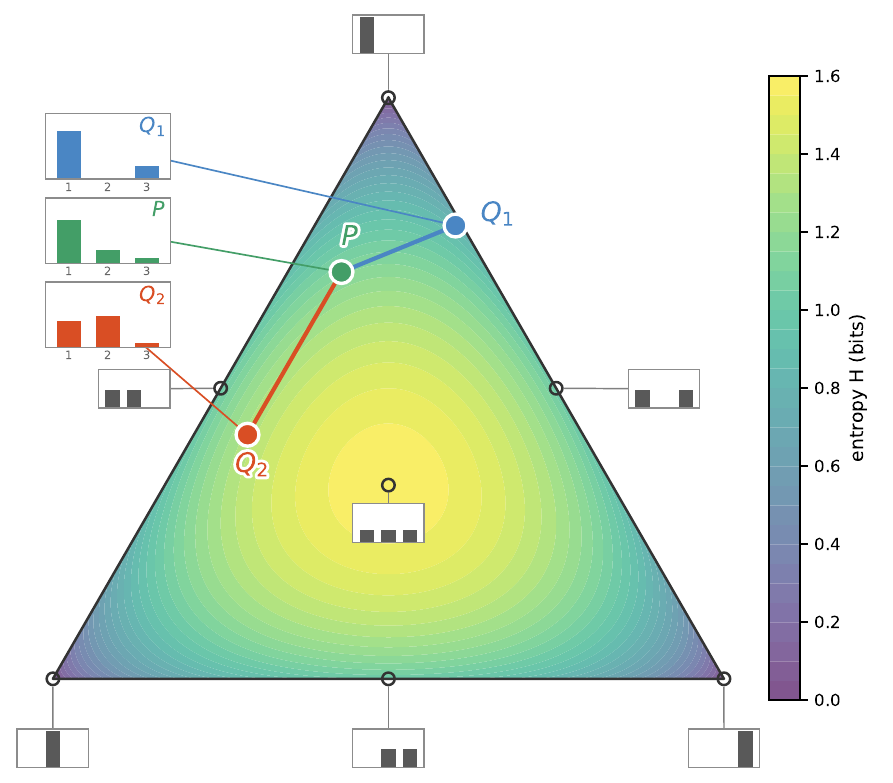}
    \caption{Entropy of all distributions.}
    \label{fig:jsd-entropy}
\end{subfigure}\hfill
\begin{subfigure}[b]{0.27\textwidth}\centering
    \includegraphics[width=\linewidth]{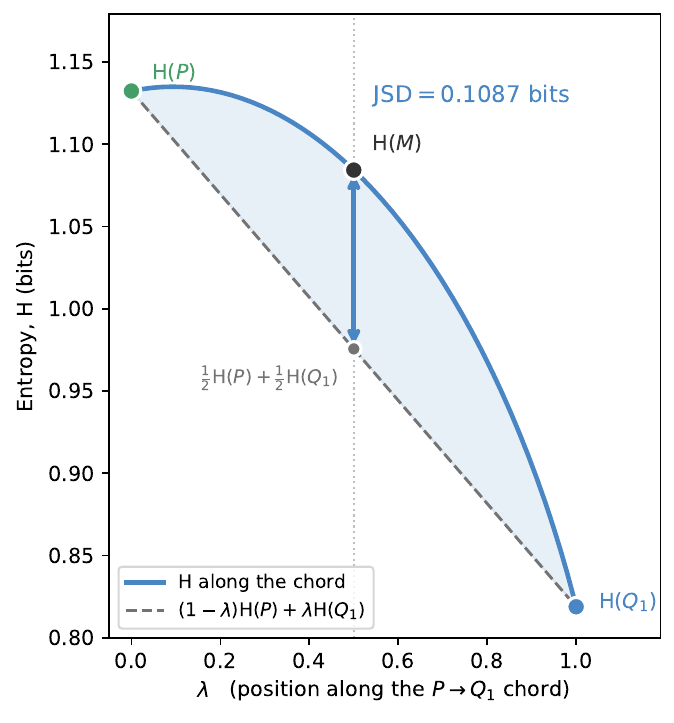}
    \caption{JSD between distr. $P$, $Q_1$.}
    \label{fig:jsd-chord}
\end{subfigure}\hfill
\begin{subfigure}[b]{0.35\textwidth}\centering
    \includegraphics[width=\linewidth]{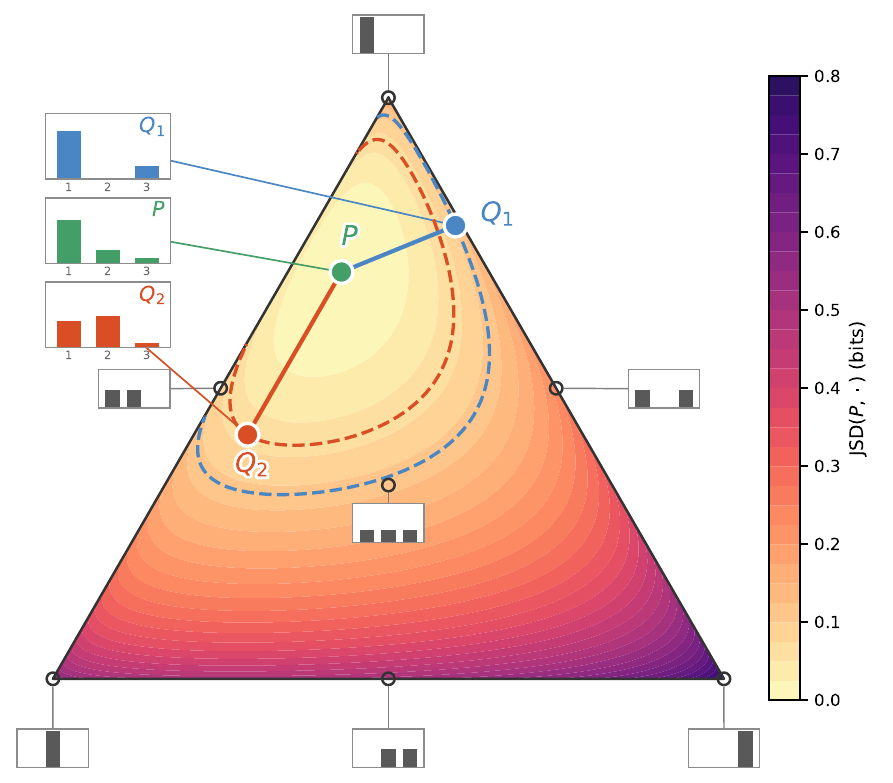}
    \caption{JSD of all distr. w.r.t. $P$.}
    \label{fig:jsd-divergence}
\end{subfigure}
\caption{%
\textbf{Jensen--Shannon divergence calculation on the 2-simplex.}
(\textbf{\subref{fig:jsd-entropy}}): The 2-simplex represents all possible distributions (summing to one) across three possible bins. The corners of the simplex correspond to distributions comprised of only a single bin, edges mixtures between two bins. The entropy of each possible distribution is shown (heatmap), with the centre point [1/3, 1/3, 1/3] having the highest entropy and vertices [1, 0, 0], etc., the lowest. We indicate the distributions for three sample distributions, P, Q1, and Q2.
(\textbf{\subref{fig:jsd-chord}}): The Jensen--Shannon divergence between distributions P and Q1 is the difference between the entropy at their midpoint, M, and the mean of their individual entropies.
(\textbf{\subref{fig:jsd-divergence}}): As with A, except the heatmap shows contours of equal JSD from distribution P. Note that distribution Q2 is closer to P than Q1 is under the JSD measurement, despite being further away in a Euclidean sense. This difference arises because Q1 underestimates bin 2 by too much; the true value of bin 2 in P is very surprising given the estimate in Q1.
}
\label{fig:jsd-calculation}
\end{figure}

\begin{itemize}
\item \textbf{Jensen--Shannon divergence (JSD)}, the primary metric, defined with the composition 
term of the objective (\cref{subsec:loss}). It turns predicted counts into proportions and measures 
how far the predicted composition sits from the true one. \Cref{fig:jsd-calculation} illustrates how this is calculated.
\item \textbf{Mean absolute error (MAE) per bottle.} Taken within a bottle and averaged across 
bottles. This is the target quantity in its own units. \Cref{tab:tuned,tab:matched} also split it by 
truth, over all test rows: MAE$_{=0}$ on rows with no individuals, MAE$_{>0}$ on the rest, because a model can win the 
aggregate purely by predicting zeros well.
\item \textbf{MAE on the log scale} (MAE-log), the same error on $\log$ counts. Counts span four 
orders of magnitude, so MAE is dominated by the abundant rows and MAE-log scores proportional rather 
than absolute error.
\item \textbf{Presence AUC}, whether truly present BINs rank above absent ones; $0.5$ is chance. The 
score ranked is the predicted count $\smash[t]{\hat{n}_i^{(b)}}$, for BIOTRAM exactly as for every baseline.
\item \textbf{Spearman correlation per bottle}, whether a bottle's BINs are put in the right order;
\cref{tab:tuned,tab:matched} compute it over all test rows at once.
\item \textbf{Shannon diversity $R^2$}, predicted against observed within-bottle Shannon diversity 
from single-specimen barcode data; a standard $\alpha$-diversity quantity, and a direct test of 
whether the predicted composition is the right shape rather than merely close row by row.
\item \textbf{Bray--Curtis Mantel correlation}, between-bottle dissimilarity on family-level 
abundances; a standard quantity in $\beta$-diversity analysis, assessing whether two bottles are 
correctly judged similar. It is computed at family level, because at the global scale of the dataset 
two bottles often share no BIN.
\end{itemize}
The last two are the quantities an ecological study would actually report, and they better represent 
the model's usefulness in practice.

\section{Full baseline comparison}\label{app:baselines}

\begin{figure}[t]\centering
\includegraphics[width=\textwidth]{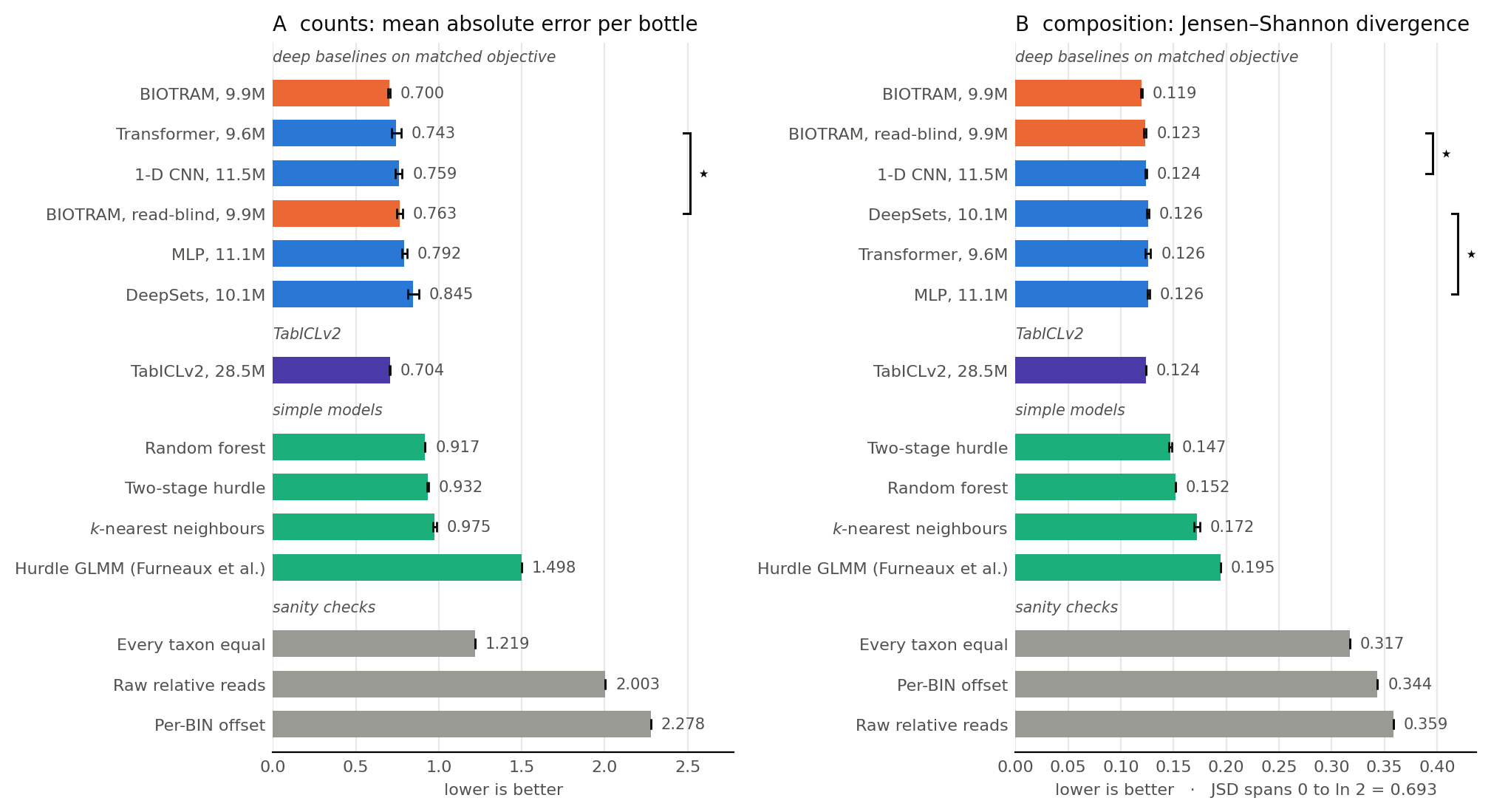}
\caption{\textbf{The comparison of \cref{tab:matched,tab:main} in bar form.} Bars start at zero, and
error bars show one standard deviation across seeds. Rows are grouped because the groups are not 
comparable to one another: only the first holds the training objective fixed with no external model 
on either side, and so only it speaks to the architecture. \textbf{A starred bracket spans rows that 
are not distinguishable} by a two-sided Welch test over seeds at $\alpha = 0.05$; it is drawn for 
the matched-objective group only, since elsewhere the objective differs between rows and such a 
statement would not be like-for-like. On the counts BIOTRAM is clear of \SepReadsMAEbtlAhead{} of the 
\NMatchedBaselines{} matched baselines and on composition of \SepReadsJSDAhead{}; the read-blind 
variant shown beside it is clear of \SepGenMAEbtlAhead{} on the counts and of \SepGenJSDAhead{} on 
composition. BIOTRAM is \TabICLCountRel{} TabICLv2 on count accuracy, \TabICLMAEGapPct\% apart, at 
\TabICLParamPct\% of its parameters.} \label{fig:baselines}
\end{figure}

\Cref{tab:tuned} in the main text trains every deep model on the objective of \cref{eq:loss}, each
with the auxiliary weights tuned for it (\cref{app:losstune}). \Cref{tab:matched} gives every deep
model the same weights, those selected during development (\cref{app:losstune}), so that only the architecture differs, and
\cref{tab:main} lets every model use the objective it does best under, which answers the more
practical question of what the best available predictor is. The two are not comparable to one
another and should not be read as a single ranking. \Cref{fig:baselines} shows both in bar form.

\mypar{Architecture only.}\label{app:matched} With the exact same loss, BIOTRAM's MAE per bottle is
\MatchedMAEbtlGainPct\% below that of the strongest baseline, against \TunedMAEGainPct\% when each
model has its own weights, and its JSD differs from the best baseline's by $\MatchedJSDGap$, against
$\TunedJSDGap$. Split by truth, over all rows, the MAE gain is \MatchedZeroGain\% on rows with no
individuals and \MatchedNonZeroGain\% on the rest, against \TunedZeroGain\% and \TunedNonZeroGain\%
with tuned weights. Of the \NMatchedBaselines{} baselines, BIOTRAM is ahead of \SepReadsMAEbtlAhead{} on MAE and
\SepReadsJSDAhead{} on JSD; on rank correlation it is ahead of \SepReadsSpearAhead{} and level with
\SepReadsSpearTied{}, and on presence AUC it is level with \SepReadsAUCTied{} of them (\cref{tab:main}).

\begin{table}[t]\centering\footnotesize
\caption{\textbf{Architecture only: every deep model at the same loss weights.}
As \cref{tab:tuned}, but every deep model is trained with the same loss weights, $(\lambda_{\mathrm{comp}},\lambda_{\mathrm{pres}},\lambda_{\mathrm{rps}}) = (\LossWeightComp,\LossWeightPres,\LossWeightRps)$, rather than its own tuned ones, so that only the architecture differs.
The read-blind row is BIOTRAM without the BIN's own read in the amplification module (\cref{sec:methods}).
The first two rows are the fixed reference predictors of \cref{tab:main}, and Params is each model's number of parameters.
We highlight the \best{best}
and \runnerup{second best} model. %
}
\label{tab:matched}
\adjustbox{max width=\textwidth}{
\footnotesize
\begin{tabular}{l r ll lll}
\toprule
Model & Params & JSD $\downarrow$ & Spearman $\uparrow$ & MAE/bottle $\downarrow$ & MAE$_{=0}$ $\downarrow$ & MAE$_{>0}$ $\downarrow$ \\
\midrule
Every taxon equally abundant & -- & 0.3174\phantom{\tabpm{0.000}} & --\phantom{\tabpm{0.000}} & 1.219\phantom{\tabpm{0.000}} & 1.000\phantom{\tabpm{0.000}} & 1.723\phantom{\tabpm{0.000}} \\
Raw relative reads & -- & 0.3586\phantom{\tabpm{0.000}} & 0.504\phantom{\tabpm{0.000}} & 2.003\phantom{\tabpm{0.000}} & 0.544\phantom{\tabpm{0.000}} & 3.633\phantom{\tabpm{0.000}} \\
\midrule
Hurdle GLMM \citep{FurneauxRoslin_FlyInSoup} & -- & 0.1947\phantom{\tabpm{0.000}} & 0.609\phantom{\tabpm{0.000}} & 1.498\phantom{\tabpm{0.000}} & 0.730\phantom{\tabpm{0.000}} & 2.286\phantom{\tabpm{0.000}} \\
\midrule
Transformer & 9.6M & 0.1257\tabpm{0.0025} & 0.737\tabpm{0.004} & 0.743\tabpm{0.029} & 0.306\tabpm{0.026} & 1.449\tabpm{0.084} \\
1-D CNN & 11.5M & 0.1238\tabpm{0.0008} & 0.734\tabpm{0.002} & 0.759\tabpm{0.020} & 0.312\tabpm{0.010} & 1.459\tabpm{0.048} \\
MLP & 11.1M & 0.1261\tabpm{0.0011} & 0.736\tabpm{0.001} & 0.792\tabpm{0.016} & 0.317\tabpm{0.013} & 1.533\tabpm{0.037} \\
DeepSets & 10.1M & 0.1257\tabpm{0.0011} & 0.738\tabpm{0.002} & 0.845\tabpm{0.034} & 0.363\tabpm{0.020} & 1.642\tabpm{0.104} \\
\midrule
\textbf{BIOTRAM} & 9.9M & \best{0.1194}\tabpm{0.0007} & \runnerup{0.740}\tabpm{0.001} & \best{0.700}\tabpm{0.007} & \runnerup{0.264}\tabpm{0.008} & \best{1.341}\tabpm{0.016} \\
BIOTRAM, read-blind ablation & 9.9M & \runnerup{0.1229}\tabpm{0.0012} & 0.737\tabpm{0.001} & 0.763\tabpm{0.017} & 0.286\tabpm{0.007} & 1.435\tabpm{0.031} \\
\midrule
TabICLv2 (in-context) & 28.5M & 0.1238\tabpm{0.0001} & \best{0.751}\tabpm{0.001} & \runnerup{0.704}\tabpm{0.003} & \best{0.249}\tabpm{0.002} & \runnerup{1.413}\tabpm{0.006} \\
\bottomrule
\end{tabular}}
\end{table}

\mypar{The hurdle mixed model of \citet{FurneauxRoslin_FlyInSoup}.} They relate metabarcoding reads to
individual counts in 45 Malaise-trap bottles with two generalized linear mixed models: a logistic
regression of whether a BIN detected by metabarcoding is present as an individual, and a zero-truncated
Poisson regression of how many individuals it has when it is. Both take the logarithms of the BIN's reads,
pooled over PCR replicates, and of the bottle's total read count, with random intercepts nested by order,
family and genus and no effect at a rank the BIN is not identified to. Their unit of analysis, a BIN with
reads in a bottle, is exactly one of our rows, so we fit both models to our training rows with the
software they used (glmmTMB \GLMMVersion{}) and predict each row by the hurdle's mean, the presence
probability times the truncated mean. The specification is fixed, so nothing is tuned, and the model sees
only its published predictors. Under the three shifts of
\cref{tab:shift} it scores \GLMMSpatialJSD{} (spatial), \GLMMTemporalJSD{} (temporal) and
\GLMMLaneJSD{} (lane), behind the best matched baseline in each.

\begin{table*}[t]\centering\small
\caption{Test set, occurrence-count target. Mean over seeds $\pm$ one standard deviation across seeds.
Bold is the \best{best} value in each column and underline the \runnerup{second-best} value. All models see the same rows, the same split and the same evaluation code, and are selected on
validation JSD. Every deep model is at the 10.5M rung of the capacity ladder, BIOTRAM's own size.}
\label{tab:main}
\resizebox{\textwidth}{!}{%
{\small
\begin{tabular}{@{}l lll lll l@{}}
\toprule
Model & JSD $\downarrow$ & MAE/bot. $\downarrow$ & MAE-log $\downarrow$
 & Pres. AUC $\uparrow$ & Spearman $\uparrow$ & Shannon $R^2$ $\uparrow$ & $\beta$ Mantel $\rho$ $\uparrow$ \\
\midrule
\multicolumn{8}{l}{\itshape Reference} \\
Every taxon equally abundant & 0.3174\phantom{\tabpm{0.000}} & 1.22\phantom{\tabpm{0.000}} & 0.502\phantom{\tabpm{0.000}} & 0.500\phantom{\tabpm{0.000}} & -- & 0.627\phantom{\tabpm{0.000}} & 0.724\phantom{\tabpm{0.000}} \\
Raw relative reads & 0.3586\tabpm{0.0000} & 2.00\phantom{\tabpm{0.000}} & 0.424\phantom{\tabpm{0.000}} & 0.781\phantom{\tabpm{0.000}} & 0.486\phantom{\tabpm{0.000}} & 0.369\phantom{\tabpm{0.000}} & 0.635\phantom{\tabpm{0.000}} \\
Per-BIN offset {\smaller\smaller\citep{McLaren2019-ga}} & 0.3435\tabpm{0.0000} & 2.28\phantom{\tabpm{0.000}} & 0.424\tabpm{0.000} & 0.803\phantom{\tabpm{0.000}} & 0.519\phantom{\tabpm{0.000}} & 0.443\phantom{\tabpm{0.000}} & 0.588\phantom{\tabpm{0.000}} \\
\multicolumn{8}{l}{\itshape Linear models} \\
Ridge regression & 0.2159\phantom{\tabpm{0.000}} & 1.25\phantom{\tabpm{0.000}} & 0.424\tabpm{0.000} & 0.802\phantom{\tabpm{0.000}} & 0.539\phantom{\tabpm{0.000}} & 0.622\phantom{\tabpm{0.000}} & 0.776\phantom{\tabpm{0.000}} \\
\multicolumn{8}{l}{\itshape Mixed models} \\
Hurdle GLMM {\smaller\smaller\citep{FurneauxRoslin_FlyInSoup}} & 0.1947\phantom{\tabpm{0.000}} & 1.50\phantom{\tabpm{0.000}} & 0.430\phantom{\tabpm{0.000}} & 0.842\phantom{\tabpm{0.000}} & 0.583\phantom{\tabpm{0.000}} & 0.731\phantom{\tabpm{0.000}} & 0.830\phantom{\tabpm{0.000}} \\
\multicolumn{8}{l}{\itshape Nearest neighbours} \\
$k$-nearest neighbours & 0.1721\tabpm{0.0029} & 0.98\tabpm{0.01} & 0.319\tabpm{0.002} & 0.869\tabpm{0.001} & 0.626\tabpm{0.003} & 0.777\tabpm{0.006} & 0.839\tabpm{0.007} \\
\multicolumn{8}{l}{\itshape Tree-based models} \\
Random forest & 0.1517\tabpm{0.0001} & 0.92\tabpm{0.00} & 0.288\tabpm{0.000} & 0.900\tabpm{0.000} & 0.669\tabpm{0.000} & 0.765\tabpm{0.001} & 0.854\tabpm{0.003} \\
Gradient boosting & 0.1653\tabpm{0.0040} & 0.92\tabpm{0.02} & 0.296\tabpm{0.002} & 0.884\tabpm{0.006} & 0.641\tabpm{0.011} & 0.751\tabpm{0.021} & 0.841\tabpm{0.005} \\
\multicolumn{8}{l}{\itshape Zero-inflated models} \\
Zero-inflated ridge & 0.1742\tabpm{0.0000} & 0.95\tabpm{0.00} & 0.304\tabpm{0.000} & 0.889\phantom{\tabpm{0.000}} & 0.646\phantom{\tabpm{0.000}} & 0.680\tabpm{0.000} & 0.814\tabpm{0.000} \\
Two-stage hurdle & 0.1472\tabpm{0.0013} & 0.93\tabpm{0.00} & 0.290\tabpm{0.001} & 0.906\tabpm{0.000}            & 0.678\tabpm{0.001} & 0.749\tabpm{0.004} & 0.852\tabpm{0.003} \\
\multicolumn{8}{l}{\itshape Deep models} \\
MLP & 0.1247\tabpm{0.0016}          & 0.74\tabpm{0.00} & 0.245\tabpm{0.003} & \runnerup{0.916}\tabpm{0.000} & 0.686\tabpm{0.002} & 0.850\tabpm{0.011} & \runnerup{0.894}\tabpm{0.005} \\
DeepSets & 0.1261\tabpm{0.0013} & 0.74\tabpm{0.01} & 0.244\tabpm{0.001} & 0.913\tabpm{0.000} & 0.686\tabpm{0.001}          & 0.881\tabpm{0.007} & 0.876\tabpm{0.001} \\
1-D CNN on the barcode & 0.1250\tabpm{0.0010} & 0.73\tabpm{0.02} & \runnerup{0.230}\tabpm{0.004} & 0.916\tabpm{0.002} & 0.686\tabpm{0.003} & 0.865\tabpm{0.012}          & \best{0.895}\tabpm{0.006} \\
Transformer over the bottle & 0.1256\tabpm{0.0008} & \runnerup{0.71}\tabpm{0.01} & \runnerup{0.229}\tabpm{0.003} & 0.913\tabpm{0.002} & 0.682\tabpm{0.002}          & \runnerup{0.886}\tabpm{0.007} & \runnerup{0.889}\tabpm{0.008} \\
\midrule
\textbf{BIOTRAM} & \best{0.1194}\tabpm{0.0007} & \best{0.70}\tabpm{0.01} & 0.232\tabpm{0.002} & 0.914\tabpm{0.001} & \runnerup{0.688}\tabpm{0.001} & \best{0.894}\tabpm{0.007} & \runnerup{0.890}\tabpm{0.006} \\
\midrule
\multicolumn{8}{l}{\itshape BIOTRAM, read-blind ablation (no BIN's own read in the amplification module)} \\
BIOTRAM, read-blind (ablation) & \runnerup{0.1229}\tabpm{0.0012} & 0.76\tabpm{0.02} & 0.242\tabpm{0.002} & 0.913\tabpm{0.000} & 0.686\tabpm{0.001} & 0.879\tabpm{0.007} & 0.886\tabpm{0.005} \\
\midrule
TabICLv2 (in-context, 28.5M) & 0.1238\tabpm{0.0001} & \runnerup{0.70}\tabpm{0.00} & \best{0.226}\tabpm{0.001} & \best{0.924}\tabpm{0.000} & \best{0.699}\tabpm{0.001} & 0.820\tabpm{0.005} & 0.885\tabpm{0.002} \\
\bottomrule
\end{tabular}}}
\end{table*}

\mypar{TabICLv2 and foundation-model pretraining.}
TabICLv2 \citep{qu2026_TabICLv2} is kept apart from the other baselines in \cref{tab:tuned,tab:matched,tab:main},
partly because of its size: its \TabICLParams{} parameters are almost three times as many as
BIOTRAM's, but also because TabICLv2 is a transformer pretrained on synthetic tabular datasets and
used in context: without any gradient step on our data, it predicts each test row by attending to
the training rows. It therefore brings prior knowledge from pretraining and direct access to the
training set at inference, which no other model has. The pretraining matters: the same architecture
trained from scratch on our data reached a JSD of \TabICLScratchJSD{}, against
\TabICLPretrainedJSD{} for the pretrained model in the same setting. TabICLv2 has the best presence
AUC (\TabICLAUC), MAE-log (\TabICLMAElog) and MAE$_{=0}$ (\TabICLMAEZero) of any model in our
tables. BIOTRAM is \TunedTabICLCountRel{} it on count accuracy (MAE \TunedBioMAE{} against
\TunedTabICLMAE{}, \TunedTabICLMAEGapPct\% apart) and \TunedTabICLCompRel{} it on composition, with
\TabICLParamPct\% of its parameters, while remaining behind it on MAE$_{=0}$ and rank correlation 
and on presence AUC (\cref{tab:tuned}). Attending to the training rows also has a price at inference: a bottle costs
TabICLv2 $\CostTabICLFlops$ floating-point operations against BIOTRAM's $\CostBioFlops$, a factor of
$\CostTabICLRatio$.

\section{Error by abundance and by novelty}\label{app:errors}

\Cref{tab:abundance,tab:unseen} break the error of \cref{tab:matched} down by test row: by true
specimen count, and by whether the BIN occurs in any training bottle. Both compare BIOTRAM with the
four matched baselines only. \Cref{tab:abundance} scores MAE on $\log(1+y)$, so that errors on very
different counts can be compared; \cref{tab:unseen} scores it on the counts.

\mypar{By abundance.} BIOTRAM's structure acts most directly at the two ends of the range. Rows with
no individuals, the most frequent bucket, are the ones the presence gate of \cref{eq:inversion} can
remove outright. On counts above 100, BIOTRAM's error is \AbundTopCutPct\% below that of
\AbundTopCutVs{}, the matched baseline with the lowest overall MAE, a difference \AbundTopCutSig{}.
Few rows fall in this bucket, but they hold a large share of the specimens, so they weigh most on any
estimate of total abundance or biomass. Between 1 and 100 specimens, BIOTRAM trails at least one
matched baseline on \AbMidBucketsBehind{} of the four buckets.

\mypar{By novelty.} Because the splits are by bottle, \UnseenPct\% of test rows carry a BIN that
occurs in no training bottle. Their per-BIN quantities cannot be learned from training rows and must
be inferred from the barcode and the taxonomy. Most of these BINs are rare, so every model makes a
smaller error on them than on seen BINs. On seen BINs, BIOTRAM's MAE is \SeenMAE{} against
\SeenMAEBase{} for \AbundTopCutVs{}. On unseen BINs, where errors in the barcode-based estimates pass
directly into the count, it is behind \AbUnseenBehind{} of the matched baselines, level with
\AbUnseenTied{} and ahead of \AbUnseenAhead{}.

\begin{table*}[t]\centering\footnotesize
\caption{Error by true abundance, at the matched objective: MAE on $\log(1+n_i^{(b)})$ for BIOTRAM and for
all four baselines of \cref{tab:matched}, each trained on BIOTRAM's own loss with the
same inputs; columns ordered by overall MAE. {Bold} is the \best{best} value in each row and 
{underline} every value a two-sided Welch test over seeds \runnerup{cannot separate} from it at 
$\alpha=0.05$, or the \runnerup{second-best} value where it separates them all.}
\label{tab:abundance}
\setlength{\tabcolsep}{4pt}
\adjustbox{max width=\textwidth}{
{\small
\begin{tabular}{l r r ccccc}
\toprule
True count & Rows & \% & BIOTRAM $\downarrow$ & Transformer $\downarrow$ & CNN $\downarrow$ & MLP $\downarrow$ & DeepSets $\downarrow$ \\
\midrule
0 (absent) & 14,449 & 53.6\% & \best{0.173} & \runnerup{0.194} & 0.199 & 0.200 & 0.214 \\
1 & 8,373 & 31.1\% & \runnerup{0.214} & \best{0.211} & 0.214 & 0.216 & 0.263 \\
2--5 & 3,275 & 12.2\% & 0.428 & \runnerup{0.416} & 0.416 & 0.416 & \best{0.388} \\
6--20 & 677 & 2.5\% & 0.681 & \runnerup{0.663} & 0.694 & 0.710 & \best{0.627} \\
21--100 & 146 & 0.5\% & \runnerup{0.609} & 0.631 & 0.701 & 0.672 & \best{0.568} \\
100+ & 14 & 0.1\% & \best{0.415} & \runnerup{0.446} & 0.542 & 0.635 & 0.498 \\
\bottomrule
\end{tabular}
}
}
\end{table*}

\begin{table*}[t]\centering\footnotesize
\caption{Test rows split by whether the BIN appeared in training, at the matched objective. MAE in
occurrences, for BIOTRAM and the baselines of \cref{tab:matched}; marked as in \cref{tab:abundance}.}
\label{tab:unseen}
\setlength{\tabcolsep}{4pt}
\adjustbox{max width=\textwidth}{
{\small
\begin{tabular}{l r ccccc}
\toprule
Test rows & Rows & BIOTRAM $\downarrow$ & Transformer $\downarrow$ & CNN $\downarrow$ & MLP $\downarrow$ & DeepSets $\downarrow$ \\
\midrule
Seen in training & 21,903 & \best{0.853} & \runnerup{0.941} & 0.955 & 1.004 & 1.068 \\
Never seen & 5,031 & 0.373 & 0.376 & \runnerup{0.357} & \best{0.343} & 0.467 \\
\bottomrule
\end{tabular}
}}
\end{table*}

\section{What a prediction looks like}\label{app:scatter}
\begin{figure*}[t]\centering
\includegraphics[width=\textwidth]{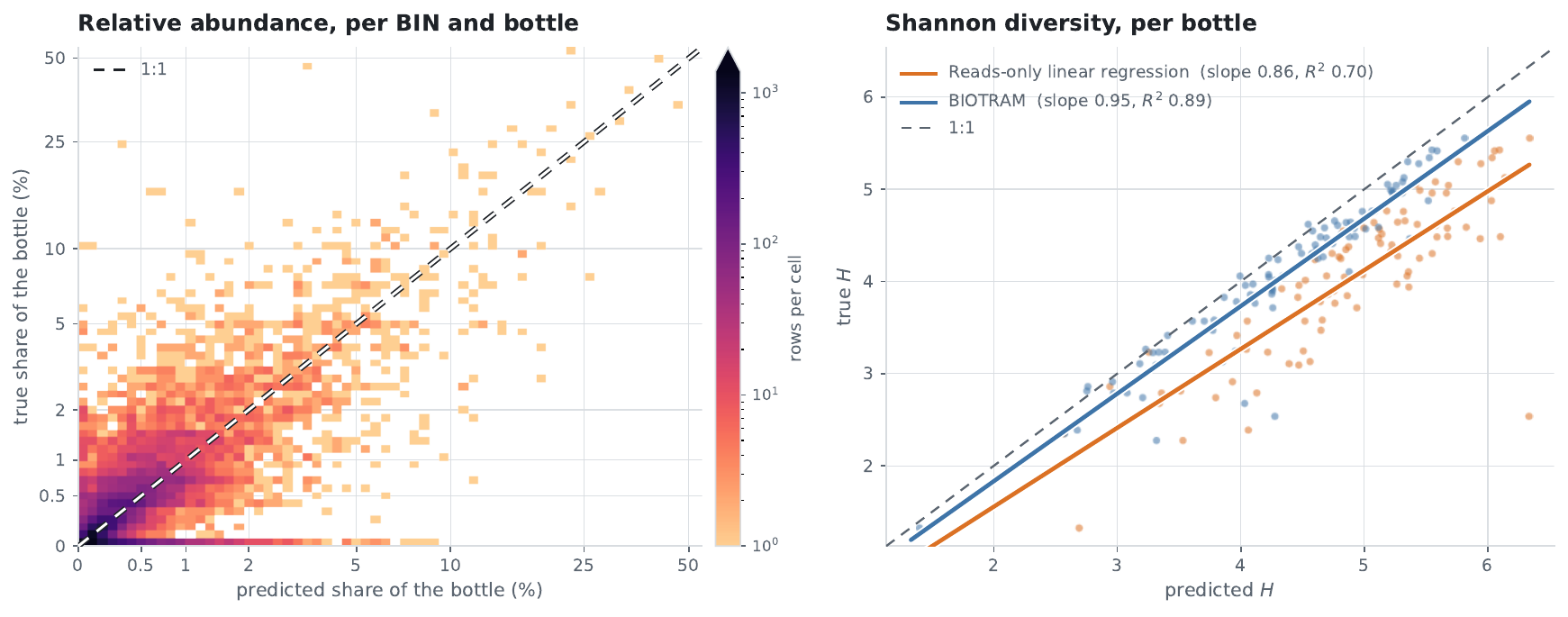}
\caption{\textbf{Predictions, row by row and bottle by bottle.} One seed
(\ScatterSeed{}) of the reported arm on the test split: \ScatterRows{} rows across
\ScatterBottles{} bottles. \textbf{Left:}~relative abundance per BIN and bottle, predicted against true, as a
two-dimensional histogram on a $\log(1+x)$ axis in percent; colour is the number of rows in a cell,
on a logarithmic scale, and the dashed line is 1:1. The cell at the origin holds every row that is
zero in both, more than half the dataset; the colour scale is pinned to the largest cell beside it,
so that one saturates (the arrow on the bar) rather than pushing the rest into the dark end.
\textbf{Right:}~Shannon diversity per bottle for BIOTRAM (blue, slope \ScatterSlopeBio{},
$R^2$~\ScatterRtwoBio{}) and for an ordinary least-squares regression of the specimen count on
log read count and the log of the bottle's total read count (orange, slope \ScatterSlopeRead{}, $R^2$~\ScatterRtwoRead{}),
each with its own fitted line against the 1:1 diagonal.}
\label{fig:scatter}
\end{figure*}

\Cref{fig:scatter} shows the predictions themselves for one seed of the reported arm. The left panel 
groups every test row by prediction and truth. The quantity is each BIN's share of the bottle it is in.
The right panel aggregates each bottle to its Shannon diversity and compares BIOTRAM against a linear
regression that sees only read counts, the BINs and the bottle's total; each carries its own
least-squares fit. The slope is the part worth reading: an estimator can have a respectable $R^2$ 
while systematically flattening diversity toward the mean, so we are looking for the model closest to the 1:1 line.

\mypar{Calibration of the predictive distribution.} The negative-binomial head gives a full predictive
distribution for each count, not only a point estimate, so we check and score the distribution itself 
on the held-out rows over \CalibSeeds{} seeds. For a discrete target, the plain probability integral 
transform $P(X \le y)$ is biased upward when about half of the rows are exactly zero (its mean here 
is \CalibNaivePIT{}), so we use its randomised mid-point version, whose mean is 0.500 under 
calibration. It averages \CalibMidPIT{}: the predictive distribution is centred close to the 
observations. It is too wide, however, as nominal 90\% intervals contain \CalibCov{} of the 
observations. Part of this excess is unavoidable: on counts that are mostly zero or single-digit, a 
discrete interval cannot cover exactly 90\%.

\section{Distribution shift}\label{app:shift}

\mypar{Regimes.} Each shift regime of \cref{tab:shift} replaces the random split by one that holds
out a whole stratum of bottles for testing: the bottles of a region never sampled in training
(spatial), those collected after every training bottle (temporal), or those of a sequencing batch
absent from training (lane). Every deep model is retrained on the remaining bottles with the matched
objective, so the rows of \cref{tab:shift} read against \cref{tab:matched}. The bottles with excess
that fall inside the held-out stratum are dropped as well, which leaves 694 (spatial), 702 (temporal)
and 684 (lane) training bottles instead of \DataVolMax{}.

\mypar{Reading the table.} Each row compares BIOTRAM with the best baseline of that row alone,
chosen among the four matched deep baselines and the hurdle mixed model, so different rows can name
different models: the best baseline in distribution need not be the best one under a given shift.
The baseline of a row is chosen on JSD, and its MAE per bottle is shown beside BIOTRAM's.
$\Delta$ is each model's change in JSD from its own random-split score (\cref{tab:matched,tab:main}), so
the $\Delta$ of a row's baseline is not measured from the Random row whenever the two rows name
different models.

\mypar{Results.} The shifts cost every model far more than the differences between models: BIOTRAM
loses \DeltaSpatial{}, \DeltaTemporal{} and \DeltaLane{} JSD on the spatial, temporal and lane
regimes, while its margin over the best baseline of each regime is $\OODSpatialGap$,
$\OODTemporalGap$ and $\OODLaneGap$, each tested with the same two-sided Welch test over seeds
($\alpha = 0.05$) as in \cref{sec:results}. The hurdle mixed model of \citet{FurneauxRoslin_FlyInSoup}
scores \GLMMSpatialJSD{}, \GLMMTemporalJSD{} and \GLMMLaneJSD{} JSD under these three regimes, behind
the best matched baseline in each. On counts, BIOTRAM has the lower MAE per bottle in every row of
\cref{tab:shift}, including the collaterals dataset, where the hurdle mixed model is closer on
composition.

\section{Validation on external ``collaterals'' dataset}\label{app:external}

Our main results are evaluations using splits of data collected from one metabarcoding pipeline in 
one laboratory only.
We also scored model performance on a ``collaterals'' dataset, comprised of \ExtBottles{} Malaise-trap 
bottles from another country (Sweden), processed by a different laboratory.
The dataset is unpublished and was shared by collaborators.
Each bottle was metabarcoded in a single PCR replicate, without location, date or wet-biomass records, 
from two DNA extractions of the same material scored separately: a mild lysis (L), as in our data, and a homogenate (H).
The specimens were also barcoded individually, so each bottle's true composition is known.
\Cref{tab:external} compares BIOTRAM, one of its variants, the four matched deep baselines, TabICLv2 
and the hurdle mixed model of \citet{FurneauxRoslin_FlyInSoup} on these bottles; with \ExtBottles{} of 
them, only large differences are interpretable.

\begin{table}[t]\centering\footnotesize
\caption{\textbf{Extended results on external Collaterals dataset.} 15 Malaise-trap bottles from a
different country, processed by a different laboratory, with one PCR replicate per bottle and no
bottle-level covariates (\cref{app:external}). Lysis (Lys.) and homogenate (Hom.) are two DNA extractions of
the same bottles, scored separately. JSD is the mean over seeds $\pm$ one standard deviation across seeds; MAE/bottle
(as in \cref{tab:shift}) and presence AUC (over all rows) are means over seeds. Every model sees identical rows.
Extends bottom half of \cref{tab:shift}.
Bold: \best{best} model; underlined: \runnerup{second-best}.
}
\label{tab:external}
\setlength{\tabcolsep}{4pt}
\begin{tabular}{l ll cc cc}
\toprule
& \multicolumn{2}{c}{JSD $\downarrow$} & \multicolumn{2}{c}{MAE/bottle $\downarrow$}
& \multicolumn{2}{c}{Presence AUC $\uparrow$} \\
\cmidrule(lr){2-3}\cmidrule(lr){4-5}\cmidrule(lr){6-7}
Model & Lys. & Hom. & Lys. & Hom. & Lys. & Hom. \\
\midrule
\emph{Read share} (sanity check) & 0.277\phantom{\tabpm{0.000}} & 0.237\phantom{\tabpm{0.000}} & 3.38 & 2.86 & \best{0.772} & \best{0.839} \\
\emph{Uniform} (sanity check) & 0.255\phantom{\tabpm{0.000}} & 0.255\phantom{\tabpm{0.000}} & 3.12 & 2.87 & 0.540 & 0.504 \\
\midrule
\textbf{BIOTRAM} & \runnerup{0.219}\tabpm{0.019} & 0.218\tabpm{0.015} & 2.78 & 2.54 & 0.722 & 0.740 \\
BIOTRAM trained one replicate at a time & 0.224\tabpm{0.024} & \runnerup{0.217}\tabpm{0.023} & 2.98 & 2.48 & 0.745 & 0.760 \\
Transformer over the bottle & 0.220\tabpm{0.018} & 0.229\tabpm{0.022} & 2.68 & 2.37 & 0.730 & 0.723 \\
DeepSets & 0.290\tabpm{0.020} & 0.296\tabpm{0.013} & 3.04 & 2.62 & 0.704 & 0.695 \\
1-D CNN on the barcode & 0.301\tabpm{0.007} & 0.321\tabpm{0.007} & \runnerup{2.53} & \runnerup{2.34} & 0.681 & 0.658 \\
MLP & 0.308\tabpm{0.005} & 0.335\tabpm{0.009} & 2.59 & 2.38 & 0.692 & 0.650 \\
TabICLv2 (in-context) \citep{qu2026_TabICLv2} & 0.223\tabpm{0.016} & 0.232\tabpm{0.014} & \best{2.40} & \best{2.19} & 0.735 & 0.739 \\
Hurdle GLMM \citep{FurneauxRoslin_FlyInSoup} & \best{0.203}\phantom{\tabpm{0.000}} & \best{0.197}\phantom{\tabpm{0.000}} & 4.15 & 3.98 & \runnerup{0.770} & \runnerup{0.823} \\
\bottomrule
\end{tabular}
\end{table}

\mypar{Scoring.} 
\ExtNewBins{} of the BINs are absent from the \ExtVocab{}-BIN vocabulary the models were trained on:
they receive BarcodeMamba embeddings, taxonomy mapped through the training vocabulary, and body-size
and primer-efficiency predictions from the same models as every other BIN (\cref{subsec:anchors}).

\mypar{Results.} Assigning each bottle's specimens in proportion to its reads gives a JSD of
\ExtReadShareJSDL{} on L and \ExtReadShareJSDH{} on H, and spreading them uniformly over the 
detected BINs gives \ExtUniformJSDL{}. BIOTRAM reaches \ExtBioJSDL{} and \ExtBioJSDH{} over 
\ExtBioSeeds{} seeds, and its seed-averaged prediction is closer to the truth than the read shares 
in \ExtBioBetterL{} and \ExtBioBetterH{} of the \ExtBottles{} bottles. By a two-sided Welch test 
over seeds, it is level with \ExtBioLevelWith{} on both extractions and ahead of \ExtBioAheadOf{}, 
which do worse than the read shares. It is behind \ExtBioBehind{} on composition (\ExtGLMMJSDL{} and 
\ExtGLMMJSDH{}) but clearly ahead on count error (MAE/bottle \ExtGLMMMAEL{} and \ExtGLMMMAEH{} 
against BIOTRAM's \ExtBioMAEL{} and \ExtBioMAEH{}). With fifteen bottles that ordering is not firm: 
a bootstrap over bottles puts the mixed model's margin over BIOTRAM at $p=\ExtGLMMPairPL$ and 
$p=\ExtGLMMPairPH$. A variant trained with each PCR replicate as a separate sample, the format of 
this dataset, is level with BIOTRAM (\ExtTubeJSDL{} and \ExtTubeJSDH{}).

\section{The latent quantities}\label{app:latents}

\mypar{Do the predicted quantities mean what they are named?} The model's distinguishing
property is its intermediate quantities that are supposed to correspond to physical quantities and
give meaningful explanations of the model's prediction. Since the primer efficiency and body mass
receive the corresponding external estimates as input features, we cannot correlate them against
those estimates without circularity. Instead, we can check that the empirical per-BIN
$\log(\mathrm{reads}/\mathrm{specimen})$, measured from the metabarcoding pipeline, correlates with the predicted
$\bar{a}_i + \log m_i$ (\cref{eq:rate}), averaged per BIN over the test bottles. The module also sees the BIN's own read, so part of that
association is mechanical rather than discovered, and the read-blind variant is the cleaner test of
the same question.

\begin{figure*}[t]\centering
\includegraphics[width=0.74\textwidth]{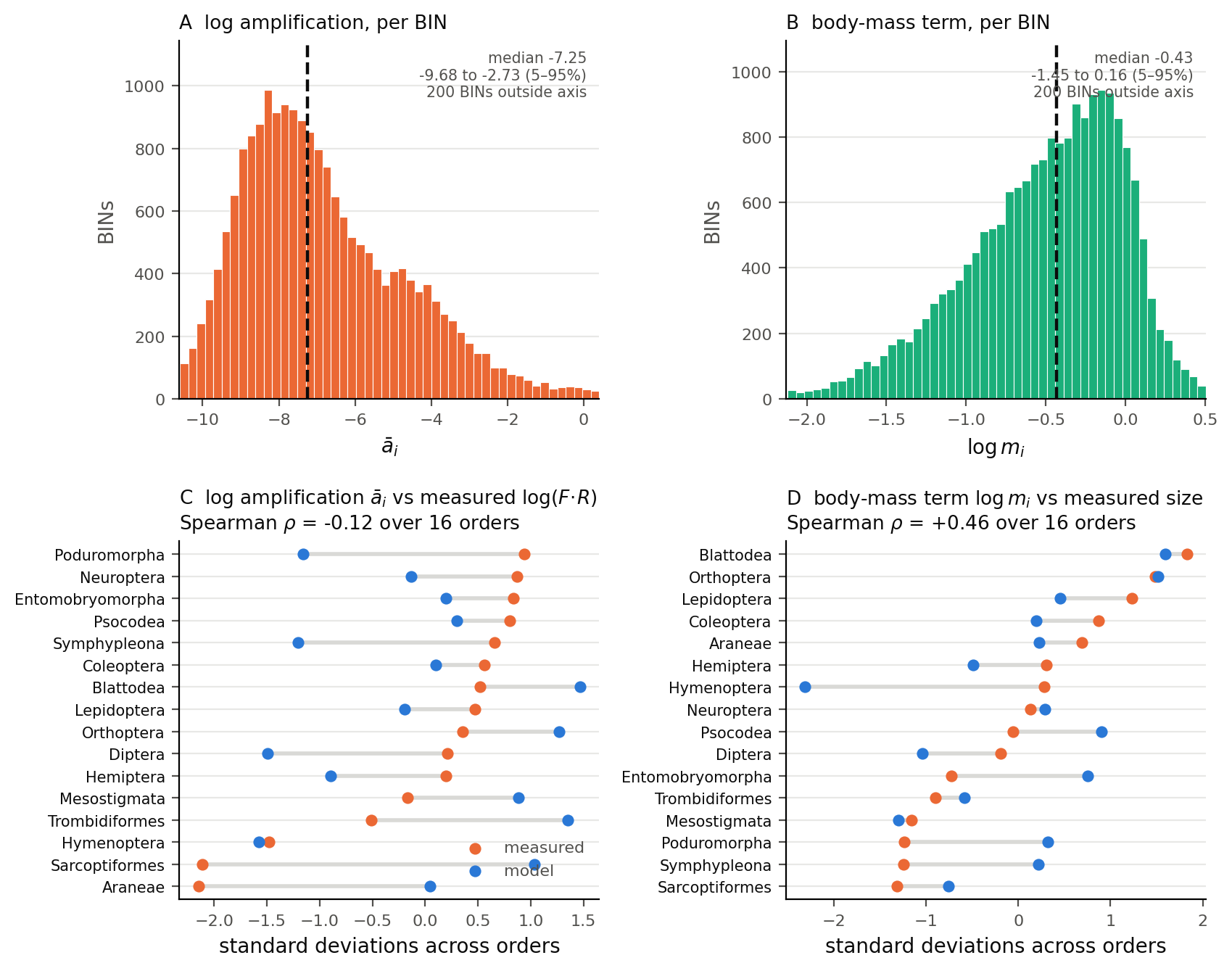}
\caption{\textbf{The latent quantities of the reported model.}
\textbf{Left:} log DNA amplification;
\textbf{right:} body mass.
\textbf{Top:}~their distributions across the test set; axes are clipped to the central 99\% and the count outside is stated.
\textbf{Bottom:}~the same latent quantities (blue) against external measurements (orange), aggregated to taxonomic order and expressed in standard deviations across orders so that both sit on one axis.}
\label{fig:latents}
\end{figure*}

Over \MechSeeds{} seeds of BIOTRAM at the development weights (\cref{tab:matched}), it reaches Pearson $+\MechSumR \pm \MechSumSd$, against
$+\MechFrR \pm \MechFrSd$ for $\bar{a}_i$ alone and $+\MechMassR \pm \MechMassSd$ for $\log m_i$ alone. The
read-blind variant, whose amplification term never sees the read, still reaches
$+\MechTwinSumR \pm \MechTwinSumSd$ ($+\MechTwinFrR$ and $+\MechTwinMassR$ for the two terms alone,
\MechTwinSeeds{} seeds). \Cref{fig:latents} shows the distribution of both quantities across the
test set and their agreement with the external measurements at the level of taxonomic order.
Note that the objective supervises $\smash[t]{\bar{a}_i^{(b)}} + \log m_i$ toward this quantity at weight $\LossWeightRps$, so
the number is a generalization measure over held-out bottles rather than an unsupervised discovery.

\section{Loss-weight tuning}\label{app:losstune}

\mypar{Principle.} A baseline comparison is only fair if every model's objective is tuned with the same
care, since an under-tuned baseline makes the proposed model look better than it is \citep{Xin2022_MTO}. The three auxiliary weights
$(\lambda_{\mathrm{comp}},\lambda_{\mathrm{pres}},\lambda_{\mathrm{rps}})$ of \cref{eq:loss} are therefore tuned
for each of the \LossTuneModels{} models separately, under one protocol fixed before any run and with
the same budget. The likelihood is the negative binomial of \cref{eq:loss-nb} for every model.

\mypar{Search.} Every model runs the same four stages. (i) A full $3^3$ factorial over the three weights,
on the logarithmic grid $\lambda_{\mathrm{comp}} \in \LossTuneGridComp$,
$\lambda_{\mathrm{pres}} \in \LossTuneGridPres$ and $\lambda_{\mathrm{rps}} \in \LossTuneGridRpo$. (ii) At the
best of these, each weight set to zero in turn, and KL in place of JSD for the composition term. (iii) A
second $3^3$ factorial at half the step, around the best configuration so far; a weight set to zero in
(ii) is tested at zero against two values below the smallest one of (i), so that zero is found to be
optimal rather than assumed. (iv) The learning rate at a third of and at three times the model's tuned
value. Stage (iii) is a factorial rather than one weight at a time because the weights interact:
\LossTuneInteractN{} of the \LossTuneModels{} models have a two-way interaction that is significant in a
factorial ANOVA of stage (i), which a search moving one weight at a time can walk past. A best
configuration on the edge of the range explored for a weight receives up to two further steps outward,
under the same rule for every model.

\mypar{Budget.} Every model is given the same \LossTuneConfigs{} configurations, plus between zero and
\LossTuneBoundaryMax{} boundary steps granted by the rule described above: \LossTuneSpendLo{} to \LossTuneSpendHi{}
configurations per model. BIOTRAM's \LossTuneSpendBio{} configurations lie inside the range of the baselines
(\cref{tab:losstune}). Each configuration is trained on the same \LossTuneSearchSeeds{}
seeds, \LossTuneSearchRuns{} runs in all. The budget is counted in configurations and seeds.

\mypar{Selection.} A configuration is scored on validation JSD, MAE and presence AUC together. Each
metric is divided by its own seed-to-seed standard deviation, pooled over the stage-(i) factorial, so
that the three enter on a common scale, and they are averaged per seed and then over seeds into a
composite $S$; the configuration with the lowest $S$ is selected. A configuration is admissible only 
if none of the three metrics is worse than at the centre of the stage-(i) grid by more than one noise 
band, so the composite cannot trade one metric for another.

\mypar{Confirmation.} The selected weights of each model are retrained on \LossTuneConfirmSeeds{} seeds
disjoint from every seed used in the search, and from those on which the model's architecture was
selected. The test set plays no part in any selection; \cref{tab:losstune} and \cref{tab:tuned} in
the main text report it on these retrained models.

\begin{table}[t]\centering\footnotesize
\caption{\textbf{Every model at its own loss weights.} The three auxiliary weights of
\cref{eq:loss} are tuned separately for each model, with the same protocol and budget
(\cref{app:losstune}): every model is given the same 61 configurations, plus up to two additional steps for any weight whose best value lies at the edge of the explored range, each trained on the 
same five seeds. Test metrics at the selected weights, mean $\pm$ standard deviation over eight 
seeds disjoint from the search; the test set played no part in selection. We highlight the 
\best{best} and \runnerup{second best} model.}
\label{tab:losstune}
\begin{tabular}{l r lll lll}
\toprule
 & \multirow[b]{2}{\widthof{Budget}}[-\aboverulesep-\belowrulesep-\cmidrulewidth]{Budget\\ configs} & \multicolumn{3}{c}{Selected weights} & \multicolumn{3}{c}{Test} \\
\cmidrule(lr){3-5}\cmidrule(l){6-8}
Model & & $\lambda_{\mathrm{comp}}$ & $\lambda_{\mathrm{pres}}$ & $\lambda_{\mathrm{rps}}$ & JSD $\downarrow$ & MAE/bottle $\downarrow$ & AUC $\uparrow$ \\
\midrule
Transformer & 63 & 1 & 0 & 0.1 & 0.1257\tabpm{0.0019} & \runnerup{0.723}\tabpm{0.020} & 0.9156\tabpm{0.0017} \\
1-D CNN & 61 & 2 & 0.009375 & 0.2 & 0.1235\tabpm{0.0008} & 0.762\tabpm{0.016} & 0.9145\tabpm{0.0008} \\
MLP & 63 & 1 & 0 & 0.00625 & \runnerup{0.1234}\tabpm{0.0006} & 0.737\tabpm{0.025} & 0.9154\tabpm{0.0007} \\
DeepSets & 65 & 1 & 0.0375 & 0.00625 & 0.1256\tabpm{0.0010} & 0.752\tabpm{0.036} & \runnerup{0.9162}\tabpm{0.0013} \\
\midrule
\textbf{BIOTRAM} & 62 & 1 & 0.15 & 0 & \best{0.1200}\tabpm{0.0009} & \best{0.699}\tabpm{0.017} & \best{0.9213}\tabpm{0.0006} \\
\bottomrule
\end{tabular}
\end{table}

\mypar{Result.} With every model at its own weights, BIOTRAM is ahead of the best baseline on the three
test metrics: JSD \LossTuneBioJSD{} against \LossTuneBaseJSD{} for \LossTuneBaseJSDModel{}
($p = \LossTuneJSDP$), MAE per bottle \LossTuneBioMAE{} against \LossTuneBaseMAE{} for \LossTuneBaseMAEModel{}
($p = \LossTuneMAEP$), and presence AUC \LossTuneBioAUC{} against \LossTuneBaseAUC{} for
\LossTuneBaseAUCModel{} ($p = \LossTuneAUCP$; two-sided Welch tests over seeds). The selected weights
differ between architectures (\cref{tab:losstune}): the MLP and the Transformer do best with no presence
term, zero beating both smaller values, whereas the 1-D CNN and DeepSets keep a small one, and BIOTRAM keeps its
presence term and does best without $\mathcal{L}_{\mathrm{rps}}$. No single weighting suits every model,
which is why each is tuned separately. Everywhere else in the paper, except in \cref{tab:tuned}, 
BIOTRAM and the matched baselines use the weights selected during development,
$(\lambda_{\mathrm{comp}},\lambda_{\mathrm{pres}},\lambda_{\mathrm{rps}}) = (\LossWeightComp,\LossWeightPres,\LossWeightRps)$, rather than those tuned here. BIOTRAM's tuned weights reach \LossTuneBioJSD{} test JSD, 
slightly behind the \BioJSD{} its development weights reach in \cref{tab:matched}. That is expected 
rather than contradictory: a configuration is selected on a composite of validation JSD, MAE and 
presence AUC, never on test JSD, so the returned configuration need not be the one with the lowest 
test JSD --- and the comparison in this appendix holds every model to that same rule.

\section{Ablation study}\label{app:ablation}

We ran a hierarchical ablation over \AblBlocks{} capability blocks. Each block is removed from the
reported configuration on its own, and the model retrained, paired
seed-for-seed against a control trained on the identical split; the screen runs at four seeds and
survivors are confirmed at eight. \Cref{tab:ablationfull} reports the four-seed screen on the test set, 
grouped by the kind of component: for each metric, the mean paired difference from the control and 
the significance of a paired $t$-test.

\mypar{What carries the model.} Removing 17 of the 20 blocks in \cref{tab:ablationfull}
raises test JSD significantly ($p<0.05$, uncorrected). The two largest effects are data and chemistry 
rather than architecture: the primer binding efficiencies (\AblPrimerTestJSD{} test JSD) and the 
extra training bottles with excess (\AblExcessTestJSD{}). The learned components on top of the 
frozen encoder follow (\AblLearnedEncoderTestJSD{}), then the cross-BIN attention of the amplification 
module (\AblRateTransformerTestJSD{}), and giving it the BIN's own reads is worth 
\AblReadsInModuleTestJSD{} on its own.

\mypar{Presence supervision.} Removing either form of presence supervision, the presence loss or the
read-informed presence heads, moves test JSD and the presence AUC of \cref{tab:ablationfull} by less
than 0.003, but costs a large part of the AUC of the presence probability $\sigma(\smash[t]{z_i^{(b)}})$
($-0.084$ and $-0.053$).

\mypar{Identity is redundant.} Removing the frozen barcode embedding while leaving the taxonomy
embedding in place costs \AblBarcodeTestJSD{} test JSD ($p=\AblBarcodeTestJSDP$), and removing the 
taxonomy embedding while keeping the barcode costs \AblTaxonomyTestJSD{} ($p=\AblTaxonomyTestJSDP$). 
The first result is consistent with the taxonomy being able to stand in for the barcode: genus is 
unknown for \PctNoGenus{}\% of BINs and species for \PctNoSpecies{}\%, so for three quarters of the BINs a
taxonomy lookup resolves only to family or coarser, yet what it does resolve is apparently enough
to cover most of what the barcode carries.

\begin{table*}[t]\centering\footnotesize
\def\ablsig#1#2{{\boldmath$#1$}\rlap{$^{#2}$}}%
\caption{\textbf{Model component ablations.} Each row removes one component of BIOTRAM, retrains the
model and compares it with the full model on the test set, paired seed for seed over \AblFullSeeds{}
seeds (full model: JSD \AblAnchorTestJSD{}, MAE/bottle \AblAnchorMAEbtl{}, presence AUC
\AblAnchorCountAUC{}). $\Delta$ is the mean paired difference, ablated minus full, so a positive $\Delta$ JSD or
MAE/bottle, or a negative $\Delta$ AUC, means that the component helps. Values in bold differ from 
zero at $p<0.05$ in a two-sided paired $t$-test (${*}: p<0.05$, ${**}: p<0.01$, ${\mathord{*}\mathord{*}\mathord{*}}: p<0.001$). 
Rows are ranked by the $p$-value of $\Delta$ JSD within each group. \na{}: a compositional head 
predicts no counts to score.}
\label{tab:ablationfull}
\adjustbox{max width=\textwidth}{
\small
\begin{tabular}{@{}l@{\hspace{2em}}r@{\hspace{2.4em}}r@{\hspace{2.4em}}r@{\hspace{1.4em}}}
\toprule
Component removed & $\Delta$ JSD & $\Delta$ MAE/bottle & $\Delta$ presence AUC \\
\midrule
\textit{Architecture} \\
\quad Barcode CNN and embedding adapter & \ablsig{+0.0085}{***} & \ablsig{+0.040}{*} & \ablsig{-0.0031}{***} \\  %
\quad Cross-BIN attention (each BIN processed alone) & \ablsig{+0.0054}{**} & \ablsig{+0.046}{***} & \ablsig{-0.0024}{**} \\  %
\quad Taxonomy embedding & \ablsig{+0.0034}{*} & \ablsig{+0.012}{*} & \ablsig{-0.0016}{*} \\  %
\quad Read-informed presence heads & $+0.0019$ & \ablsig{+0.023}{*} & \ablsig{-0.0029}{***} \\  %
\quad Frozen barcode embedding & $+0.0014$ & $+0.004$ & $-0.0011$ \\  %
\addlinespace
\textit{Inputs} \\
\quad Primer binding efficiencies & \ablsig{+0.0127}{***} & \ablsig{+0.026}{**} & \ablsig{-0.0242}{***} \\  %
\quad Extra training bottles with excess & \ablsig{+0.0117}{***} & \ablsig{+0.020}{**} & \ablsig{-0.0043}{**} \\  %
\quad Bottle's total specimen count $\nu^{(b)}$ & \ablsig{+0.0037}{**} & \ablsig{+0.073}{***} & \ablsig{-0.0052}{**} \\  %
\quad BIN's own read count in the amplification module (read-blind variant) & \ablsig{+0.0039}{**} & \ablsig{+0.048}{**} & \ablsig{-0.0019}{*} \\  %
\quad Bottle and primer inputs to the presence head & \ablsig{+0.0018}{**} & $+0.004$ & \ablsig{-0.0012}{*} \\  %
\quad Barcode-predicted body size & \ablsig{+0.0015}{*} & $+0.008$ & $-0.0006$ \\  %
\quad Per-replicate sequencing depth & \ablsig{+0.0008}{*} & $+0.014$ & \ablsig{-0.0011}{*} \\  %
\quad Pretrained geographic encoder (sinusoidal encoding instead) & \ablsig{+0.0012}{*} & $-0.006$ & $0.0000$ \\  %
\addlinespace
\textit{Loss} \\
\quad Composition loss $\mathcal{L}_{\mathrm{comp}}$ & \ablsig{+0.0023}{**} & \ablsig{+0.022}{*} & \ablsig{-0.0015}{*} \\  %
\quad Count objective (compositional head instead) & \ablsig{+0.0027}{*} & \na{} & \na{} \\  %
\quad Presence loss $\mathcal{L}_{\mathrm{pres}}$ & \ablsig{+0.0013}{*} & $-0.003$ & \ablsig{-0.0018}{*} \\  %
\quad Reads-per-specimen loss $\mathcal{L}_{\mathrm{rps}}$ & \ablsig{+0.0007}{*} & $+0.005$ & $-0.0003$ \\  %
\quad Per-BIN dispersion (one global value instead) & $+0.0007$ & $+0.007$ & $-0.0001$ \\  %
\addlinespace
\textit{Regularization} \\
\quad Exponential moving average of the weights & \ablsig{+0.0052}{**} & $+0.051$ & \ablsig{-0.0032}{**} \\  %
\quad Online augmentation & \ablsig{+0.0014}{**} & \ablsig{+0.027}{*} & \ablsig{+0.0010}{*} \\  %
\bottomrule
\end{tabular}
}
\end{table*}

\section{Model scaling against baselines}\label{app:scaling}

\begin{figure}[t]\centering
\includegraphics[width=\textwidth]{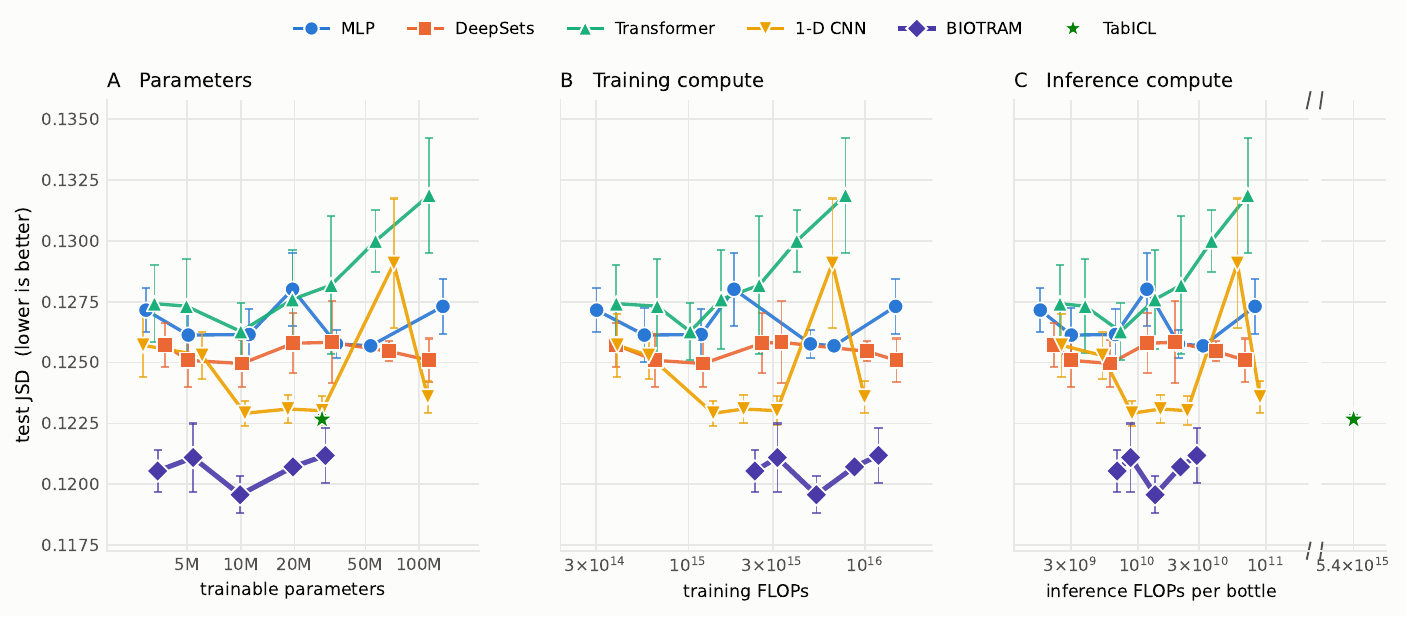}
\caption{\textbf{Accuracy against cost.} Test JSD of each architecture against its training/inference demands:
(\textbf{A}) number of trainable parameters;
(\textbf{B}) the compute requirements of one training run;
(\textbf{C}) and the compute requirements of one prediction (note the discontinuity in the x-axis here).
FLOPs are measured on the real forward and backward passes.
We explore baseline models up to \LadderMax{} parameters to ensure we explore a compute budget at least as high as that used for BIOTRAM, but find benefits from increased size are saturated before reaching this.
TabICLv2 does not require any additional training and is deployed with in context learning only; consequently its inference cost is much larger---more than 100,000 %
times that of the largest BIOTRAM model we explore.
}
\label{fig:cost}
\end{figure}

\mypar{Capacity.}
Model capacity is not what separates the models. Every architecture is retrained at each rung of a
capacity ladder, \LadderMin{} to \LadderMax{} parameters for the baselines and \LadderBioMin{} upward
for BIOTRAM, and tuned again at every rung (\cref{fig:cost}). Capacity moves each of them by at
most \LadderSpanMax{} JSD across its own ladder, while BIOTRAM's margin over the best baseline of the
same rung stays between \LadderGapMin{} and \LadderGapMax{}: BIOTRAM at its worst size
(\LadderBioWorst{}) is still ahead of \LadderBaseBestModel{} at its best (\LadderBaseBest{}).

\mypar{Compute.} \Cref{fig:cost} also places the same models against the cost of training one
model and of predicting one bottle, both counted in floating-point operations rather than in wall
clock, so that they do not depend on the hardware. At the \LadderPromotedSize{} rung BIOTRAM trains
for \CostTrainRatio$\times$ the compute of the deep baselines. Predicting one bottle costs it 
$\CostBioFlops$ operations, inside the range the deep baselines span at that rung ($\CostDeepFlopsLo$ 
to $\CostDeepFlopsHi$), so cost per prediction does not follow the parameter count. TabICLv2 has no 
training cost on our data and instead attends to every training row for each prediction, at 
$\CostTabICLFlops$ operations per bottle.

\begin{figure}[t]\centering
\includegraphics[width=0.8\textwidth]{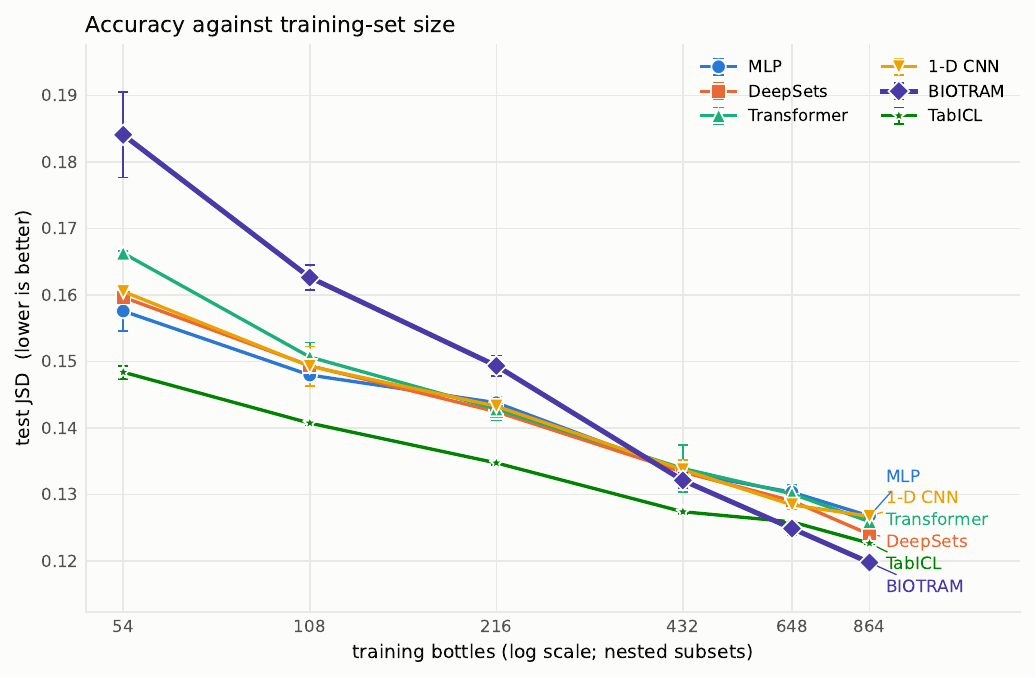}
\caption{\textbf{Accuracy against training data volume.} Test JSD of every model retrained
on nested subsets of the training bottles (log scale), each with the configuration tuned on the full
training set and at a single size: BIOTRAM's own for the four deep baselines, as in \cref{tab:main},
and \TabICLParams{} parameters for TabICLv2. Error bars show one standard deviation across
\DataVolSeeds{} seeds. With all \DataVolMax{} bottles, BIOTRAM reaches \DataVolBioMax{} against TabICLv2's
\DataVolTabICLMax{} ($p = \DataVolTabICLP$).}
\label{fig:datavolume}
\end{figure}

\mypar{Training-set size.} Every model is also retrained on nested subsets of the \DataVolMax{}
training bottles: the first $k$ bottles of one fixed random permutation, for \DataVolSteps{} values
of $k$ from \DataVolMin{} to \DataVolMax{}, so that each subset contains the smaller ones and all
models at a given $k$ see the same bottles. Validation and test bottles are unchanged, and everything
fitted on the training data, such as the feature standardisation or TabICLv2's context, is refitted on
the subset. Each model keeps a single size and the configuration tuned on the full training set
(\cref{fig:datavolume}).
Every curve is still falling at the full training set, and BIOTRAM's most steeply, from a JSD of
\DataVolBioMin{} to \DataVolBioMax{}: it is behind every other model up to \DataVolBehindTo{}
bottles, and from \DataVolDeepAheadFrom{} bottles on it is ahead of the four deep baselines, level
there with TabICLv2 and ahead of it on the full training set as well (\DataVolBioMax{} against
\DataVolTabICLMax{}, $p = \DataVolTabICLP$). Its structure therefore does not
make BIOTRAM more data-efficient at this size; it pays off once there are enough bottles to learn the
per-BIN quantities it relies on.

\section{How a read reaches the count}\label{app:elasticity}

\mypar{Read elasticity.} We measure how a change in a BIN's reads propagates to its predicted count,
as the elasticity $\partial \log \hat{n}_i^{(b)} / \partial \log \mathrm{reads}_i^{(b)}$. In the read-blind
variant, \cref{eq:inversion} passes a change in reads to each replicate estimate one for one
(measured: \ElastGenStackU{}), and the learned replicate merge takes back \ElastGenMergeU{}, for an
end-to-end elasticity of \ElastGenFinalU{} (\ElastGenSeeds{} seeds). The merge sees only the three
replicate estimates and the bottle's specimen total, so this discount cannot depend on the bottle's
total read count or on the other BINs. In the reported model, the amplification term also responds to the BIN's
own read: its derivative with respect to the BIN's own read inputs (log read count and read share)
has a per-seed mean absolute value between \JacOwnLo{} and \JacOwnHi{} over \JacSeeds{} seeds, at least
\JacRatioLo{} times its derivative with respect to another BIN's read (\JacCross{}). On average, when a
BIN's reads double, its predicted amplification rises with an elasticity of \ElastReadsModuleU{}, so
part of the extra reads is attributed to stronger amplification rather than to more individuals. The
replicate estimate then grows with an exponent of \ElastReadsStackU{} instead of one, the merge takes
back \ElastReadsMergeU{}, and the end-to-end elasticity is \ElastReadsFinalU{}, below the read-blind
variant's \ElastGenFinalU{}. With the read in view, the discount is larger and can vary between BINs
and with the bottle's total read count. Every baseline can fit such an interaction, since each row it receives
carries the log read count, the BIN's share and the bottle's total read count; the read-blind variant 
gives it up by construction.

\mypar{What the per-BIN read contributes.} Estimating the amplification without the BIN's own read
keeps the two sides of \cref{eq:rate} strictly separate, at a cost of \TrackJSDGap{} JSD
(\cref{tab:matched}). The read-blind variant is still ahead of \SepGenMAEbtlAhead{} of the matched
baselines and level with \SepGenMAEbtlTied{} on MAE per bottle, and ahead of \SepGenJSDAhead{} and
level with \SepGenJSDTied{} on JSD.

\mypar{Attention across BINs.} The amplification module is the only component in which BINs attend
to one another; it was designed to express competition between BINs for shared reagents and
sequencing budget. To test whether it uses this attention or mainly the BIN's own read, we make five comparisons,
each paired over six seeds. Three compare the module with one of identical width, depth and
feed-forward structure in which BINs do not attend to one another, varying which reads both see;
every variant receives the total read count of the three replicates. The other two remove that
total: one repeats the first comparison without it, the other measures what it adds to the module
without attention. When the BIN's own read and those of the other BINs are
visible, attention lowers JSD by \XbinAttnJSD{} (\XbinAttnSeeds{} seeds, $p = \XbinAttnP$; each $p$ 
in this paragraph is a paired $t$-test, Holm-corrected across the five paired comparisons). Masking only the BIN's own read removes this gain 
($\XbinSelfMaskedJSD$ JSD, \XbinSelfMaskedSeeds{} seeds, $p = \XbinSelfMaskedP$), and with no reads 
at all, attention has no measurable effect (\XbinNoReadsSeeds{} seeds, $p = \XbinNoReadsP$).
Without the replicate total read count, attention over BINs with reads changes JSD by $\XbinNoDepthJSD$
(\XbinNoDepthSeeds{} seeds, $p = \XbinNoDepthP$), and the total read count without attention changes it by
$\XbinDepthOnlyJSD$ (\XbinDepthOnlySeeds{} seeds, $p = \XbinDepthOnlyP$). 
Attention therefore pays off only when the module sees the BIN's own read, and most when it also sees
the replicate's total read count, which does nothing without attention. This suggests that it exploits
an interaction between the per-BIN reads and the total reads rather than competition between taxa.

\end{document}